Titre :
« Traduction commentée de *la Conférence Bakerienne sur le Mécanisme de l'Œil de Thomas Young (1800)* ».

Auteur :
MORIZOT Olivier,
Aix-Marseille Université, CNRS, Centre Gilles Gaston Granger, Aix-en-Provence, France

Coordonnées :
Olivier.morizot@univ-amu.fr

Date de création :
22 juillet 2021

**Traduction commentée en français de *La Conférence Bakerienne. Sur le Mécanisme de l'Œil* de Thomas Young (1800).**

Le problème posé par la capacité de l'œil à voir à différentes distances ne va pas forcément de soi. Je ne parle pas de la solution à ce problème, mais bien de l'urgence même de se considérer ce problème et de s'y confronter. De fait, nombreux sont les théoriciens de la vision qui ont légitimement pu estimer que l'œil était capable de voir nettement à différentes distances tout simplement parce qu'il en était capable, ou parce que quelque faculté psychique le suppléait dans ce processus [Helmholtz, 1867, I : 163 ; 169]. Dans les théories du rayon visuel, ou celui-ci à la rencontre des objets et dispose de la sensation de sa propre longueur, le problème ne peut manifestement pas se poser. Et probablement nécessite-t-il préalablement la découverte de l'image rétinienne [Kepler, 1604] afin de pouvoir se poser la question des conditions de sa netteté ; mais encore le problème peut-il être contourné en invoquant les spécificités ou perfection de l'appareil visuel humain.

Plus le fait que l'œil soit susceptible de voir nettement à différentes distances, c'est donc probablement le fait qu'il ne le soit pas toujours que nous puissions voir tantôt nettement, tantôt confusément un même objet immobile par rapport à nous – ou qu'en déplaçant le lieu de notre attention indépendamment du lieu de notre mise au point mise au point l'on soit capable de voir ce même objet tantôt simple, tantôt double – qui rend ce problème véritablement saillant et incontournable.

Or ce n'est en considérant à quel point ce problème aujourd'hui si évident est loin de l'avoir été de tout temps que l'on peut comprendre la portée et le sens du travail de Young sur la question. Car de fait, saillant, incontournable, extrêmement disputé et susceptible d'une multitude de solutions controversées et encore loin d'être départagées : voilà bien l'état dans lequel se trouvait le problème de l'accommodation de la vision à différentes distances à l'époque où Thomas Young présente et publie sa Conférence Bakerienne *Sur le Mécanisme de l'Œil* [Young, 1801]. Lui-même avait déjà abordé le sujet quelques années plus tôt dans le tout premier article qu'il avait soumis à la Royal Society [Young, 1793] ; mais depuis, la question avait été le sujet de trois conférences Crooniennes successives à la Royal Society[1] [Hunter, 1794 ; Home, 1795 ; Home, 1795], d'une multitude d'articles et ouvrages dont l'épaisse bibliographie de cet article témoignera, et elle serait encore le sujet le conférence Croonienne de l'année suivante [Home, 1802]. Ce que révèle toutefois la retour régulier de cette question comme sujet des conférences Crooniennes – et que confirme la comparaison du contenu de ces articles en général – c'est que plus qu'un problème de physique, d'optique, d'ophtalmologie ou d'anatomie, c'est un problème fondamental de myologie qui se pose à leurs auteurs, et auquel leurs lecteurs attendent une solution : le mystère qu'il s'agit de résoudre alors n'est en pas tant celui du mécanisme permettant l'accommodation, que celui de la découverte du ou des muscles qui la permettent, et de leur fonctionnement[2].

Ce qui fait que malgré la résolution optique exceptionnellement élaborée et convaincante du problème de l'accommodation de l'œil proposée par Young dans l'article que nous allons lire maintenant, la solution qu'il propose ne pourra être définitivement acceptable qu'avec la découverte du mécanisme musculaire qui la permet. Question à laquelle on le verra naturellement s'attacher ardemment, mais avec insuccès, en fin de ce texte ; mais qui ne pourra de toute façon être pleinement résolue qu'après la découverte des muscles lisses qui n'aura lieu qu'au milieu du siècle suivant [Kölliker, 1847] et le classement au nombre de ceux-ci de ce que l'on appelle aujourd'hui muscles ciliaires, responsables de la déformation du cristallin : avant cela les anatomistes, ophtalmologues, opticiens et physiciens avaient comme nous le verrons sous les yeux un muscle qu'ils ne savaient pas reconnaître. C'est probablement ce qui expliquera en partie le peu considération que reçut déjà ce texte de Thomas Young, malgré son extrême richesse. C'est en tout cas une hypothèse qui nous semble avoir le mérite de dépasser le mythe traditionnel du génie incompris de son temps, et au moins compléter l'opinion exprimée par Hermann von Hemholtz dans l'hommage à la fois éclairé et obscurci par plus d'un demi-siècle de distance qu'il rend à ces travaux dans la bibliographie de son propre travail sur l'accommodation : « Th. Young, On the Mecanism of the Eye. Travail fait avec une perspicacité et un esprit d'invention merveilleux, qui était parfaitement suffisant pour terminer le débat sur l'accommodation, mais souvent difficile à comprendre à cause de sa concision, et donc la lecture exige la parfaite connaissance de l'optique mathématique. » [Helmholtz, 1867, I : 170]. Gageons que plus qu'un défaut de concision – nous verrons que l'article qui suit est loin d'être court, surtout lorsqu'on le compare aux autres textes de l'époque sur la question –, ou de lourdeur mathématique – l'intégration au problème d'un traitement optico-mathématique est effectivement l'une des spécificités de ce texte de Young à l'époque, mais aussi complexe soit-il, il renforce bien plus considérablement la proposition qu'il ne l'affaiblit – le plus grand défaut de ce texte de Thomas Young est probablement qu'il détenait la bonne solution, mais pas le bon problème !

---

[1] Depuis 1738, la médaille Croonienne (du nom de William Croone qui en est le mécène original) récompense l'auteur des travaux les plus illustres de l'année réalisés spécifiquement sur la question des « mouvements locaux » ; c'est-à-dire sur le mouvement musculaire. Cette médaille s'accompagne d'une dotation, ainsi que d'une présentation de ces travaux lors d'une leçon, ou conférence (la conférence Croonienne), donnée en fin de chaque année devant la Royal Society ou le Collège Royal de Médecine et publiée en tout début d'année suivante.

[2] Voir par exemple la très longue série de thèses sur le sujet passée en revue en introduction de [Young, 1793, 169-172].

**Anatomie de l'œil**

Pour aborder sereinement la lecture de ce texte, probablement vaut-il mieux avoir quelques notions élémentaires d'anatomie de l'œil. Et pour cela il semble que la courte description fournie par Young lui-même en introduction de sa leçon de philosophie naturelle dédié à la vision puisse être une base suffisante et idéalement contextualisée, qui peut être rapprochée de la Figure 17 de la Planche IV :
« L'œil est un sphéroïde irrégulier, ne différant pas bien largement d'une sphère ; il est principalement composé de substances transparentes de densités réfractives variées, conçues pour collecter les rayons de lumière qui divergent depuis chaque point d'un objet en un foyer situé sur sa surface postérieure, qui est capable de transmettre à l'esprit une impression de la couleur et de l'intensité de la lumière, en même temps qu'une distinction de la situation du point focal déterminée par la position angulaire de l'objet.
La première réfraction a lieu à la surface de la cornée, ou cette membrane transparente qui se projette en avant du globe de l'œil ; mais le pouvoir réfringent de la cornée a peu d'effet, celle-ci étant d'épaisseur presque uniforme, et elle sert seulement à donner une forme convenable à l'humeur aqueuse qui remplit sa concavité et la distend. Cette humeur est partiellement divisée par l'uvée, ou iris, qui est de différentes couleurs chez les différentes personnes, présentant une perforation en son centre appelée pupille. Immédiatement derrière l'uvée, et connectés de très près à sa base, se trouvent les procès ciliaires, dont les sommets pendent comme une courte frange devant la lentille cristalline, une substance bien plus réfringente que l'humeur aqueuse et augmentant en densité en direction de son centre. La cavité restante est remplie d'un fluide aqueux, logé dans une structure cellulaire de membrane extrêmement fine, et appelé humeur vitrée. La rétine recouvre toute la partie postérieure de cette cavité ; elle est semitransparente et soutenue par la choroïde ou membrane chorioïde, une membrane noire ou brune très opaque, prolongeant l'uvée et les procès ciliaires : mais immédiatement là où la rétine est connectée au nerf optique, la choroïde est nécessairement perforée ; et en cette partie une petite portion de la rétine est presque insensible. Le tout est entouré par une prolongation opaque de la cornée appelée sclérotique » [Young, 1807, I : 447-448].

{P.23}[3] *La Conférence Bakerienne. Sur le Mécanisme de l'Œil. Par* Thomas Young, *M.D. F.R.S.*

Lue le 27 Novembre 1800.

I. En l'an 1793, j'ai eu l'honneur d'exposer devant la Royal Society quelques remarques sur la faculté par laquelle l'œil s'accommode à la perception d'objets situés à des différentes distances (Phil. Trans. for 1793, p. 169)[4]. L'opinion que j'entretenais alors, quoique n'elle n'eût jamais été placée exactement sous cette lumière, n'était ni aussi nouvelle ni aussi oubliée que je le supposais moi-même, de même que la plupart des personnes avec lesquelles j'avais pu m'entretenir sur le sujet. M. HUNTER, qui depuis longtemps avait élaboré une opinion similaire, était encore moins conscient d'avoir été précédé, et s'était engagé au moment de sa mort dans une étude des faits relatifs à celle-ci (Phil. Trans. for 1794, p. 21)[5] ; une étude pour laquelle, pour autant qu'elle concernait la physiologie, il était indubitablement qualifié. M. HOME, avec l'aide de M. RAMSDEN dont la Society ne peut que regretter la perte récente, poursuivit l'enquête que M. HUNTER avait commencée ; et les résultats de ses expériences semblaient réfuter l'hypothèse de la muscularité de la lentille cristalline de manière très satisfaisante (Phil. Trans. For 1795, p. 1)[6]. Je pensai donc qu'il m'incombait de me saisir de la première opportunité pour témoigner ma conviction de la justesse des conclusions de M. Home, que je mentionnai par conséquent dans une thèse publiée à {P.24} Gottingen en 1796 (De Corporis humani Viribus conservatricibus, p.68)[7], de même que dans un essai présenté l'an dernier à cette Société (Phil. Trans. For 1800, p. 146)[8]. Il y a trois mois environ, je fus amené à reprendre le sujet en passant en revue le papier du Dr Porterfield sur les mouvements internes de l'œil (Edim. Med. Essays, Vol. IV. p. 124)[9], et je fis très inopinément quelques observations qui, je pense pouvoir m'aventurer à le dire, paraissent conclure définitivement en faveur de mon opinion première, pour autant que cette opinion attribuait à la lentille de l'œil le pouvoir de changer sa forme. En même temps, je dois remarquer que toute personne s'étant trouvée engagée dans des expériences de cette nature aura conscience de l'extrême délicatesse et précaution requises non seulement pour les conduire, mais aussi pour en tirer des conclusions ; de même qu'elle admettra volontiers qu'aucune excuse n'est nécessaire pour les erreurs en lesquelles nombreux ont été induits, tout comme moi, au moment d'appliquer ces expériences à la détermination de propriétés optiques et physiologiques.

---

[3] On indiquera entre accolades le numéro de la page sur laquelle se trouve la partie du texte qui les suit, dans la version originale de la conférence publiée dans les Philosophical Transactions of the Royal Society of London, vol. 91, 1801, 23-88 [Young, 1801]. C'est cette version originale du texte lue, devant la Royal Society, qui sera traduite et commentée ici. Cette conférence est néanmoins de nouveau publiée dans [Young, 1807, II : 573-606] avec quelques modifications dont les plus significatives seront signalées. Par ailleurs, la version originale du texte se conclue sur une série de « corrections » [Young, 1801, 83] qui ont ici été directement implémentées dans le corps du texte, mais qui seront signalées par une note et entre deux symboles « / \ ».

[4] [Young, 1793]. Dans ce premier article soumis aux Philosophical Transactions et suite auquel il intègre la Royal Society, Young discute les spéculations antérieures relatives à la manière dont est produite l'accommodation. Puis il donne quelques arguments en faveur de l'idée qu'elle est le produit du changement de courbure du cristallin, permis par le fait que celui-ci serait un muscle. Le résultat immédiat de la publication de ce texte fut malheureusement une réclamation publique quant à la priorité de cette découverte par John Hunter. La rumeur circula même alors que Young se serait lancé dans ces recherches après avoir entendu parler des travaux de Hunter au cours d'un dîner mondain.

[5] [Hunter, 1794]. John Hunter (1728-1793) est un chirurgien britannique, considéré à son époque comme la principale autorité en matière de maladies vénériennes. Il est lauréat systématique de la médaille Croonienne de 1775 à 1782. Et Hunter s'apprête à la recevoir de nouveau en 1792, lorsqu'il décède. Les travaux qu'il menait alors portaient sur l'accommodation de l'œil et reposaient sur l'idée qu'une excitation du cristallin d'un animal fraîchement mort trempé dans une eau chaude mettrait en évidence sa capacité à se contracter, comme le font d'autres muscles ; et prouverait dès lors sa muscularité et le rôle de sa contraction dans l'accommodation. Ils se déroulent donc à peu près simultanément avec les premiers travaux de Young sur la question [Young, 1793]. Les travaux de Hunter, bien qu'inachevés, sont néanmoins publiés dans une version très sommaire en 1794 par Everard Home, frère de son épouse ; qui prolongera lui-même ces travaux par la suite et n'aura de cesse de défendre la primauté de la découverte de Hunter sur Young.

[6] [Home, 1795]. Everard Home (1756-1832) est lui-même chirurgien et lauréat à son tour de la médaille Croonienne tous les ans de 1790 à 1801 (sauf 1791, 1792 et 1797). Comme annoncé, il poursuivra les travaux de Hunter sur le cristallin avec l'aide de Jesse Ramsden (1735-1800), constructeur d'instruments de précision (en particulier optiques), et lui aussi membre de la Royal Society. Mais contrairement à ce à quoi l'on aurait pu s'attendre, tous deux concluent que le cristallin, bien que de constitution fibreuse comme l'annonçait Hunter, n'était pas un muscle, ne changeait pas de forme, et ne jouait aucun rôle dans l'accommodation. Cette conclusion reposant en bonne partie sur l'étude qu'ils avaient réalisée sur un jeune patient du nom de Benjamin Clerk (auquel il sera fait référence plus tard dans ce texte), dont le cristallin de l'œil droit seulement avait été retiré par Home pour le soigner d'une cataracte, et dont une série d'expériences semblaient indiquer qu'il était toujours en mesure d'accommoder avec cet œil – et même mieux qu'avec son œil sain. Ces observations menèrent Ramsden et Home à considérer que l'hypothèse d'un quelconque rôle du cristallin dans

II. Outre l'investigation concernant l'accommodation de l'œil à différentes distances, j'aurai l'occasion d'attirer l'attention sur quelques autres particularités relatives à ses fonctions ; et je commencerai par une réflexion générale sur le sens de la vision. J'énumèrerai ensuite quelques propositions dioptriques utiles à mes propos et décrirai un instrument pour déterminer facilement la distance focale de l'œil. Sur ces fondations, j'étudierai les dimensions et les pouvoirs réfringents de l'œil humain dans son état de repos ; et la forme et la grandeur[10] de l'image qui se dessine sur la rétine. J'examinerai ensuite l'amplitude des changements auxquels l'œil est soumis et le degré d'altération de ses proportions qui sera nécessaire pour ces changements, sur la base des principales hypothèses qui {P.25} méritent comparaison. Je poursuivrai en relatant une variété d'expériences qui paraissent être ces plus propres à décider de la vérité de chacune de ces hypothèses, et en examinant les arguments qui ont été mis en avant contre l'opinion que je tenterai de confirmer. Et je conclurai par quelques illustrations anatomiques de la faculté des organes de diverses classes d'animaux de remplir les fonctions qu'on leur attribue.

---

l'accommodation était définitivement invalidée, et que celle-ci était alors nécessairement due à un changement de courbure de la cornée – dont la suite de leurs expériences tente de rendre compte [Home, 1795, 10-19]. Il est à noter que la conférence Croonienne donnée par Home l'année suivante –mentionnée plus loin – sera elle aussi dédiée à la démonstration du rôle de la cornée dans l'accommodation [Home, 1796]. Et surtout que la dernière qu'il donnera, fin 1801 – soit précisément un an après la lecture par Young du présent texte – sera explicitement intitulée *Sur le pouvoir de l'œil de s'ajuster à différentes distances, même dépourvu de lentille cristalline* [Home, 1802].

[7] [Young, 1796, 68]. Young profite en effet de la publication de sa thèse en 1796 pour signaler que sa théorie de la muscularité du cristallin n'est « ni nouvelle, ni vraie ».

[8] [Young, 1800, 146-147]. Young conclut effectivement cet article sur les propriétés du son et de la lumière en rappelant avoir abandonné sa théorie du cristallin dès l'instant où la preuve de son invalidité avait été apportée par Home ; et en déclarant s'apprêter à faire de même si ses conclusions sur le son et la lumière venaient également à être réfutées. Mais le présent article montrera que l'histoire ne s'est pas arrêtée là.

[9] [Porterfield, 1738]. William Porterfield (vers 1696-1771), bibliothécaire, fut aussi président du Collège Royal des Médecins d'Édimbourg. Il publia en 1737 et 1738 deux longs articles sur les mouvements de l'œil ; le premier sur ses mouvements externes, le second sur ses mouvements internes, parmi lesquels il inclut l'accommodation, terme qu'il forge à l'occasion. Ses expériences mettent en œuvre un optomètre – instrument destiné à mesurer entre autres l'amplitude d'accommodation – auquel Young fera référence plus tard dans ce texte.

[10] Le mot employé ici en anglais étant « magnitude ». Il reviendra plusieurs fois dans le texte et nous le traduirons systématiquement par « grandeur », plus indéterminé que les mots « tailles » communément employés aujourd'hui en optique géométrique pour traduite les termes « size » ou « height ».

III. On suppose généralement que l'œil est de loin le mieux compris de tous les sens externes ; cependant ses pouvoirs sont si compliqués et si diversifiés que nombre d'entre eux n'ont pas été examinés jusqu'à présent ; et qu'au sujet des autres, beaucoup de recherche laborieuse a été réalisée en vain. Certes, il ne peut être nié que nous sommes capables d'expliquer l'usage et le fonctionnement de ses différentes parties de manière beaucoup plus satisfaisante et intéressante que ceux de l'oreille, qui est le seul organe qui puisse lui être strictement comparé ; Car dans l'odorat, le goût et le toucher les objets parviennent presque sans préparation en contact immédiat avec les extrémités des nerfs ; et la seule difficulté est de concevoir la nature de l'effet qu'ils produisent et sa communication au sensorium[11]. Mais l'œil et l'oreille sont seulement des organes préparatoires, conçus pour transmettre les impressions de la lumière et du son à la rétine et à la terminaison du délicat nerf auditif. Dans l'œil, la lumière est transmise jusqu'à la rétine sans aucun changement dans la nature de sa propagation : dans l'oreille, il est très probable que les petits os ne transmettent pas les vibrations du son par le mouvement successif des différentes particules d'un même milieu élastique, mais comme des corps passifs durs et inélastiques obéissant au même instant dans toute leur étendue aux mouvements de l'air. Dans l'œil, nous jugeons très précisément de la direction de {P.26} la lumière d'après la partie de la rétine sur laquelle elle frappe. Dans l'oreille, nous n'avons d'autre critère que la faible différence de mouvement dans les petits os, selon la partie du tympan que le son, concentré par différentes réflexions, heurte en premier ; dès lors, l'idée de direction est nécessairement très indistincte et il n'y a pas de raison de supposer que différentes parties du nerf auditif soient exclusivement affectées par des sons provenant de directions différentes. [12]<En supposant que l'œil soit capable de transmettre une idée distincte de deux points sous-tendant un angle d'une minute, qui est peut-être l'intervalle le plus petit auquel deux objets peuvent être distingués, bien qu'une ligne dont l'épaisseur sous-tend un dixième de minute seulement puisse être parfois perçue comme un seul objet ; Selon cette supposition, il doit y avoir environ 360 mille points sensitifs pour un champ visuel de 10° de diamètre, et plus de 60 millions pour un champ de 140°.[13] Mais du fait de la sensibilité variable de la rétine, que l'on expliquera plus tard, il n'est pas nécessaire de supposer qu'il y a plus de 10 millions de points sensitifs, non plus qu'il puisse aisément y en avoir moins d'un million : le nerf optique peut par conséquent être considéré comme étant constitué de plusieurs millions de fibres distinctes[14]. Par une expérience un peu grossière, je trouve que je peux distinguer deux sons similaires, provenant de deux points sous-tendant un angle d'environ cinq degrés. Mais l'œil peut discriminer environ 90 mille points différents dans un espace sous-tendant cinq degrés en chaque direction. Il y a plus de mille espaces de la sorte dans une hémisphère : de sorte que l'oreille est capable de transmettre des impressions provenant d'environ mille directions différentes. Cependant, ce n'est pas dans tous les cas que l'oreille dispose d'une si belle discrimination des directions : la raison de cette différence entre l'œil et l'oreille est évidente ; chaque point de la rétine n'a que trois couleurs principales à percevoir, puisque les autres sont probablement composées de diverses proportions de ces trois[15] ; mais comme il y a plusieurs milliers ou millions de variétés de sons dans chaque direction, il était impossible que le nombre de directions qu'il est possible de distinguer soit bien grand.> Chaque point sensible de la rétine est capable de recevoir des impressions distinctes, autant de la couleur que de la puissance[16] de la lumière[17] ; mais il n'est pas absolument certain que toutes les parties du nerf auditif soient capables de recevoir l'impression de chacun des tons beaucoup plus divers que l'on peut distinguer ; bien qu'il soit extrêmement probable que toutes les différentes parties de la surface exposée au fluide du vestibule soient plus ou moins affectées par tous les sons, mais à différents degrés et moments selon la direction et la qualité de la vibration. Il n'est pas aisé de déterminer si en toute rigueur nous pouvons, oui ou non, entendre deux sons ou voir deux objets au même instant ; mais il est suffisant que nous puissions faire les deux sans interruption de quelque intervalle de temps perceptible à l'esprit. Et en effet, nous ne pourrions former aucune idée de grandeur, sans perception comparative, et par conséquent presque simultanée, de deux parties ou plus du même objet. L'étendue du champ de vision parfaite pour chaque position de l'œil n'est certainement pas très grande ; mais il apparaitra par la suite que ses pouvoirs réfringents sont conçus pour saisir une vue modérément distincte de tout un hémisphère : Le sens de l'ouïe est uniformément parfait dans presque toutes les directions.

---

[11] On remarquera que la comparaison « stricte » que Young s'apprête à développer entre l'oreille et l'œil – ou pour commencer entre l'ouïe et la vision, à l'exclusion des autres sens – fait directement écho à la comparaison entre le son et la lumière qu'il a développée dans [Young, 1800, 125-130] ; mais aussi qu'elle reprend exactement l'axe argumentatif emprunté par Leonhard Euler dès la première page de sa *Nouvelle Théorie de la Lumière et des Couleurs* [Euler, 1746, 169-170].

[12] Tous les passages signalés dans cette traduction entre <> n'apparaissent que dans la version de 1807 du texte. En l'occurrence on trouvera le long passage qui suit à la page : [Young, 1807, II : 575].

[13] 1 minute correspond à un soixantième de degré. Et cette valeur est effectivement celle qui est aujourd'hui communément admise comme quantifiant le pouvoir de résolution moyen de l'œil humain. Par conséquent, Young associe cet écart angulaire à la distance entre deux points sensibles contigus de la rétine et évalue logiquement que l'on devrait avoir 600 points de la sorte sur un arc de 10°, 360000 sur une section carrée de la rétine de 10° par 10° de côté, et environ 70 millions sur toute l'étendue du champ visuel, estimé à 140° pour un œil. On estime aujourd'hui que la *macula* – partie centrale de la rétine d'environ 10° de diamètre angulaire où la résolution angulaire est maximale – contient quelques 7 millions de photorécepteurs (essentiellement des cônes), et que la totalité de la rétine en compte 120 millions environ (essentiellement des bâtonnets). Signe que l'espacement entre deux photorécepteurs contigus n'est pas le critère limitant la résolution angulaire de l'œil.

[14] Les photorécepteurs étant connectés dans la rétine même à des cellules ganglionnaires intégrant l'activité d'un plus ou moins grand nombre d'entre eux, le nerf optique n'est au final constitué que d'un million de fibres nerveuses environ.

---

[15] Ce que suggère ici Young, c'est que contrairement à l'idée de Newton selon laquelle il la lumière solaire serait composée de sept types de rayons, causant chacun une sensation colorée différente, et dont les mélanges produiraient toute la diversité des autres couleurs [Newton, 1730, 134-137], il n'y aurait que trois types de lumière « colorée », qui suffiraient à produire toutes les autres sensations colorées par leurs mélanges. Le sens de cette affirmation – directement liée à la structure de la rétine supposée par Young – sera quelque peu développé dans [Young, 1802a] et [Young, 1802b].

[16] Ici le mot employé par Young est « strength » régulièrement employé dans ses autres textes sur la lumière et dont l'homonymie du mot *puissance* choisi ici pour le traduire avec le concept de la physique moderne représentant une énergie divisée par un temps ne doit pas nous faire croire que Young attribuait le même sens à ce mot. L'idée vague d'une grandeur qui quantifierait pour Young l'intensité, l'amplitude, la force ou la puissance d'une onde lumineuse est régulièrement appelée « strength » et elle est encore très flottante.

[17] Malgré la grande variété et la précision de ses expériences, Young n'a manifestement pas remarqué l'incapacité de l'essentiel de la surface de la rétine à produire des sensations colorées ; à savoir de toute la rétine à l'exclusion de la *macula* située au centre de la rétine, et qui n'occupe finalement qu'un champ visuel de moins de 10° de diamètre. C'est probablement que n'ayant pas même envisagé cette possibilité, il n'était évidemment pas en mesure d'élaborer d'expérience susceptible de la mettre évidence, ni même d'être attentif à de potentiels signes pouvant la trahir.

{P.27} IV. PROPOSITIONS DIOPTRIQUES.[18]

*Proposition* I. *Phénomène.*

En toute réfraction le rapport du Sinus de l'angle d'incidence au Sinus de l'angle de réfraction est constant[19] (NEWTON, Opt. I, Ax. 5., Opt. de SMITH, 13., Opt. de WOOD, 24.)[20].

*Scholie 1*. Nous l'appellerons le rapport de $m$ à $m\pm1$ et $m\pm1$, $n$.[21] Pour les réfractions de l'air dans l'eau, $m$ = 4 et $n$ = 3 ; depuis l'air dans le verre, le rapport est approximativement de 3 à 2.

---

[18] Les « propositions dioptriques » qui suivent (I à VIII) sous toutes là pour préparer la résolution géométrique ultérieure de certains des problèmes posés par l'œil. Elles sont absentes de la réédition de cette conférence dans le cours de philosophie naturelle de Young [Young, 1807, II : 573-606]. En partie parce que leur contenu se trouve dans le corps même du cours [Young, 1807, II : 70-76 ; 80-81]. On pourra trouver un équivalent des plus élémentaires de ces propositions – dans une formulation plus actuelle et pédagogique, mais aussi plus longue – dans [Morizot, 2016, 45-68 ; 93-117]. Elles sont aussi retranscrites en notations plus proches de celles employées actuellement dans la traduction française de ce texte réalisée par Marius Tscherning [Tscherning, 1894, 90-112].

[19] Il s'agit ni plus ni moins que de la loi de la réfraction, aujourd'hui communément écrite sous la forme $n.\sin i = n'.\sin r$ ; $i$ étant l'angle d'incidence mesuré entre le rayon incident et la droite orthogonale à la surface séparant les deux milieux au point d'impact du rayon lumineux considéré ; $r$ étant l'angle de réfraction mesuré entre le rayon réfracté et cette même droite normale ; $n$ et $n'$ étant les indices de réfraction de la lumière respectivement dans le milieu depuis lequel arrive la lumière et dans le milieu dans laquelle elle est transmise, et qui sont chacun égaux au rapport de la vitesse de la lumière dans le vide $c$ à la vitesse de la lumière $v$ dans le milieu considéré : $n = \frac{c}{v}$. Pour sa part Young écrit cette même loi sous la forme $\frac{Sin\ i}{Sin\ r} = constante$, soit de la même manière que l'écrivait Descartes lui-même [Descartes, 1637, 101] ; tant du fait que la formulation originale cartésienne ne s'exprime pas immédiatement sur la nature de la constante et l'envisage plus comme étant caractéristique du couple de milieux traversés que de leur propriétés individuelles [Young, 1807, II : 70-71] ; que du fait de l'utilisation d'un rapport de longueurs : le Sinus de l'angle $i$ étant la longueur de la corde sous-tendue par l'angle $i$ dans un cercle de rayon R, soit $Sin\ i = R.\sin i$. La valeur de R n'a pas à être donnée ici, dans la mesure où $Sin\ r = R.\sin r$ pour une valeur de R implicitement supposée identique. Young ne s'exprimant pas immédiatement sur la nature de cette constante, puisque celle-ci est encore alors sujette à débat, comme le montre les lignes qui vont suivre.

[20] A noter que Newton ajoute dans sa formulation de la loi de la réfraction [Newton, 1730, 5-8] que la valeur de cette constante pour un passage de l'air à l'eau ou de l'air au verre, dans le cas précis d'une « lumière rouge » ; précisant que ce rapport est très légèrement différent pour des « lumières d'autres couleurs ». Newton présente cette loi comme un axiome (le cinquième) de l'*Opticks*, ce qui lui permet de laisser en suspens le problème consistant à justifier la cause de cette loi. Problème pourtant jugé central par Huygens, qui dès lors s'efforcera de le résoudre par un modèle vibratoire [Huygens, 1690, 33-36], de même que le fera Euler [Euler, 1746, 211-213]. Young surenchérit pourtant dans un premier temps sur cette proposition en citant les formulations très similaires à celles de Newton qu'en font deux autres auteurs britanniques : Robert Smith (1689-1768) [Smith R., 1738, I : 3] et James Wood (1760-1839) [Wood J., 1799, 9-10]. Ces deux ouvrages sont des manuels d'optique de référence, destinés principalement aux étudiants des universités anglaises. Profondément ancrés dans une interprétation de l'optique newtonienne qui associe la lumière à un flux de corpuscules dont les déviations sont expliquées par des forces s'exerçant à courte distance, ces textes ont pour mérite d'illustrer à la fois la manière dont le monde académique anglais a intégré les propositions optiques de Newton en enseignant un corpus dominé sans partage par la théorie des projectiles [Cantor, 1983], et le fait que le contenu officiel de l'optique a très peu évolué depuis les découvertes newtoniennes de la fin du siècle précédent.

[21] $m$, $m\pm1$ et $n$ désignent trois valeurs numériques qui peuvent prendre des valeurs quelconques, mais dont les rapports sont égaux à la constante présente dans la loi des Sinus ; leurs valeurs varieront donc avec les milieux considérés. En somme, en appelant $i$ l'angle que fait le rayon incident avec la droite normale au dioptre au point d'impact et $r$ l'angle que fait le rayon réfracté avec cette même droite, Young propose d'écrire : $\frac{Sin\ i}{Sin\ r} = \frac{m}{m\pm1} = \frac{m}{n}$. Par cette formulation en deux temps, Young suggère sans le démontrer que la constante à laquelle on a identifié le rapport des Sinus depuis Descartes, peut toujours être écrite comme le rapport de deux nombres différents d'une unité seulement : c'est pourquoi il l'écrit $\frac{m}{m\pm1}$ dans un premier temps, pour remplacer immédiatement et

*Scholie 2*. Selon Barrow (*Lect. Opt.* ii, 4.), Huygens, Euler (*Conject. phys. circa prop. soni et luminis, Opusc. t. ii*) et l'opinion que j'ai récemment soumise à la Royal Society (Phil. Transactions 1800, p. 128), la vitesse de la lumière est d'autant plus grande que le milieu est rare[22] : d'après Newton (Scol. Prop. 96. l. i. Princip. Prop. 10. p. 3. l. ii. Opt.) et la doctrine la plus généralement admise, c'est le contraire[23]. Dans les deux hypothèses, elle est toujours la

---

définitivement $m\pm1$ par la lettre $n$, uniquement pour plus de commodité d'écriture dans la suite du texte. Pour les deux exemples indiqués ensuite, qui sont le passage de l'air à l'eau ($m$ = 4 et $n$ = 3) et de l'air au verre ($m$ = 3 et $n$ = 2), $m$ est à chaque fois supérieur d'une unité à $n$, mais le cas inverse se présente évidemment dès que l'on considère le trajet inverse de l'eau vers l'air ou du verre vers l'air. Ce choix n'est pas arbitraire ; il relève certainement de l'observation d'une série de cas de réfraction simples (réfraction air-eau, air-verre...) dont Young tente de faire émerger par induction une règle mathématique générale ; mais une règle qui est manifestement aussi imprégnée néo-pythagorisme, en ce qu'elle doit être aussi simple que possible, et pouvoir être écrite comme un rapport de nombres entiers– en cela le rapport de deux entiers successifs était un candidat presque trop beau pour ne pas être retenu (même si l'on verra que la nécessité qu'il s'agisse de nombres entiers sera abandonnée dès la prochaine proposition). Ce faisant, Young suggère néanmoins que le rapport des Sinus ne peut qu'être inclus entre les valeurs 1/2 et 2, ce que l'on sait aujourd'hui incorrect. Mais peu importe ; il faut bien comprendre en tout cas que les grandeurs $m$ et $n$ n'ont pas vocation ici à représenter ce que nous appelons aujourd'hui les indices de réfraction des deux milieux traversés ; mais que le fait d'exprimer le rapport des Sinus par celui de deux nombres $m$ et $n$ n'est pour lui, on le verra, qu'un procédé pratique pour faciliter le calcul de certaines réfractions, notamment celles présentées en Propositions VII et VIII.

[22] Ce mot « rare », doit être entendu dans les textes de Young présentés ici au sens de « subtil », ou « très peu dense ».

[23] Isaac Barrow (1630-1677) est un théologien, philologue et mathématicien anglais qui occupa la chaire lucasienne de mathématiques à Cambridge, avant de laisser la place à son ancien élève Isaac Newton. Il y développe une série de cours, dont des leçons d'optique qui seront publiées en 1669 et dans lesquelles il explique notamment la réfraction de la lumière passant d'un milieu moins dense vers un milieu plus dense [Barrow, 1669, 14-16]. Le « rayon » lumineux de Barrow est un objet tridimensionnel, composé de corpuscules matériels, dont le front de propagation doit toujours être orthogonal aux côtés : dans un milieu homogène cette sorte de faisceau peut donc ressembler à un cylindre long et fin, qui peut cependant se déformer au contact de surfaces réfléchissantes ou réfringentes. Il se trouve que Barrow suppose que la lumière se déplace plus lentement dans les milieux denses, car ces milieux s'opposent plus à la propagation des corpuscules de lumière ; et le passage progressif des corpuscules du faisceau vers un milieu où leur vitesse est plus faible, alors qu'une autre partie du faisceau – se trouvant toujours dans le premier milieu – continue à se propager à grande vitesse, entraine une rotation sans déformation du front de propagation, s'accompagnant nécessairement d'une courbure locale du faisceau lumineux, qui se conclue donc en sa réfraction en direction de la normale. Pour sa part, Christian Huygens justifie que les vibrations de l'éther – qui est pour lui le substrat de la propagation de la lumière – se propagent plus lentement dans les corps les plus denses par l'idée que cet éther pénètre la matière, mais que les particules de celle-ci imposent « des petits détours » aux vibrations lumineuses, ralentissant forcément leur progression [Huygens, 1690, 30]. Leonhard Euler considère quant à lui que les vibrations de l'éther se communiquent plus ou moins bien à la matière et que c'est cela qui différencie un corps opaque ou réfléchissant d'un corps transparent. Les particules d'un corps transparent vont donc vibrer sous le coup des vibrations de l'éther, mais d'une vibration qui se propagera plus lentement du fait d'une densité et d'une élasticité différentes de celle de l'éther lui-même [Euler L., 1750, 11]. Pour Young enfin, l'idée que les particules des corps matériels elles-mêmes vibreraient ne tient pas. Rappelons donc qu'il défend l'idée que l'éther pénètre les corps matériels. Mais que du fait d'une attraction qu'exerce la matière sur l'éther, la densité de ce-dernier dans un corps matériel est d'autant plus grande que ce corps est dense. C'est cette augmentation de la densité d'éther qui justifie mécaniquement selon lui une vitesse de la lumière plus faible dans les corps denses [Young, 1800, 128]. Au contraire Newton, dans ses *Principia* [Newton, 1687, 231-232] comme dans son *Opticks* [Newton, 1730, 245-251], soutient que la lumière est composée de corpuscules se propageant plus rapidement dans les milieux denses que dans le vide ou dans l'air. Et Young souligne qu'il s'agit encore de l'opinion la plus généralement admise en 1800. Probablement parce que ce choix permet de justifier de la réfraction de la lumière par la mécanique newtonienne : elle est la conséquence de l'action d'une force de très courte portée, perpendiculaire à la surface séparant les milieux et s'exerçant directement sur les corpuscules lumineux. Ainsi, le passage de la lumière de l'air à l'eau produit une trajectoire se rapprochant de la normale qui est comparable à celle qu'aurait un corps soumis à une force attractive ; Newton en conclue que la lumière est accélérée en pénétrant dans l'eau,

même dans le même milieu et varie avec le rapport des Sinus des angles. Cette circonstance peut être utilisée pour faciliter le calcul de quelques réfractions très compliquées.

*Proposition* II. *Phénomène.*

Si l'on interpose entre deux milieux réfringents un troisième milieu limité par des surfaces parallèles, la réfraction globale reste inchangée (Opt. de Newton, l. i. p. 2. Prop. 3. Smith. r. 399. Wood, 105).[24]

*Corollaire.* D'où, si les réfractions depuis deux milieux différents dans un troisième sont données, on peut trouver par conséquent la réfraction à la surface commune entre ces deux milieux. Soient les réfractions données {P.28} comme $m : n$ et comme $m^1 : n^1$ ; alors le rapport recherché sera celui de $mn^1 : m^1n$. Par exemple soient les trois milieux le verre, l'eau et l'air ; alors $m = 3$, $n = 2$, $m^1 = 4$, $n^1 = 3$, $mn^1 = 9$ et $m^1n = 8$. Si les rapports sont 4 : 3 et 13 : 14, nous avons $mn^1 : m^1n :: 39 : 56$ ; et en divisant par 56 − 39 on obtient 2,3 et 3,3 pour $m$ et $m+1$, dans la Schol. 1, Prop. 1.[25]

---

et donc qu'elle s'y propage plus rapidement que dans l'air. Ce qui est intéressant ici, c'est bien entendu que l'hypothèse d'une décélération ou d'une accélération de la lumière lors de son passage de l'air à l'eau est invérifiable à l'époque où Young écrit ces mots. Mais que d'aucuns ayant eu le sentiment que la première option était une conséquence nécessaire d'un modèle vibratoire (comme ceux de Huygens, Euler et Young) et la seconde une conséquence nécessaire d'un modèle corpusculaire (comme celui de Newton), il a été envisagé qu'une mesure des vitesses respectives de la lumière dans l'air et dans l'eau pourrait constituer une expérience cruciale permettant de trancher définitivement entre ces deux modèles [Foucault, 1853]. Or non seulement Pierre Duhem a-t-il exposé les raisons épistémologiques pour lesquelles une tel espoir avait été vain [Duhem, 1914, 285-289] ; mais le fait historique, mis en évidence ici, que la théorie de Barrow était une sorte de théorie des projectiles envisageant un ralentissement de la lumière passant de l'air à l'eau, ou que la théorie vibratoire de Hooke considérait pour sa part une accélération de celle-ci [Hooke, 1665], montrait déjà bien la versatilité des théories scientifiques et leur capacité à combiner modèles et hypothèses sans qu'aucune combinaison ne revête de nécessité absolue.

[24] Le phénomène décrit ici est que la direction d'un rayon réfracté dans ne change pas si l'on interpose une épaisseur quelconque d'un troisième milieu (à faces parallèles) entre les deux milieux initialement considérés. Cet effet se démontre très simplement avec la formulation moderne de la loi de la réfraction $n.\sin i = n'.\sin r$, où un indice de réfraction est associé en propre à chaque milieu. Mais la démonstration n'est pas triviale quand le rapport des sinus est considéré comme dépendant des deux milieux conjointement et définissant donc leur relation. C'est pourquoi Young présente la chose comme un phénomène dont il déduira seulement ensuite le corollaire qui suit, selon lequel les valeurs des constantes donnant $\frac{\sin i}{\sin r}$ étant connues et égales à $a$ et $b$ pour les réfractions de deux milieux différents 1 et 2 vers un milieu 3, il est possible de déduire que la constante est égale à $\frac{a}{b}$ pour une réfraction directe du milieu 1 vers le milieu 2, c'est-à-dire au rapport des deux constantes connues. Le calcul est simple et nécessite seulement d'exprimer la loi de la réfraction dans les deux cas et de faire le produit des deux expressions obtenues – étant entendu que la réfraction de 3 vers 2 sera régie par la constante $\frac{1}{b}$. Ceci dit, ce calcul n'est conclusif que si l'on reconnait le phénomène décrit en premier lieu selon lequel la direction de sortie d'un rayon traversant trois milieux successifs séparés par des surfaces parallèles reste inchangée si l'on retire le milieu intermédiaire. C'est donc un procédé similaire qu'emploie Newton en partant de l'observation expérimentale de la direction et de la blancheur conservée d'un rayon lumineux traversant successivement divers milieux pour en déduire le même théorème sur les réfractions composées dans la référence citée par Young [Newton, 1730, 112-113]. C'est enfin assez logiquement ce même cheminement déductif que reprennent Smith [Smith R., 1738, I : 158-159] et Wood [Wood J., 1799, 50-52].

[25] Young retombe donc sur le résultat annoncé dans la scholie 1 en remarquant qu'en multipliant le produit de deux rapports de deux nombres $m$ et $m\pm 1$ séparés d'une unité seulement – l'un compris entre 1 et 2 et l'autre en ½ et 1 – on obtient un nouveau rapport $a : b$ forcément inclus entre ½ et 2, que l'on peut donc toujours parvenir à exprimer comme un nouveau rapport de deux nombres séparés par une unité seulement, à condition de ne pas considérer uniquement des rapports de nombres entiers. Ce en multipliant le numérateur et le

*Proposition* III. *Problème*. (Plaque II. Fig. 1.)

Au sommet d'un triangle donné (CBA), placer une surface réfringente donnée (B) de façon que les rayons incident et réfracté puissent coïncider avec les côtés du triangle (AB et BC).[26]

Nommons ces côtés $d$ et $e$, /la base étant prise pour unité\[27] ; puis prenons dans la base, du côté de $d$ (ou AB), une partie (AE) égale à $\frac{nd}{nd+me}$, ou (AD=) $\frac{md}{md+ne}$ ; traçons une ligne (EB, ou DB) jusqu'au sommet, et la surface doit être perpendiculaire à cette ligne, chaque fois que le problème est physiquement possible. Quand $e$ devient infini et parallèle à la base, prendre $\frac{nd}{m}$ ou $\frac{md}{n}$ du côté de $d$ comme intersection du rayon de courbure[28].

*Proposition* IV. *Théorème*. (Fig. 2.)[29]

Dans le cas des réfractions obliques sur des surfaces sphériques, la droite (AI ou KL) joignant les foyers conjugués (A, I ou K, L) passe par le point (G) où la droite (EF), bissectant les cordes (BC, BD) découpées par la surface sphérique dans les rayons incidents et réfractés, rencontre sa perpendiculaire passant par le centre (H).[30]

---

dénominateur conjointement par leur différence ($b$ - $a$), puisque $\frac{\frac{b}{b-a}-1}{\frac{b}{b-a}} = \frac{\frac{a}{b-a}}{\frac{b}{b-a}} = \frac{a}{b}$. On remarquera plus loin que les rapports 4/3 et 13/14 choisis comme exemple par Young correspondent aux valeurs du rapport $m/n$ pour les réfractions successives de l'air vers l'humeur aqueuse, puis de l'humeur aqueuse au cristallin.

[26] Le problème géométrique décrit ici consiste donc à déterminer la forme de la surface réfringente, étant donnés les directions des rayons incident et réfracté, la position du point d'incidence et la valeur $m/n$ du rapport des Sinus. Le problème étant destiné à la détermination des réfractions par des lentilles ou des dioptres courbes, il est sous-entendu que la forme de la surface réfringente est sphérique. Et que ce sont donc la courbure et le centre D ou E de la sphère qui sont à déterminer ici, selon qu'elle est supposée convexe ou concave par rapport à la direction d'incidence du rayon lumineux. La solution de ce problème, donnée dans le paragraphe qui suit, permettra à Young d'estimer l'aberration sphérique de l'œil, ainsi que la courbure des faces du cristallin accommodé à partir de la détermination des positions relatives des points conjugués A et C pour des rayons pénétrant dans le cristallin en différents points B de sa surface.

[27] Correction demandée dans [Young, 1801, 83].

[28] En s'appuyant sur la Figure 1 et en posant avec Young AB = $d$, BC = $e$ et AC = 1 ; puis en supposant le problème effectivement résolu et menant aux expressions de AD et AE données par Young comme positions par rapport à A des deux centres de courbure possibles D et E pour résoudre le problème (en effet DB et EB sont dits perpendiculaires à la surface réfringente en B) ; on peut aisément vérifier la véracité de la solution. En notant que par la loi des sinus : $\frac{AB}{\sin BDA} = \frac{AD}{\sin(\pi-i)} = \frac{AD}{\sin i}$ et $\frac{BC}{\sin BDA} = \frac{BC}{\sin(\pi-BDA)} = \frac{AC-AD}{\sin r}$. Donc que $\frac{\sin i}{\sin r} = \frac{AD.BC}{AB.(AC-AD)} = \frac{AD.e}{d.(1-AD)} = \frac{\frac{md.e}{md+ne}}{\frac{d.ne}{md+ne}} = \frac{m}{n}$, ce qui démontre que la solution est correcte. Le même raisonnement vaut pour vérifier la solution donnée pour AE. Et dans le cas où $e$ tendrait vers l'infini, donc où l'on chercherait la forme de la surface sphérique passant par B dont A serait le foyer, la nouvelle expression donnée pour AD peut être vérifiée en remarquant que l'angle $BDA = r$, donc que $\frac{AB}{\sin BDA} = \frac{AD}{\sin i}$ implique $\frac{\sin i}{\sin r} = \frac{AD}{AB} = \frac{md}{n.d} = \frac{m}{n}$.

[29] Cette proposition part d'un théorème très général sur la réfraction des rayons lumineux parvenant obliquement sur une surface réfringente sphérique, pour élaborer ensuite toute une série de résultats d'optique géométrique relatifs aux surfaces réfringentes sphériques (Corollaires 1 à 6, puis 8 et 9), puis aux lentilles minces sphériques (Corollaires 7, 10, 11). Ils serviront par la suite de support au calcul des réfractions par la cornée, le cristallin, par chacune des faces de celui-ci prise séparément, ou par l'œil pris dans sa globalité.

[30] Ce théorème est démontré dans [Young, 1807, II : 74]. Il offre un moyen de construire l'image périphérique d'un point objet A situé hors de l'axe d'une surface réfringente sphérique ; mais comme cette image n'est pas ponctuelle, Young la définit pour commencer comme l'intersection I de deux rayons issus d'un même point A

*Corollaire 1.* Soient $t$ et $u$ les Cosinus d'incidence et de réfraction, le rayon étant égal à 1, et $d$ et $e$ les distances respectives des foyers des rayons incidents et réfractés ; alors $e = \frac{mduu}{mdu-ndt-ntt}$.[31]

*Corollaire 2.* Pour une surface plane, $e = \frac{mduu}{-ntt}$.[32]

{P.29} *Corollaire 3*. Pour des rayons parallèles, $d = \infty$, et $e = \frac{muu}{mu-nt}$.[33]

*Scholie 1.* On peut observer que la caustique par réfraction s'interrompt brutalement à sa surface, non pas géométriquement mais physiquement, du fait de la réflexion totale.[34]

---

d'inclinaisons « infiniment proches » après réfraction. Dès lors, Young montre que cette intersection (ou « foyer périphérique » [Young, 1807, II : 73]) se trouve nécessairement sur la droite (AG) liant le point objet A au « centre relatif » du dioptre sphérique G, qui est le projeté orthogonal du centre véritable du dioptre H sur la droite (EF) joignant les milieux des deux cordes de la surface réfringente dessinées par les rayons incident et réfracté. Les rayons incident et réfracté étant tracé, G peut être déterminée, et l'image périphérique I de A se trouve à l'intersection de (AG) et du rayon réfracté. Bien que la Figure 2 représente une sphère presque complète afin justement de pouvoir construire les cordes (BC) et (BD) des rayons incident et réfracté (égales à deux fois le Cosinus de leurs angles), il faut bien entendre que la réfraction n'a lieu qu'au point B et que le rayon ne ressort jamais du second milieu réfringent.

[31] Le résultat annoncé de ce corollaire est démontré à deux reprises dans [Young, 1807, II : 73-74], notamment comme une conséquence du théorème précédent. Il repose sur des définitions importantes pour la suite : « 411. Définition : Le point d'intersection des directions de deux rayons quelconques ou plus est appelé leur foyer ; et le foyer est soit réel, soit virtuel, selon qu'ils s'y rencontrent ou qu'ils tendent seulement vers lui ou de lui. 412. Définition : quand la divergence ou la convergence de rayons est modifiée par réfraction ou réflexion à une surface quelconque, les foyers des rayons incidents et réfractés ou réfléchis sont dits conjugués l'un de l'autre ; et le nouveau foyer est appelé image du précédent foyer » [Young, 1807, II : 71]. Ainsi, Young parle ici de « foyer des rayons incidents » pour désigner ce que l'on appelle aujourd'hui « point objet » en optique géométrique ; mais la seconde définition est plus large que la notion d'image de l'optique géométrique paraxiale, puisqu'elle intègre l'idée qu'à un point objet peuvent correspondre plusieurs points images (qu'il appelle « foyers »), selon les rayons choisis pour la tracer : en somme, qu'aucun système optique à l'exclusion du miroir plan n'est jamais rigoureusement stigmatique. Sur la Figure 2, $t$ (= Cos $i$) désigne la distance BE ; $u$ désigne la distance BF ; $d$ désigne la distance du point objet A au point d'impact du rayon sur la surface réfringente B ; et $e$ désigne la distance du point d'impact B au point image D ; $d$ et $e$ étant conventionnellement considérées comme positives si elles sont mesurées dans le sens de propagation de la lumière, et négatives sinon. Ce corollaire nous donne donc la distance $e$ (= BI) de l'intersection de deux rayons lumineux inclinés infiniment proches issus d'un point A, après leur réfraction par une surface réfringente sphérique située à la distance $d$ (= AB) du point A et telle que $\frac{Sin\ i}{Sin\ r} = \frac{m}{n}$.

[32] La distance $e$ du foyer périphérique ayant été démontrée pour un surface courbe quelconque dans le Corollaire précédent, Young passe à la limite pour déterminer la position de l'image par une surface plane : $t$ et $u$ ayant précédemment été définis comme les Cosinus d'incidence et de réfraction, pour un rayon de la sphère réfringente égal à 1, il réécrit d'abord la même relation pour une sphère de rayon quelconque $a$, ce qui donne : $e = \frac{mdauau}{mdau-ndat-natat}$. Puis il fait tendre la valeur de $a$ vers l'infini pour obtenir la nouvelle expression de $e$. On remarque que la valeur obtenue est négative, donc que l'image d'un point hors d'axe par une surface réfringente plane se situe bien du même côté que le point objet.

[33] C'est à un nouveau passage à la limite que Young procède ici pour déterminer la distance focale image « périphérique » $e$ d'une surface réfringente sphérique en faisant tendre la distance $d$ de l'objet vers l'infini dans l'expression obtenue au Corollaire 1. On remarque que cette distance dépend de $t$ et $u$, c'est-à-dire de l'inclinaison des rayons considérés, et donc qu'elle n'est pas unique.

[34] La caustique désigne l'enveloppe des rayons lumineux issus d'un point source unique (situé ici à l'infini), après déviation par un instrument optique. Ici l'instrument est une sphère transparente – et l'on parle alors de diacaustique (ou caustique par réfraction) – qui dans le contexte de cet article sur l'œil pourra servir de modèle naïf pour entamer l'étude du cristallin. Ici l'expression « réflexion totale » employée ici par Young n'a pas le sens qui lui est associé aujourd'hui en optique géométrique, mais désigne ce que nous appelons « incidence rasante ». Ainsi Young suggère que si la limite de la diacaustique d'une sphère réfringente est très abruptement marquée, ce n'est pas par simplification géométrique du problème, mais du fait de la limite physique que marque l'incidence rasante au sommet de la sphère, au-delà de laquelle il n'y a tout simplement plus de rayons pénétrant

*Corollaire 4*. Appelez $\frac{muu}{mu-nt}$, $b$, et $\frac{ntt}{mu-nt}$, $c$ ; alors $e = \frac{bd}{d-c}$ et $e - b = \frac{bc}{d-c}$ ; ou, pour le dire en mots, le rectangle limité par les distances focales de rayons parallèles, traversant et retraversant toute surface par les mêmes droites, est égal au rectangle limité par les différences entre ces longueurs et les distances de deux foyers conjugués quelconques.[35]

*Corollaire 5*. Pour des rayons perpendiculaires, $e = \frac{md}{d-n} = m + \frac{mn}{d-n}$ ; ou, si le rayon de la sphère est égal à $a$, $e = \frac{mad}{d-na}$ ; et si $d$ et $e$ sont donnés, on trouve le rayon $a = \frac{de}{md+ne}$.[36]

*Corollaire 6*. Pour des rayons perpendiculaires et parallèles, $e = m$, ou $e = ma$.[37]

*Corollaire 7*. Pour une lentille biconvexe, en négligeant son épaisseur, appelons le rayon de courbure de la première face $g$, celui de la seconde $h$ et $e = \frac{ndgh}{dg+dh-ngh}$. D'où $n = \frac{de}{d+e} \cdot \frac{g+h}{gh}$ ; et pour des rayons parallèles $e = \frac{ngh}{g+h}$ et $n = e \cdot \frac{g+h}{gh}$. Si $g = h = a$, $e = \frac{nad}{2d-na}$ ; et pour des rayons parallèles $e = \frac{na}{2}$ : en appelant $b$ cette distance focale principale, $e = \frac{bd}{d-b}$, comme dans le Cor. 4 ; d'où nous avons le foyer de deux lentilles jointes ; et aussi, $b = \frac{de}{d+e}$.[38]

---

[35] Toujours pour des rayons obliques, $b$ désigne la distance focale « périphérique » image (déterminée au Corollaire 3) et $c$ la distance focale « périphérique » objet du dioptre sphérique pour des rayons obliques (c'est-à-dire la distance au dioptre d'un point A objet tel que les rayons issus de lui que l'on considèrera ressortiront parallèles entre eux après réfraction ; distance $c$ (=AB) que l'on obtiendra en déduisant une expression de $d$ à partir de celle de $e$ donnée au Corollaire 1 ; puis en faisant tendre $e$ vers l'infini. Young conclue en donnant ici le cas général de la relation de conjugaison du dioptre sphérique dite « de Newton » – c'est-à-dire la relation associant la distance d'un point objet A au foyer objet « périphérique » F, à la distance de son image « périphérique » A' au foyer image « périphérique » F', et aux distances focales « périphériques » objet $f$ et image $f'$ – aujourd'hui communément formulée : $\overline{FA}.\overline{F'A'} = f.f'$ (soit dans l'écriture de Young : $(e-b).(d-c) = b.c$).

[36] Young traite à présent le cas particulier de rayons arrivant « perpendiculairement » à la surface sphérique, c'est-à-dire les rayons peu inclinés et peu éloignés de l'axe du système, pour lesquels on peut considérer que $i = r = 0$ (optique paraxiale). Le résultat est donc obtenu à partir de la formule du Corollaire 1, en posant $t = u = 1$ et en se souvenant que par définition $n = m \pm 1$. Ce Corollaire sera utilisé plus tard pour déterminer la position de l'image d'un point lumineux par la cornée de l'œil.

[37] En considérant les rayons incidents comme étant peu inclinés et parallèles, le point objet est renvoyé à une distance $d$ infinie, et on obtient la distance focale paraxiale : $e = ma$. Soit en écriture actuelle : $\overline{SF'} = \frac{n'}{n'-n}.\overline{SH}$, où H est le centre de courbure de la surface réfringente, $n'$ son indice, S l'intersection de la surface avec la droite allant de H au point objet (sommet), et $n$ l'indice du milieu environnant. Ce Corollaire sera utilisé plus tard pour évaluer le changement de rayon de courbure de la cornée ou le déplacement de la rétine qui seraient nécessaires à l'accommodation dans le cas d'un cristallin de forme fixe, et à disqualifier ces hypothèses.

[38] Ici Young extrapole ses résultats obtenus au Corollaire 5 pour des rayons peu inclinés réfractés par une surface réfringente à la détermination des propriétés paraxiales de la lentille mince sphérique biconvexe : c'est-à-dire au cas élémentaire de l'enseignement classique actuel en optique géométrique. Ce Corollaire sera utilisé plus loin pour déduire la position du point d'accommodation d'un œil presbyte regardant à travers l'optomètre de Young combiné à une lentille, et à l'estimation de la distance focale du cristallin. Pour obtenir la première formule, il suffit d'écrire la relation obtenue au Corollaire 5 successivement pour les deux dioptres de rayons de courbure $g$ et $-h$ (en inversant $m$ et $n$ pour le second dioptre) ; puis de considérer que les sommets des deux dioptres sont si proches qu'ils sont indiscernables, donc que la distance $d$ de l'objet pour le second dioptre est égale à la distance $-e$ de l'image par le premier dioptre. Dès lors, $e$ tend vers la distance focale image paraxiale $f' = \frac{n}{n'-n} \cdot \frac{R_1 R_2}{R_1 + R_2}$ lorsque $d$ tend vers l'infini (où $R_i$ désigne les rayons de courbures des faces 1 et 2 de la lentille), ou $f' = \frac{n}{n'-n} \cdot \frac{R}{2}$ lorsque la lentille est symétrique. L'avant-dernière formule du Corollaire est simplement la relation de conjugaison paraxiale avec origine au centre de la lentille mince sphérique, où $e$ est la distance

*Corollaire 8.* Dans une sphère, $e = ma.\frac{d+a}{2d-(m-2)a}$, pour la distance au centre de la sphère, et $b = \frac{ma}{2}$.[39]

{P.30} *Scholie 2.* Dans tous ces cas, si les rayons lumineux convergent, $d$ doit être considérée comme négative. Par exemple, pour trouver le foyer de deux lentilles jointes, convexes ou concaves, l'expression devient $e = \frac{bd}{d+b}$.[40]

*Corollaire 9.* Dans le Cor. 3, le diviseur devient en fin de compte constant ; et quand l'inclinaison est faible, le foyer varie comme $uu$.[41]

*Corollaire 10.* Pour des rayons parallèles arrivant obliquement sur une lentille biconvexe ou biconcave d'épaisseur négligeable, le rayon de courbure étant 1, on a /$e = \frac{ntt}{2(mu-nt)}$\[42] ; qui varie en fin de compte comme le /carré du Cosinus d'incidence\[43], ou comme $\frac{m+n}{nn}t + t^2$.[44]

---

lentille-image paraxiale, $d$ la distance lentille-objet et $b$ la distance focale image paraxiale (que Young nomme « distance focale principale ») ; relation communément formulée : $\frac{1}{e} + \frac{1}{d} = \frac{1}{b}$. La dernière formule du Corollaire donne alors accès à ce que Young appelle « the joint focus of two lenses » qui, dit-il, peut être trouvé « en ajoutant au soustrayant les réciproques de leurs distances focales prises à part, [...] ou en divisant leur produit par leur somme ou différence » [Young, 1807, II : 72]. Ce qui revient donc à la détermination de la distance focale paraxiale équivalente $f'_{eq}$ de deux lentilles accolées de distances focales respectives $f'_1$ et $f'_2$ (en considérant dans notre cas que $f'$ peut être négative aussi bien que positive) : $\frac{1}{f'_{eq}} = \frac{1}{f'_1} + \frac{1}{f'_2}$. Formule que l'on obtient cette fois en écrivant deux fois la formule de conjugaison de la lentille mince, puis en considérant que la distance objet pour la seconde est la distance image de la première. Cette dernière formule permettra à Young à déterminer la focale des deux lentilles à accoler pour reproduire artificiellement le pouvoir convergent de la cornée [Young, 1801, 58].

[39] Young donne ici la distance du centre H de la sphère transparente (de rayon $R$ et d'indice $n$) à son foyer image F' pour des rayons lumineux proches de l'axe : $\overline{HF'} = \frac{n.R}{2}$ ; dont on pourra trouver une démonstration complète dans [Morizot, 2016, 64-66]. Ce Corollaire sera utilisé plus tard pour évaluer la déformation nécessaire du cristallin au cours de l'accommodation dans le cas où il prendrait une forme sphérique, et à disqualifier cette hypothèse.

[40] Le cas envisagé ici est celui de rayons incidents convergents, donc d'un objet virtuel, situé après la lentille. Dans ce cas, et selon la convention de signe choisie par Young, la distance objet-lentille $d$ est alors considérée comme négative, car mesurée dans le sens opposé au sens de propagation de la lumière. La formule est alors déduite de l'expression de $e$ donnée au Corollaire 7, en remplaçant $d$ par $-d$.

[41] Si les angles d'incidence est faible, son cosinus $t$ est proche de 1, et celui de l'angle de réfraction $u$ aussi ; dès lors $mu - nt \approx m - n = 1$. Et la distance $e$ donnée dans le Corollaire 3 peut alors raisonnablement être approximée par $muu$.

[42] Correction demandée dans [Young, 1801, 83].

[43] *Idem.*

[44] Ce Corollaire traite de rayons parallèles arrivant obliquement sur une lentille mince sphérique symétrique, et sera utilisé plus tard pour justifier le défaut d'astigmatisme de l'œil par une inclinaison du cristallin par rapport à l'axe visuel. Ici la lentille est envisagée comme une succession de deux surfaces sphériques accolées de rayons de courbure normalisés à 1 (donc -1 pour le second). Donc $d$ tend vers l'infini pour le premier dioptre mais il s'agit à nouveau de considérer l'effet de l'inclinaison des rayons : c'est le « foyer périphérique » de la lentille mince biconvexe ou biconcave que l'on va déterminer. Le résultat est donc obtenu en injectant le résultat du Corollaire 3 dans la formule du Corollaire 1 : la première surface, impactée par un faisceau parallèle oblique, va faire converger les rayons en un foyer image « périphérique » situé à une distance $e$ qui sera derrière la seconde surface et qui sera alors considéré comme le point objet (situé à une distance -$d$) pour celle-ci. Il faut également prendre soin pour la seconde surface d'inverser $m$ et $n$ dans la formule du Corollaire 1 (on sort du milieu réfringent) et de considérer que le Cosinus du rayon incident $t$ est identique au Cosinus du rayon réfracté $u$ par la première surface.

*Scholie 3.* Pour la lentille biconvexe, l'épaisseur réduit l'effet de l'obliquité à proximité de l'axe ; pour la biconcave, elle l'augmente.

*Scholie 4.* Il n'est pas de surface sphérique, excepté un cas particulier (Wood, 155), qui puisse réunir un pinceau oblique de rayons lumineux, même en un point physique. Les rayons obliques que nous avons jusque-là considérés sont seulement ceux qui sont situés dans cette section du pinceau réalisée par un plan passant par le centre et par le point lumineux. Ils restent dans ce plan malgré la réfraction et par conséquent ils ne rencontreront pas les rayons des sections collatérales avant de parvenir à l'axe. La remarque a été faite par Sir Isaac Newton et développée par le Dr. Smith (Smith r. 493,494), elle semble cependant avoir été trop peu remarquée (Wood, 362). Ainsi le foyer géométrique devient une ligne, un cercle, un ovale, ou une autre figure, selon la forme du pinceau, la nature de la surface, et la position du plan recevant l'image. Quelques-unes des variétés de l'image focale d'un pinceau cylindrique réfracté obliquement sont exposées en Planche VI. Fig. 28.[45]

{P.31} *Corollaire 11.* D'où la droite joignant les foyers conjugués les plus éloignés passera toujours par le centre. La distance du foyer le plus éloigné de rayons parallèles sera exprimée par $f = \frac{m}{mu-nt}$ ; et le plus petit cercle d'aberration sera à une distance /$\frac{2mu}{(1+uu).(mu-nt)}$\[46], divisant la longueur de l'aberration dans le même rapport que celui des distances qui séparent ses deux extrémités de la surface. Dans le cas du Cor. 10, $f = \frac{n}{2(mu-nt)}$.[47]

---

[45] Cette scholie rappelle qu'il n'y a pas de surface sphérique réfringente qui soit parfaitement – ni même approximativement (« physiquement ») – stigmatique. Sauf l'exception – traitée par Wood [Wood J., 1799, 87] – d'un faisceau de rayons lumineux convergeant vers un point Q, mais interceptés avant d'y parvenir par un dioptre sphérique de sommet C et de centre E tel que $\frac{QE}{EC} = \frac{\sin i}{\sin r}$ (où $i$ et $r$ sont les angles d'incidence et de réfraction de chaque rayon) : Wood démontre en effet que dans ce cas tous les rayons sont déviés vers un point unique après réfraction. Cette scholie nous ramène donc à la constatation que l'image d'un point par un système constitué de surfaces sphériques n'est jamais un point, puisque comme nous l'avons vu dans les Corollaires précédents, deux rayons inclinés infiniment proches issus de A se croiseront en un point image I – dit « périphérique » – dépendant de leur inclinaison et situé sur (AG), donc hors d'axe ; mais que par symétrie axiale du problème, ces mêmes rayons croiseront forcément en un point situé sur l'axe de la sphère l'ensemble des rayons issus de A et présentant la même inclinaison qu'eux par rapport à l'axe de la sphère (donc tous les rayons situés à la surface d'un cône de sommet A, d'axe passant par le centre H de la sphère et d'angle au sommet $\widehat{HAB}$). Ce point est nommé « foyer radial » par Young [Young, 1807, II : 73], dépend lui aussi de l'inclinaison des rayons considérés et se distingue donc du « foyer périphérique » étudié jusqu'ici. C'est le sens du renvoi très elliptique que fait ici Young à certaines remarques de Newton, Smith [Smith R., 1738, II : 82] et Wood [Wood J., 1799, 193-194]. L'image d'un point objet par une sphère réfringente ne sera donc pas un point, mais « une ligne, un cercle, un ovale, ou une autre figure » selon la configuration du problème. Les images décrites dans ce cas serviront plus loin – par analogie avec l'image d'un point lumineux que Young observe à l'œil nu à différentes distances – à légitimer la justification de l'astigmatisme de l'œil par une inclinaison du cristallin.

[46] Correction demandée dans [Young, 1801, 83].

[47] La scholie précédente a souligné la nécessité de considérer la multiplicité des points images (ou « foyers ») d'un même point objet. Ce Corollaire repose implicitement sur l'étude d'une surface sphérique réfringente de rayon 1, et sur le fait que le plus éloigné des points images pour une telle sphérique correspond à un « foyer radial ». « La droite joignant les foyers conjugués les plus éloignés » est donc la droite joignant le point objet à un point image radial, et par définition de celui-ci elle passe forcément par le centre de la surface réfringente sphérique. L'expression de la distance $f$ correspond à la distance focale radiale d'un faisceau de rayons parallèles mais inclinés par rapport à l'axe (résultat démontré dans [Young, 1807, II : 74-75]). Elle est donc différente de la distance focale « périphérique » $e$ donnée au Corollaire 3. Le « plus petit cercle d'aberration » évoque alors le lieu où la tache image du point objet A sera la plus petite. Enfin, la dernière valeur de $f$ renvoie au cas de la lentille mince biconvexe de rayon 1 et doit être comparée à la distance focale « périphérique » donnée au Corollaire 10.

*Corollaire 12*. Cette proposition s'applique également aux rayons réfléchis ; et dans ce cas, la droite tirée depuis le centre passe par le point d'incidence.

*Proposition V. Problème.*

Trouver le lieu et la grandeur de l'image d'un petit objet, après réfraction par un nombre quelconque de surfaces sphériques.[48]

*Construction*. (Planche II. Fig. 3) D'un point quelconque (B) de l'objet (AB) dessiner les lignes allant vers (C) le centre de la première surface, et vers (D) le foyer de rayons parallèles arrivant en direction opposée : depuis l'intersection de la deuxième droite (BD) avec la tangente (EF) au sommet, dessiner une ligne (EH) parallèle à l'axe et elle coupera la première droite (BC) en (H), la première image du point (B). Poursuivre avec cette image comme nouvel objet et répéter l'opération pour chaque surface, et le dernier point sera l'image exigée. Pour le calcul, trouver la place de l'image par le Cor. 5. Prop. IV et sa grandeur sera à celle de l'objet comme leurs distances respectives au centre.[49]

*Corollaire.* /Si une image floue est reçue sur un plan quelconque, pour déterminer sa grandeur il sera nécessaire de se référer à l'ouverture donnant accès aux rayons. Si l'ouverture est supposée infiniment petite, elle peut être considérée comme un point lumineux afin de trouver la direction des rayons émergeant.\[50]

{P.32} *Proposition VI. Problème.*

Déterminer la loi par laquelle la réfraction par une surface sphérique doit varier, de manière à collecter des rayons parallèles en un foyer parfait.[51]

*Solution*. Soit $v$ le Sinus verse[52] pour un rayon de 1 ; alors, en chaque point en-dehors de l'axe, $n$ restant le même, $m$ doit devenir égal à $\sqrt{mm \pm 2nv}$ ; et tous les rayons seront collectés au foyer principal.

*Corollaire*. La même loi servira pour une lentille biconvexe, dans le cas où les foyers conjugués seront équidistants[53], en substituant $n$ à $m$.

---

[48] Ce problème doit évidemment introduire à l'étude de l'image formée par la succession de la cornée et des deux faces du cristallin.

[49] Les images successives d'un point objet par une série de surfaces réfringentes sphériques se construisent en considérant à chaque fois que le point image produite par une surface servira de point objet pour la suivante. La position de chaque image peut être retrouvée par le calcul grâce au Corollaire 5 de la Proposition IV (donc en se plaçant dans le cas de l'approximation paraxiale). Et la taille GH de l'objet AB est donnée par le rapport $\frac{GH}{AB} = \frac{CG}{CA}$, car les rayons lumineux passant par le centre C du dioptre ne sont pas déviés.

[50] Correction demandée dans [Young, 1801, 83 ; 60]. Ce Corollaire permettra plus loin d'exclure la possibilité que l'accommodation se fasse par un déplacement de la rétine, puisque la taille de l'image rétinienne des objets du champ visuel, même floue, s'en trouverait fortement affectée.

[51] Ce qui se prépare sous les traits de ce problème purement théorique – cherchant à déterminer la variation de l'indice d'un milieu transparent inhomogène permettant la focalisation parfaite de rayons lumineux parallèles à l'axe par une surface sphérique – c'est la justification optique de la nécessité de l'indice variable du cristallin de l'œil pour l'amélioration de l'image rétinienne.

[52] Le Sinus verse désigne la différence entre le rayon du cercle et le Cosinus, donc $v = 1 - \cos i$ quand ce rayon est pris égal à 1.

[53] C'est-à-dire dans le cas où le point image serait symétrique du point objet par rapport à la lentille, donc dans le cas où la distance objet lentille serait égale à deux fois la distance focale de la lentille dans le cas d'une lentille mince sphérique biconvexe dans l'approximation paraxiale. Dans ce cas, par symétrie du problème, les rayons

*Proposition VII. Problème.*

Trouver le foyer principal d'une sphère, ou d'une lentille, dont la parties internes sont plus denses que les externes.[54]

*Solution.* Afin que la distance focale puisse être finie, la densité d'une partie finie autour du centre doit être uniforme : appelez $\frac{1}{l}$ le rayon de cette partie, celui de la sphère étant pris pour unité ; posez que toute la réfraction depuis le milieu environnant vers cette sphère centrale est comme $m$ à $n$ ; prenez $r = \frac{\log l}{\log m - \log n}$, et supposez que la densité varie partout inversement comme la puissance $\frac{1}{r}$ de la distance au centre : alors la distance focale principale depuis le centre sera $\frac{r-1}{2} \cdot \frac{m}{nl-m}$. Quand $r = 1$, elle devient $\frac{1}{2(H.L.M - H.L.n)}$. Pour une lentille, déduire un quart de la différence entre son axe[55] et le diamètre de la sphère dont ses surfaces sont des portions.

*Corollaire.* Si l'on suppose que la densité varie soudainement à la surface, $m$ doit exprimer la différence des réfractions au {P.33} centre et à la surface ; et la distance focale ainsi déterminée doit être diminuée en conséquence en fonction de la réfraction à la surface.

*Proposition VIII. Problème.*[56]

/{P.83} Trouver le chemin d'un rayon de lumière tombant obliquement sur une sphère de densité réfringente variant comme une puissance quelconque de la distance au centre.

La densité réfringente, au sens de ces propositions, varie comme le rapport des Sinus, et comme la vélocité de la lumière dans le milieu (Schol. 2. Prop. I.). Soit $x^{-\frac{1}{r}}$ la vélocité à la distance $x$ ; alors, en considérant la force réfringente comme étant une espèce d'attraction,

---

seront parallèles à l'axe à l'intérieur de la lentille, et le résultat obtenu précédemment pour des rayons parallèles à l'axe parvenant sur une surface sphérique peut aisément être extrapolé.

[54] Le problème de la réfraction par un milieu à gradient d'indice se complexifie ici pour parvenir à la détermination de la distance focale de la lentille à indice variable ; et à l'indice global que devrait avoir une lentille uniformément réfringente de même profil pour posséder la même distance focale. Et Young appliquera la solution de ce problème au cas du cristallin un peu plus loin. Tscherning propose une démonstration complète de la solution donnée par Young à ce problème [Tscherning, 1894, 106-112]

[55] C'est-à-dire « son épaisseur ».

[56] La liste des corrections proposées en [Young, 1801, 83] demande une modification intégrale de cette proposition. Cette demande de modification est introduite par la justification suivante : « Du fait d'une erreur de signe, la huitième proposition se trouve erronée ; aucun usage n'ayant été fait de cette proposition, elle a été insérée sans avoir été proprement révisée. Elle devrait se présenter ainsi, avec sa démonstration ». Il est à remarquer enfin que sous cette dernière forme, la Proposition VIII permet essentiellement de démontrer la VII et pourrait donc logiquement être placée avant elle. C'est pourquoi d'ailleurs Tscherning démontre la proposition VIII [Tscherning, 1894, 98-106] avant la VII. Par ailleurs cette correction est plus importante qu'il n'y parait, car Young utilisera la proposition ainsi corrigée dès l'année suivante dans *la Conférence Bakerienne sur la Théorie de la Lumière et des Couleurs* [Young, 1802a, 42-44] pour modéliser le mécanisme d'« inflexion » de la lumière passant à proximité des corps matériels – si bien que c'est probablement son travail sur ce second texte qui lui aura inspiré cette correction. Dans ce texte, Young fait en effet l'hypothèse que la densité de l'éther à la surface des corps matériels décroit progressivement au fur et à mesure que l'on s'en éloigne, entrainant une variation progressive de l'indice de réfraction, et donc la légère déviation des rayons lumineux qui les frôlent – aujourd'hui expliquée par la théorie de la diffraction.

nous trouvons dans la Prop. 41. L. 1. Princip.[57] $\sqrt{ABFD} = x^{-\frac{1}{r}}$, $Q = s$, le Sinus d'incidence, le rayon étant pris pour unité, $Z = sx^{-1}$, $Dc = \frac{s}{2xx\sqrt{x^{-\frac{2}{r}}-s^2x^{-2}}} = \frac{1}{2}sx^{\frac{1}{r}-2}.(1-s^2x^{\frac{2}{r}-2})^{-\frac{1}{2}}$, et la fluxion[58] de l'aire décrite par le rayon[59] $= -\frac{1}{2}sx^{\frac{1}{r}-2}\dot{x}.(1-s^2x^{\frac{2}{r}-2})^{-\frac{1}{2}}$. Appelons $y$ le rapport au rayon du Sinus d'inclinaison {P.84} en chaque point ; alors $y = sx^{\frac{1}{r}-1}$, $\dot{y} = \frac{1-r}{r}sx^{\frac{1}{r}-2}.\dot{x}$, et la fluxion de l'aire $= \frac{r}{r-2}\dot{y}.(1-yy)^{-\frac{1}{2}}$, dont la fluente est $\frac{r}{2r-2}Y$, $y$ étant le sinus de l'arc $Y$ ; et l'angle correspondant est $\frac{r}{r-1}Y$.[60] Ayant obtenu la valeur de cet angle pour deux valeurs quelconques de $x$ ou $y$, la différence est l'angle décrit par le rayon. Cet angle est donc toujours

---

[57] Young renvoie ici à la Proposition 41 du Livre 1 des *Principia* [Newton, 1687, 127-132] dédiée au calcul de la trajectoire d'un corps dans un champ de force centripète variant avec la distance au centre de la force. Les notations employées ci-après font donc directement référence à la manière dont Newton paramètre et résout ce problème.

[58] « fluxion » est le terme utilisé par Newton [Newton, 1736] pour désigner la vitesse à laquelle une quantité variable (appelée fluente) varie au cours du temps. Si $x$ désigne une quantité variable, Newton désigne par $\dot{x}$ sa fluxion. Et $\frac{\dot{y}}{\dot{x}}$ correspond alors, au sens moderne, à la dérivée de la fonction $y$ par rapport à la variable $x$.

[59] Ici et dans la suite de cette proposition il s'agit du « rayon vecteur », c'est-à-dire du segment joignant le centre de la sphère au point où se trouve la lumière à l'instant considéré.

[60] La ligne de calculs qui s'enchaînent dans cette phrase repose sur l'application implicite du théorème de Laplace selon lequel, pour un rayon lumineux se mouvant dans un milieu dont l'indice $n$ dépend de la distance $x$ du point M considéré à un centre C (Figure jointe), le produit $n.x.\sin i$ reste constant, $i$ désignant l'angle que forme le rayon lumineux avec la droite CM (l'angle d'incidence).

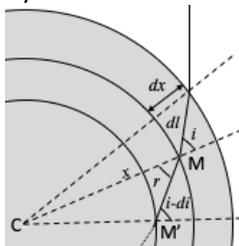

De fait, en modélisant le milieu comme une sphère inhomogène de rayon 1 composée d'un empilement de couches concentriques de densité décroissante, on peut poser que la lumière est réfractée au point M sous l'effet d'un changement infinitésimal de sa vitesse passant de $v$ à $v$-$dv$. La loi de la réfraction au point M s'écrit donc : $\frac{\sin i}{\sin r} = \frac{v}{v-dv}$ (Proposition I). La formule des sinus appliquée au triangle CMM' permet par ailleurs d'écrire : $\frac{\sin r}{x-dx} = \frac{\sin[\pi-(i-di)]}{x} = \frac{\sin(i-di)}{x}$. Par conséquent : $\frac{x.\sin i}{v} = \frac{(x-dx).\sin(i-di)}{v-dv}$, ce qu'il fallait démontrer puisque $n = \frac{c}{v}$. La première égalité posée par Young est donc la simple application de ce résultat au point d'entrée de la lumière dans le gradient d'atmosphère éthérée, puis à un point quelconque situé à une distance $x$ du centre : $\frac{1.\sin i_0}{1} = \frac{x.\sin i}{x^{\frac{1}{r}}}$. Que l'on peut réécrire : $s = y.x^{1-\frac{1}{r}}$, Young ayant posé $s = 1.\sin i_0$ et $y = \frac{x.\sin i}{x} = \sin i$. Dès lors, on obtient effectivement sa « fluxion » $\dot{y}$ par dérivation, puis la nouvelle expression de fluxion de la surface balayée par le rayon vecteur en injectant $y$ et $\dot{y}$ dans son expression donnée précédemment. Et puisque $y = \sin i$, cette surface élémentaire $\frac{r}{2r-2}.\frac{\dot{y}}{\sqrt{1-y^2}}$ balayée par le rayon vecteur peut aussi s'écrire : $\frac{r}{2r-2}.\frac{d(\sin i)}{\cos i} = \frac{r}{2r-2}.di$. Dès lors, la « fluente » de la surface est sa primitive $\frac{r}{2r-2}.i$, où $i$ est l'angle entre le rayon vecteur et le rayon lumineux au point considéré ; ou $\frac{r}{2r-2}.Y$, si $Y$ est la longueur de l'arc de cercle de rayon 1 décrit par l'angle $i$. Enfin, la section d'un disque de rayon 1 découpée par un angle $i$ étant égale à $1^2.i/2$, la surface ainsi décrite par le rayon vecteur est égale à la surface découpée dans un disque de rayon 1 par un angle $\frac{r}{r-1}.i$, ou $\frac{r}{r-1}.Y$.

à la différence des inclinaisons comme $r$ à $r$-1, et la déviation est à cette différence comme 1 à $r$-1.[61]

*Corollaire*. Ainsi, après passage à l'apside[62] et retour à la surface, la déviation est toujours proportionnelle à l'arc découpé par le prolongement du rayon incident : une telle sphère ne pourrait donc jamais réunir des rayons parallèles en quelque foyer que ce soit, la densité latérale étant trop faible vers la surface.[63]\

{P.33} *Scholie générale*. Les deux premières propositions se rapportent à des phénomènes bien connus ; la troisième peut difficilement être nouvelle ; la quatrième s'approche au plus près de la construction de MACLAURIN[64], mais elle est beaucoup plus simple et pratique ; les cinquième et sixième ne présentent aucune difficulté ; /la septième peut être déduite de la huitième, ou démontrée indépendamment d'elle\[65]. L'une est abrégée par une propriété des logarithmes ; l'autre découle des lois des forces centripètes, en supposant que les vélocités sont directement comme les densités réfringentes, en corrigeant la série pour l'emplacement de l'apside et en faisant varier le Sinus d'incidence pour déterminer la fluxion de l'angle de déviation.

V. Le Dr. PORTERFIELD a appliqué une expérience réalisée en premier par SCHEINER[66] à la détermination de la distance focale {P.34} de l'œil ; et il a décrit sous le nom d'optomètre un très excellent instrument fondé sur le principe de ce phénomène (Edim. Med. Essays, Vol. IV. p. 185)[67]. Mais l'appareil est susceptible d'être considérablement amélioré ; ainsi demanderai-je la permission de décrire un optomètre, simple par sa construction, et tout aussi commode et précis dans son utilisation.

Soit un obstacle interposé entre un point lumineux (R, Planche II. Fig. 4) et une surface réfringente quelconque, ou une lentille (CD), et soit cet obstacle perforé en deux points (A et B) seulement. Que les rayons réfractés soient interceptés par un plan, de manière à former une image sur celui-ci. Il est alors évident que lorsque ce plan (EF) passe au foyer des rayons réfractés, l'image formée sur lui sera un point unique. Mais si ce plan est approché vers l'avant (vers GH), ou repoussé vers l'arrière (vers IK), les petits pinceaux passant à travers les perforations ne se rencontreront plus en un point unique, mais tomberont sur deux endroits distincts du plan (G, H ; I, K) et formeront, dans un cas comme dans l'autre, une image double de l'objet.

---

[61] L'angle $\alpha$ balayé par le rayon vecteur entre deux points $M_1$ et $M_0$ est donc $\alpha = \frac{r}{r-1} \cdot (i_1 - i_0)$, où $(i_1 - i_0)$ est la « différence » des angles d'incidence (ou « inclinaisons ») mesurés en $M_1$ et $M_0$. Young nommant par ailleurs « déviation » l'angle $d$ formé entre les directions du rayon lumineux ainsi courbé aux points $M_1$ et $M_0$ (donc entre les tangentes au rayon courbe), on peut montrer que $d = i_0 - i_1 + \alpha$, donc que $d = \frac{(i_1 - i_0)}{r-1}$.

[62] C'est-à-dire le point de la trajectoire du rayon lumineux courbé le plus proche du centre de la sphère réfringente.

[63] Puisque les rayons périphériques d'un faisceau de rayons parallèles arrivant sur une telle sphère découpent des arcs plus petits que les rayons plus proches de son centre, leur déviation sera bien plus faible et la sphère ne pourra donc réunir des rayons parallèles en un foyer unique.

[64] Il semble qu'ici Young fasse référence au traité sur les fluxions de Maclaurin, auquel il fait référence à trois reprises dans son article publié en avril 1800 sur les courbes cycloïdes, et dont un chapitre traite le problème des caustiques par réflexion, mais aussi par réfraction [Maclaurin, 1742, I : 344-345].

[65] Correction demandée dans [Young, 1801, 83].

[66] Christoph Scheiner (1575-1650) est un jésuite allemand qui s'est illustré notamment en mathématiques et en astronomie (en co-découvrant les taches solaires), mais aussi en optique. Le traité de Scheiner auquel Young fait référence est un ouvrage dédié à l'anatomie et la physiologie de l'œil, fortement inspiré des découvertes optiques réalisées par Kepler au début du XVIIème siècle. Scheiner y propose des expériences en chambre noire, développe une lunette astronomique, positionne correctement le nerf optique, observe directement l'image rétinienne dans un œil dont il a décalotté le fond et, plus spécifiquement, décrit un montage expérimental qui inspirera la longue série des instruments que l'on appellera ensuite optomètres, utilisés pour mesurer les propriétés optiques de l'œil [Scheiner, 1619, 112-113].

[67] Le mot « optomètre » est utilisé pour la première fois par Porterfield [Porterfield, 1738, 185].

Ajoutons maintenant deux points lumineux supplémentaires (S et T, Fig. 5), l'un plus proche de la lentille que le premier point, l'autre plus éloigné ; et quand le plan recevant les images passe par le foyer des rayons venant du premier point, les images du second et troisième point doivent toutes deux être doubles (*ss*, *tt*), puisque le plan (EF) est au-delà de la distance focale des rayons venant du point le plus éloigné, et en-deçà de celle des rayons venant du plus proche. C'est sur ce principe que fut fondé l'optomètre du Dr. PORTERFIELD.

Mais si l'on suppose les trois points joints par une droite, et que cette droite est un peu inclinée par rapport à l'axe de la lentille, {P.35} chaque point de la ligne à l'exception du premier point (R, Fig. 6) aura une image double ; et chaque paire d'images, étant contiguë à celle des points lumineux avoisinants, formera avec eux deux lignes continues, et les images étant d'autant plus séparées que le point qu'elles représentent est éloigné du premier point lumineux, les lignes (*st*, *st*) convergeront de chaque côté vers l'image (*r*) de ce point et s'intersecteront à cet endroit.

La même chose se produit lorsque nous observons quelque objet à travers deux trous d'épingles disposés à l'intérieur des limites de la pupille[68]. Si l'objet est au point de vision parfaite, l'image sur la rétine sera unique ; mais dans tout autre cas, l'image étant double, nous semblerons voir un objet double : et si nous observons une ligne approximativement pointée vers notre œil, elle apparaitra comme deux lignes se croisant au point de vision parfaite. Pour cela, les trous peuvent être remplacés par des fentes, qui rendent l'image presque aussi distincte en même temps qu'elles laissent passer plus de lumière. Le nombre[69] peut être augmenté de deux à quatre, ou plus, chaque fois que des investigations particulières le rendent nécessaire.

<Cet instrument a l'avantage de mettre correctement en évidence la distance focale par simple examen, sans avoir à déplacer l'objet d'avant en arrière, ce qui est une opération susceptible d'incertitudes considérables, en particulier du fait que la focale de l'œil peut changer pendant ce temps.>[70]

L'optomètre peut être réalisé à partir d'un bout de carton ou d'ivoire, d'environ huit pouces (*20,32 cm*) de long et un (*2,54 cm*) de large, divisé dans le sens de sa longueur par une ligne noire qui ne doit pas être trop accentuée. Le contour de cette carte doit être découpé comme indiqué en Planche III. Fig. 7, de manière à ce qu'elle puisse être relevée et fixée en position inclinée au moyen des épaulements : ou bien, comme cela est ici représenté sur la gravure, on peut appliquer à l'optomètre une pièce détachée de cette forme environ. Un trou d'environ un demi-pouce carré (*1,61 cm$^2$*) doit être réalisé dans cette partie ; et les côtés découpés de manière à recevoir une réglette de papier épais munie de fentes de différentes tailles, d'un quarantième à un dixième de pouce (*0,6 mm à 2,5 mm*) de large, séparées par des espaces un peu plus grands ; afin que chaque observateur puisse choisir celle qui est la mieux adaptée à l'ouverture de sa pupille. {P.36} Pour adapter l'instrument à l'usage d'yeux presbytes, l'autre extrémité doit être équipée d'une lentille de quatre pouces (*10,16 cm*) de distance focale ; et le long de la ligne, une échelle doit être tracée de part et d'autre, graduée en pouces depuis une extrémité, et depuis l'autre selon la table calculée ici[71] d'après le Cor. 7. Prop. IV, au moyen de laquelle non seulement les rayons divergents, mais aussi ceux qui sont parallèles ou convergents depuis la lentille, sont rapportés à leur foyer virtuel.

---

[68] Il s'agit de percer deux trous dans une feuille de papier à l'aide d'une épingle, séparés d'une distance inférieure au diamètre de la pupille, et de placer ces deux trous juste devant l'œil.
[69] Sous-entendu, « de fentes parallèles ».
[70] [Young, 1807, II : 576].

L'instrument est aisément applicable à la détermination de la distance focale de lunettes requises pour des yeux myopes ou presbytes. M. Cary[72] a eu l'amabilité de me fournir les numéros et les distances focales des verres communément réalisés ; et j'ai calculé les distances auxquelles ces numéros doivent être placés sur l'échelle de l'optomètre, afin qu'un œil presbyte puisse être capable de voir à une distance de huit pouces (*20,32 cm*) en utilisant les verres de la distance focale indiquée à l'endroit du croisement des lignes le plus proche ; et un œil myope les rayons parallèles en utilisant les verres indiqués par le numéro qui se trouve à l'endroit de leur croisement le plus éloigné. |Pour faciliter l'observation, j'ai également placé ces numéros en face du point qui sera le croisement le plus proche pour des yeux myopes ; mais ceci selon la supposition arbitraire d'une capacité de changement de focalisation égale en chaque œil, ce qui je dois le confesser est souvent loin de la vérité.|[73] On ne peut s'attendre à ce que toute personne s'arrête précisément sur la puissance la mieux adaptée au défaut de son œil au premier essai. Peu sont ceux qui peuvent mener à loisir leurs yeux à l'état d'action complète ou de parfait repos ; et l'on trouvera qu'une puissance inférieure de deux ou trois degrés à celle qui est ainsi déterminée est suffisante pour les besoins ordinaires. J'ai également ajouté au second tableau les numéros qui désigneront les lunettes nécessaires à un œil presbyte pour voir à douze et à dix-huit pouces respectivement (*30 cm* et *45 cm*) : peut-être les numéros de la série du milieu seront-ils les plus {P.37} adaptés à placer sur l'échelle. L'optomètre devrait être appliqué à chaque œil ; et au moment d'observer, l'œil opposé ne doit pas être fermé, mais l'instrument doit être masqué à sa vue. Le lieu d'intersection peut être précisément déterminé au moyen d'un pointeur coulissant le long de l'échelle.[74]

L'optomètre est représenté en Planche III, Fig. 8 et 9 ; et la manière dont les lignes paraissent en Fig. 10.

---

[71] Tableau I.

[72] William Cary (1759-1825) a ouvert à Londres aux alentours de 1785 un atelier de production de lunettes, mais aussi de toute sorte d'instruments scientifiques (microscopes, lunettes astronomiques, baromètres…), qui furent rapidement réputées pour leur qualité.

[73] La phrase qui précède, comme les passages qui par la suite seront encadrées par les symboles « || », n'apparait plus dans [Young, 1807]. A ce même endroit du texte, Tscherning indique qu' « Il faut remarquer que les numéros des verres concaves sont arbitraires ; on peut voir la distance focale à laquelle correspond chaque numéro sur la table III. — Dans sa première forme, l'optomètre contenait encore une échelle, correspondant au proximum des yeux myopes. Elle était construite en faisant la supposition que l'amplitude d'accommodation était à peu près la même chez tout le monde et correspondait à 10 dioptries. Mais en travaillant avec l'optomètre Young remarqua bientôt que cette supposition était absolument fausse » [Tscherning, 1894, 117].

[74] Tscherning ajoute : « Très utile pour l'étude physiologique de l'œil, l'optomètre de Young l'est moins pour l'examen des malades, étant donné qu'on n'a aucun moyen de contrôler leurs réponses. […] On voit que Young a parfaitement constaté le défaut dont souffrent les optomètres, surtout ceux qui sont faits pour des observations à petite distance. Ils sollicitent un effort d'accommodation, qui fait paraître la myopie plus forte qu'elle ne l'est en réalité » [Tscherning, 1894, 118].

*Tableau* I. *Pour prolonger l'échelle avec une lentille de 4 pouces* (10,16 cm) *de focale.*[75]

| 4  | 2,00 | 11 | 2,93 | 30  | 3,52 | 200  | 3,92 | -35 | 4,51 | -12  | 6,00 |
|----|------|----|------|-----|------|------|------|-----|------|------|------|
| 5  | 2,22 | 12 | 3,00 | 40  | 3,64 | ∞    | 4,00 | -30 | 4,62 | -11  | 6,29 |
| 6  | 2,40 | 13 | 3,06 | 50  | 3,70 | -200 | 4,08 | -25 | 4,76 | -10  | 6,67 |
| 7  | 2,55 | 14 | 3,11 | 60  | 3,75 | -100 | 4,17 | -20 | 5,00 | -9,5 | 6,90 |
| 8  | 2,67 | 15 | 3,16 | 70  | 3,78 | -50  | 4,35 | -15 | 5,45 | -9,0 | 7,20 |
| 9  | 2,77 | 20 | 3,33 | 80  | 3,81 | -45  | 4,39 | -14 | 5,60 | -8,5 | 7,56 |
| 10 | 2,86 | 25 | 3,45 | 100 | 3,85 | -40  | 4,44 | -13 | 5,78 | -8,0 | 8,00 |

*Tableau* II. *Pour placer les nombres indiquant la distance focale de verres convexes.*[76]

| Foc. | VIII. | XII.  | XVIII. | Foc. | VIII. | XII.    | XVIII.  | Foc. | VIII.  | XII.   | XVIII. |
|------|-------|-------|--------|------|-------|---------|---------|------|--------|--------|--------|
| 0    | 8,00  | 12,00 | 18,00  | 20   | 13,33 | 30,00   | 180,00  | 8    | ∞      | -24,00 | -14,40 |
| 40   | 10,00 | 17,14 | 32,73  | 18   | 14,40 | 36,00   | ∞       | 7    | -56,00 | -16,80 | -11,45 |
| 36   | 10,28 | 18,00 | 36,00  | 16   | 16,00 | 48,00   | -144,00 | 6    | -24,00 | -12,00 | -9,00  |
| 30   | 10,91 | 20,00 | 45,00  | 14   | 18,67 | 84,00   | -63,00  | 5    | -13,33 | -8,57  | -5,92  |
| 28   | 11,20 | 21,00 | 50,40  | 12   | 24,00 | ∞       | -36,00  | 4,5  | -10,29 | -7,20  | -6,00  |
| 26   | 11,56 | 22,29 | 58,50  | 11   | 29,33 | -132,00 | -28,29  | 4,0  | -8,00  | -6,00  | -5,14  |
| 24   | 12,00 | 24,00 | 72,00  | 10   | 40,00 | -60,00  | -22,50  | 3,5  | -6,22  | -4,94  | -4,34  |
| 22   | 12,77 | 26,40 | 99,00  | 9    | 72,00 | -36,00  | -18,00  | 3,0  | -4,80  | -4,00  | -3,6   |

---

[75] Toutes les distances sont évidemment données en pouces. Tscherning ajoute que « Les tableaux suivants servaient à la construction de l'optomètre. Le tableau 1 a servi à construire la deuxième échelle de l'optomètre (à partir de gauche). L'œil observateur est supposé se trouver au bout supérieur du dessin, derrière la lentille de 4" de distance focale. Chaque point de la ligne paraît donc plus éloigné qu'il n'est en réalité, et les chiffres de l'échelle, qui sont les mêmes que ceux de la première colonne du tableau, indiquent la distance à laquelle paraît le point de la ligne à côté duquel ils se trouvent. La deuxième colonne du tableau indique la distance réelle du point à la lentille. — Les chiffres de la deuxième colonne se déduisent de ceux de la première au moyen de la formule ordinaire des lentilles. La distance virtuelle de 10'' correspond par exemple à la distance réelle de 2"86, puisque $\frac{1}{4} + \frac{1}{10} = \frac{1}{2,86}$ » [Tscherning, 1894, 118].

[76] Tscherning commente : « Le tableau II donne les places des chiffres de la première échelle à gauche de l'optomètre ; l'œil est supposé placé au bout supérieur du dessin et regardant à travers la lentille et les fentes. La première colonne du tableau contient la distance focale des verres usuels ; la deuxième indique la position du proximum d'un œil, qui aurait besoin du verre de la première colonne pour lire à distance de 8", la une troisième indique la position du proximum d'un œil qui aurait besoin du même verre pour lire à une distance de 12 " etc. Après avoir trouvé le chiffre du proximum par ce tableau, on cherche le même chiffre dans la colonne des distances virtuelles dans le tableau I ; le chiffre correspondant dans la colonne des distances réelles donne l'endroit où il faut inscrire le numéro du verre sur l'optomètre. Étant donné, par exemple, qu'un œil a besoin du no. 36 pour lire à 8", on trouve la position de son proximum par la formule $\frac{1}{P} = \frac{1}{36} - \frac{1}{8} = \frac{1}{10,28}$. Si cet œil regardait dans l'optomètre sans l'interposition de la lentille, il verrait les lignes s'entrecroiser à une distance de 10,28", mais comme il est supposé regarder à travers la lentille de 4" il verra l'entrecroisement à 2"86 puisqu'on a $\frac{1}{4} + \frac{1}{10} = \frac{1}{2,86}$; il faut, par conséquent, placer le no. 36 à une distance de 2"86 de la lentille » [Tscherning, 1894, 120].

*Tableau* III. *Pour des verres concaves.*[77]

| Nombre | Foyer et point le plus loin | Point le plus proche | Nombre | Foyer et point le plus loin | Point le plus proche | Nombre | Foyer et point le plus loin | Point le plus proche |
|---|---|---|---|---|---|---|---|---|
| 0 |  | 4,00 | 7 | 8 | 2,67 | 14 | 3,00 | 1,71 |
| 1 | 24 | 3,43 | 8 | 7 | 2,54 | 15 | 2,75 | 1,63 |
| 2 | 18 | 3,27 | 9 | 6 | 2,40 | 16 | 2,50 | 1,54 |
| 3 | 16 | 3,20 | 10 | 5 | 2,22 | 17 | 2,25 | 1,44 |
| 4 | 12 | 3,00 | 11 | 4,5 | 2,12 | 18 | 2,00 | 1,33 |
| 5 | 10 | 2,86 | 12 | 4,0 | 2,00 | 19 | 1,75 | 1,22 |
| 6 | 9 | 2,77 | 13 | 3,5 | 1,87 | 20 | 1,50 | 1,02 |

{P.38} VI. Étant convaincu de l'avantage de réaliser chaque observation avec aussi peu d'assistance que possible, je me suis efforcé de restreindre la plupart de mes expériences à mes propres yeux ; et en général, je baserai mes calculs en supposant un œil à peu près similaire au mien. Je m'efforcerai donc en premier lieu de déterminer toutes ses dimensions et toutes ses facultés.

Pour mesurer les diamètres, je fixe une petite clé à chaque pointe d'un compas ; et je peux m'aventurer à mener les anneaux au contact direct de la sclérotique. Le diamètre transverse externe est de 98 centièmes de pouce (*2,49 cm*).

Pour trouver l'axe, je tourne l'œil autant que possible vers l'intérieur. Et je presse l'une des clés contre la sclérotique à l'angle externe, jusqu'à ce qu'elle arrive à au point où le spectre[78] formé par sa pression coïncide avec la direction de l'axe visuel et, en regardant dans une glace, j'amène l'autre clé jusqu'à la cornée. En concédant une épaisseur de trois centièmes (*0,76 mm*) pour les couches, on trouve ainsi que l'axe optique de l'œil est de 91 centièmes de pouce (*23,11 mm*) de la surface externe de la cornée à la rétine. Avec un œil moins proéminent, cette méthode aurait pu ne pas fonctionner[79].

Le diamètre vertical, ou plutôt la corde, de la cornée est de 45 centièmes[80] (*11,4 mm*) : son Sinus verse de 11 centièmes (*2,8 mm*). Pour déterminer le Sinus verse, j'ai regardé avec mon œil droit l'image de mon œil gauche dans un petit miroir maintenu près du nez, alors que l'œil gauche était détourné de manière à ce que la marge de la cornée paraisse comme une ligne droite, et j'ai comparé la projection de la cornée avec l'image d'une échelle graduée maintenue dans une direction convenable derrière l'œil gauche, proche de la tempe gauche. La corde horizontale de la cornée est de presque 49 centièmes (*12,4 mm*).

---

[77] Tscherning ajoute : « La première colonne contient les numéros des verres usités autrefois en Angleterre ; la deuxième donne leurs distances focales, et en même temps la position qu'il faut donner aux numéros sur la quatrième échelle de l'optomètre (à partir de gauche), la distance focale indiquant directement la position du *remotum*, c'est-à-dire l'endroit le plus éloigné, où l'œil peut voir les lignes s'entre-croiser » [Tscherning, 1894, 120].

[78] Young emploie généralement dans ce texte le mot « spectrum » – qu'il rendra aussi plus loin par le terme de « fantom » – pour désigner la sensation de voir des taches lumineuses en l'absence de stimulus lumineux, notamment lorsque les yeux sont fermés. A ne pas confondre donc avec le *spectre coloré* de la lumière blanche qui sera aussi mentionné plus tard. Ces sensations, communément appelées « phosphènes » (à la manière des acouphènes qui sont des illusions auditives), peuvent par exemple être causées comme ici par une stimulation mécanique, ou comme un peu plus loin par l'observation prolongée d'un objet très lumineux (sur les phosphènes, voir aussi par exemple [Helmholtz, 1867, II : 266-275]).

[79] Tscherning ajoute : « Ceci est la seule détermination de l'axe de l'œil vivant que nous possédons. Le procédé peut sembler difficile, mais j'ai pourtant, quoique rarement, observé des yeux avec lesquels on aurait pu répéter la mensuration de Young » [Tscherning, 1894, 122].

[80] Sous-entendu « de pouces ».

D'où le rayon de courbure de la cornée est de 31 centièmes (*7,87 mm*).[81] On peut {P.39} penser que j'attribue une trop grande convexité à la cornée ; mais je l'ai contrôlée par un certain nombre d'observations concordantes qui seront énumérées ci-après.

L'œil étant dirigé vers son image ; la projection de la marge de la sclérotique est de 22 centièmes (*5,59 mm*) depuis la marge de la cornée vers l'angle extérieur, et de 27 (*6,86 mm*) vers l'angle intérieur de l'œil : de sorte que la cornée présente une excentricité d'un quarantième de pouce (*0,63 mm*) par rapport à la section de l'œil perpendiculaire à l'axe visuel.[82]

L'ouverture la pupille varie de 27 à 13 centièmes (*de 6,86 mm à 3,30 mm*) ; tout du moins est-ce sa taille apparente, qui doit être quelque peu réduite, du fait du pouvoir grossissant de la cornée, à peut-être 25 et 12 (*6,35 mm et 3,05 mm*). Lorsqu'elle est dilatée, elle est à peu près aussi excentrée que la cornée ; mais lorsqu'elle est le plus contractée, son centre coïncide avec la réflexion de l'image d'un objet tenu immédiatement devant l'œil ; et cette image coïncide de très près avec le centre de la marge apparente de la sclérotique toute entière : de sorte que la cornée est intersectée perpendiculairement par l'axe visuel.

Au repos, mon œil focalise sur la rétine les rayons qui divergent verticalement depuis un objet situé à une distance de dix pouces de la cornée, et les rayons qui divergent horizontalement depuis un objet à sept pouces de distance. Car si je tiens le plan de l'optomètre verticalement, les images de la ligne semblent se croiser à dix pouces (*25,4 cm*) ; et horizontalement, à sept (*17,78 cm*). La différence s'exprime par une distance focale de 23 pouces (*58,4 cm*)[83]. Je n'ai jamais fait l'expérience du moindre dérangement causé par cette imperfection, non plus que je l'aie jamais découverte jusqu'à ce que j'aie réalisé ces expériences ; et je crois pouvoir examiner les menus objets avec autant de précision que ceux dont les yeux sont formés différemment. L'ayant mentionné à M. C<small>ARY</small>, il m'informa qu'il avait {P.40} fréquemment remarqué pareille circonstance ; que beaucoup de personnes étaient obligées de maintenir un verre concave obliquement de manière à voir distinctement, contrebalançant par l'inclinaison du verre un pouvoir réfringent de l'œil trop important dans la direction de cette inclinaison (Cor. 10. Prop. IV) et ne trouvant que peu de secours dans des lunettes de même distance focale. La différence n'est pas dans la cornée, car elle persiste quand l'effet de la cornée est supprimé par une méthode qui sera décrite ci-après. La cause en est sans doute l'obliquité par rapport à l'axe visuel de l'uvée[84] et de la lentille cristalline qui

---

[81] La corde verticale $c$ de la base de la cornée ayant été mesurée, ainsi que son Son sinus verse $v$ (c'est-à-dire la longueur de sa protubérance), son rayon de courbure $R = \sqrt{(R-v)^2 + (\frac{c}{2})^2}$ peut être déterminé. La valeur donnée est la moyenne des deux valeurs obtenues à partir de la mesure des cordes verticale et horizontale de la cornée. Tscherning ajoute qu'« on peut s'étonner qu'il soit possible d'obtenir ainsi une mesure tant soit peu exacte du rayon de la cornée, mais le chiffre de Young concorde très bien, comme on voit, avec les résultats des mesures ophtalmométriques » [Tscherning, 1894, 124].

[82] Ici Young regarde son œil en face et mesure donc le léger décentrage de la cornée par rapport à l'axe du globe oculaire, représenté en Figures 19 et 20.

[83] L'œil de Young est donc manifestement myope et astigmate, puisqu'au repos il distingue nettement une ligne verticale située à 25,4 cm seulement de son œil. Mais qu'il distingue par ailleurs nettement une ligne horizontale lorsqu'elle se trouve à 17,8 cm ; il fait alors remarquer que la correction entre ces deux situations peut être effectuée par une lentille de 58,4 cm de focale qui renverrait à 17,8 cm l'image d'un objet situé à 25,4 cm, puisque $\frac{1}{17,8} - \frac{1}{25,4} = \frac{1}{58,4}$. Il est ainsi à noter que ce passage présente la première mention connue de l'astigmatisme, ainsi nécessairement que la première manière d'évaluer ce défaut par la distance focale de la lentille qui le corrigerait.

[84] Le mot employé ici en anglais est « uvea », que Young utilise plus fréquemment que le mot « iris », auquel il semble attribuer un sens équivalent : « [...] l'uvée ou iris, qui est de différentes couleurs chez les différentes personnes, présentant une perforation en son centre appelée la pupille » [Young, 1807, I : 447]. Toutefois dans ses *Leçons d'anatomie comparée* publiées à la même époque, Georges Cuvier (1769-1832) – anatomiste français

lui est presque parallèle : d'après les dimensions déjà données, cette obliquité paraîtra être de 10 degrés environ. Sans entrer dans un calcul très précis, on trouve (par le même corollaire) que la différence observée requiert une inclinaison d'environ 13 degrés ; et les trois degrés restants peuvent aisément être ajoutés par la plus grande obliquité de la face postérieure du cristallin opposée à la pupille. Il n'y aurait pas de difficulté à fixer les verres de lunettes, ou l'oculaire concave d'un télescope, dans une position susceptible de remédier à ce défaut.[85]

Afin de déterminer la distance focale de la lentille de l'œil[86], nous devons lui assigner sa distance probable à la cornée. Or le Sinus verse de la cornée étant de 11 centièmes (*2,79 mm*), et l'uvée étant presque plate, la surface antérieure de la lentille de l'œil doit probablement se trouver un peu en arrière de la corde de la cornée ; mais à une distance très insignifiante, car l'uvée a la substance d'une membrane fine et que la lentille de l'œil s'en approche de très près : on fixera donc cette une distance à 12 centièmes (*3,05 mm*). L'axe[87] et les proportions de la lentille de l'œil doivent être estimés par comparaison avec des observations anatomiques ; car elles affectent, dans une faible mesure, la détermination de sa distance focale. M. Petit a trouvé l'axe {P.41} toujours égal à deux lignes environ, ou 18 centièmes de pouce (*4,57 mm*). Le rayon de courbure de la face antérieure était pour sa plus grande valeur de 3 lignes (*6,35 mm*), mais plus souvent plus que moins. On supposera que celui de la mienne est de 3¼, ou presque $\frac{3}{10}$ de pouce (*7,62 mm*). Le rayon de courbure de la face postérieure était plus fréquemment de 2½ lignes, ou $\frac{2}{9}$ de pouce (*5,64 mm*) (Mem. De l'Acad. De Paris, 1730. p. 6. Ed. Amst.)[88]. Le centre optique sera par conséquent ($\frac{18\times 30}{30+22} =$) à environ un dixième de pouce (*2,54 mm*) de la face antérieure : d'où nous avons 22 centièmes (*5,59 mm*) pour la distance du centre à la cornée. En prenant maintenant un point lumineux à une distance de

---

promoteur de l'anatomie comparée et de la paléontologie – distingue subtilement les deux : « l'uvée, cette production de la choroïde qui forme un voile annulaire ou un diaphragme au-devant du cristallin, est recouverte à sa face antérieure d'une substance particulière qui porte le nom d'iris. L'iris est un tissu demi-fibreux, demi-spongieux, qui est collé de la manière la plus intime sur l'uvée, et qu'on ne peut en séparer qu'avec peine et dans les plus grands animaux. Il est plus épais et plus lâche à sa grande circonférence du côté du ligament ciliaire, où il semble se terminer. Il y est plus facile à séparer ; mais vers les bords de la pupille il va en s'amincissant, et il ne peut plus se distinguer de l'uvée qui le double » [Cuvier, 1805, II : 405-406]. Ainsi discernerons-nous les termes « uvée » et « iris » dans ce texte, en respectant simplement la terminologie employée par Young lui-même.
[85] Tscherning dément l'hypothèse de Young attribuant la cause de l'astigmatisme à l'inclinaison du cristallin. D'une part l'inclinaison aurait l'effet inverse d'après la Proposition IV. D'autre part il affirme n'avoir jamais rencontré de cristallin dont l'obliquité soit supérieure à sept ou huit degrés. Il attribue donc pour sa part cette asymétrie prioritairement à une déformation de la partie postérieure de la cornée [Tscherning, 1894, 126-131]. On sait aujourd'hui que ce défaut de la vision est essentiellement provoqué par un défaut de symétrie circulaire (autour de l'axe visuel) de la cornée ou du cristallin.
[86] Le mot employé ici est « lens », c'est-à-dire « lentille » en français. Afin de clarifier le sens de ce mot qui sera très fréquemment utilisé par la suite c'est par « lentille de l'œil » que nous le traduirons systématiquement par la suite, bien qu'il puisse être traduit aujourd'hui par « cristallin ». Exceptionnellement, Young parlera aussi de « crystalline lens » et plus rarement encore emploiera le terme « crystalline » seul. Par respect de la variété terminologique employée par l'auteur –rendant compte selon nous de l'incertitude quant à la nature exacte de ce corps, envisagé comme un capsule musculaire remplie de liquide cristallin dans [Young, 1793, 173], ou comme un muscle à fibres transparentes dans le présent article – nous traduirons ces termes respectivement par « lentille cristalline » et « cristallin ».
[87] C'est-à-dire l' « épaisseur ».
[88] François Pourfour du Petit (1664-1741), dit parfois « Petit le Médecin », est un médecin, naturaliste et anatomiste français, auteur de nombreux mémoires sur l'anatomie et la physiologie de l'œil, dont celui cité ici par Young [Pourfour du Petit, 1730, 5], qui étudie non seulement le cristallin de l'homme mais aussi de divers animaux. Au-delà de leur clarté et de leur exactitude, ces mémoires ont le mérite de faire se rencontrer les rôles du cristallin dans la vision et dans les maladies de l'œil.

10 pouces (*25,4 cm*), le foyer de la cornée sera 115 centièmes (*2,92 cm*) derrière le centre de la lentille (Cor. 5. Prop. IV). Mais le foyer conjugué réel est $(91 - 22 =)$ à 69 derrière le centre (*1,75 cm*) : d'où, en négligeant l'épaisseur de la lentille, sa distance focale principale est de 173 centièmes (*4,39 cm*) (Cor. 7. Prop. IV). Pour son pouvoir réfringent dans l'œil, nous avons (Cor. 7. Prop. IV) $n$ = 13,5 et $m$ = 14,5. En calculant à partir de ce pouvoir réfringent, en tenant compte également de l'épaisseur, nous trouvons qu'il nécessite d'être corrigé et qu'il se rapproche du rapport de 14 à 13 pour les Sinus. Il est bien connu que les pouvoirs réfringents des humeurs sont égaux à celui de l'eau ; et que l'épaisseur de la cornée est trop uniforme pour produire un effet quelconque sur la distance focale.

Afin de déterminer par une expérience directe le pouvoir réfringent de la lentille cristalline, j'ai fait usage d'une méthode que m'a suggérée le Dr WOLLASTON[89]. J'ai trouvé que le pouvoir réfringent du centre du cristallin humain encore frais par rapport à celui de l'eau est de 21 à 20. La différence de ce rapport avec celui de 14 à 13, déterminé par le calcul, est probablement due à deux circonstances. La première est que, la substance de la lentille de l'œil étant à un certain degré soluble dans l'eau, une portion du fluide aqueux contenu {P.42} à l'intérieur de sa capsule y pénètre après la mort, réduisant ainsi quelque peu la densité. Lorsqu'il est sec, le pouvoir réfringent est légèrement inférieur à celui du verre crown[90]. La seconde circonstance est la densité non uniforme du cristallin. Le rapport de 14 à 13 est basé sur l'hypothèse d'une densité uniforme : mais la partie centrale étant la plus dense, la totalité agit comme une lentille de plus petites dimensions ; et l'on peut trouver grâce à la Prop. VII que si la partie centrale d'une sphère est supposée être de densité uniforme, réfringente comme 21 à 20 jusqu'à la distance de son demi-rayon, et si la densité des parties externes décroît graduellement et devient à sa surface égale à celle du milieu environnant, la sphère ainsi constituée sera équivalente en distance focale à une sphère de même taille avec une réfringence de 16 à 15 environ. Et l'effet sera à peu près le même si la partie centrale est supposée plus petite que cela, mais que la densité est un peu plus grande à la surface que celle du milieu environnant, ou qu'elle varie plus rapidement dans les parties externes que dans les internes. /Ou bien, si l'on suppose une lentille de dimensions moyennes et de distance focale égales à celles du cristallin, constituée de deux segments de la partie externe d'une telle sphère, la densité réfringente au centre de cette lentille doit être comme 18 à 17.\[91] En somme, il est probable que le pouvoir réfringent du centre du cristallin humaine à l'état vivant est à peu près à celui de l'eau comme 18 à 17 ; que l'eau imbibée après la mort le réduit au rapport de 21 à 20 ; mais qu'à cause de la densité non uniforme de la lentille, son effet dans l'œil est équivalent à une réfringence de 14 à 13 pour son ensemble. Le Dr WOLLASTON a déterminé que la réfringence depuis l'air vers le centre du cristallin frais de bœufs et de moutons est d'environ 143 à 100 ; vers le centre du cristallin de poissons et vers le cristallin desséché de moutons de 152 à 100. D'où la réfraction du cristallin de bœufs dans l'eau doit être de 15 à 14 : mais le cristallin humain, quand il est frais, est décidément moins réfringent.

---

[89] William Hyde Wollaston (1766-1828), médecin, physicien et chimiste britannique, entré à la Royal Society la même année que Thomas Young, était aussi l'un de ses amis les plus proches dans la communauté scientifique, si bien qu'ils ont même réalisé des expériences ensemble à plusieurs reprises et que Wollaston se convertira à la théorie vibratoire de la lumière. Par ailleurs, les travaux de Wollaston sur le spectre de la lumière blanche joueront un rôle important dans les deux derniers articles de ce recueil [Young, 1802b ; 1804]. Mais pour l'heure, Wollaston publie en 1802 « une méthode pour examiner les pouvoirs réfringent et dispersif » de diverses substances [Wollaston, 1802] dont on peut imaginer que c'est celle qu'il lui a soufflée et qui a été mise à l'œuvre pour établir les mesures que Young lui attribue plus loin. Elle consiste simplement à mesurer l'angle limite de réfraction totale de la lumière $i_l$ à la face de sortie d'un prisme d'indice $n_p$ connu, au contact duquel aura été apposé la substance liquide dont on veut déterminer l'indice de réfraction $n_s < n_p$. Cet angle limite $i_l$ étant l'angle d'incidence pour lequel l'angle de réfraction sera égal à 90° – et donc à partir duquel il n'y aura pas de réfraction possible – sa mesure permet en effet une déduction immédiate du rapport $\frac{n_s}{n_p}$ grâce à la loi de la réfraction : $\sin i_l = \frac{n_s}{n_p}$. L'indice du prisme pourra avoir été déterminé au préalable par la même méthode en mesurant simplement l'angle de réflexion totale de ce prisme dans l'air, supposé d'indice égal à 1,00032. Enfin, l'indice d'une substance solide pourra être déterminé de la même manière, en interposant entre elle et le prisme un fluide d'indice supérieur à celui de la substance à mesurer, assurant un contact parfait entre elle et le prisme, et n'altérant en rien le résultat d'après la Proposition II.

[90] Les verres crown sont des verres optiques de faible indice de réfraction (aux alentours de 1,6) et de faible dispersion chromatique.

{P.43} Ces considérations expliqueront l'incohérence de différentes observations du pouvoir réfringent du cristallin ; et en particulier pourquoi la réfringence que j'ai autrefois calculée en mesurant la distance focale de la lentille de l'œil (Phil. Trans. For 1793. p. 174)[92] est tellement supérieure à celle déterminée par d'autre moyens. Mais pour des expériences directes, la méthode du Dr WOLLASTON est extrêmement exacte.[93]

Lorsque j'observe un minuscule point lumineux, telle l'image d'une chandelle dans un petit miroir concave, il apparait comme une étoile rayonnante, comme une croix, ou comme une ligne irrégulière, et jamais comme un point parfait, à moins que je n'applique une lentille concave inclinée à un angle convenable afin de corriger la réfringence inégale de mon œil. Si j'approche ce point très près, il s'étale en une surface à peu près circulaire et illuminée presque uniformément, à l'exception de quelques lignes très légères à peu près situées dans une direction radiale.[94] A ce propos, la meilleure image est celle d'une chandelle, ou d'un petit miroir, vu au travers d'une minuscule lentille à faible distance, ou vue par réflexion sur une plus grande lentille. Si l'on applique quelque pression sur l'œil, telle que celle du doigt le maintenant fermé, la vue est souvent confuse pendant un court moment après le retrait du doigt, et l'image est dans ce cas tachetée ou coagulée. Les lignes rayonnantes sont probablement occasionnées par quelques légères inégalités dans la surface de la lentille de l'œil, qui est très superficiellement sillonnée dans la direction de ses fibres : l'aspect coagulé sera expliqué ci-après. Quand le point est repoussé plus au loin, l'image devient évidemment ovale, le diamètre vertical étant le plus long et les lignes étant un peu plus distinctes qu'avant, la lumière étant la plus forte au voisinage du centre ; mais immédiatement au centre il y a une tache plus sombre, due à une légère dépression au sommet telle que celle qui est souvent {P.44} observable en examinant la lentille de l'œil après la mort[95]. La situation des raies est constante bien qu'irrégulière ; les plus notables sont au nombre de sept ou huit ; parfois on peut en compter une vingtaine de plus faibles. Repoussant le point un peu plus loin, l'image prend la forme d'une courte ligne verticale ; les rayons qui divergeaient horizontalement étant parfaitement réunis, alors que les rayons verticaux sont encore séparés. Dans la phase suivante, qui est celle de plus parfaite focalisation, la ligne s'étale en son milieu et s'approche presque d'un carré aux angles saillants, mais elle est marquée de lignes plus sombres vers les diagonales. Le carré s'aplatit ensuite en un losange, et le losange en une ligne horizontale inégalement brillante. A distance plus grande, la ligne s'allonge et acquiert également une épaisseur du fait de rayonnements émis depuis elle, mais ne devient pas une surface uniforme, la partie centrale restant toujours considérablement plus brillante en conséquence du même aplatissement du sommet qui plus tôt la rendait moins visible. Certaines de ces figures présentent une analogie considérable avec les images découlant de la réfraction de rayons obliques (Schol. 4. Prop. IV) et ressemblent plus fortement encore à une combinaison de deux d'entre elles dans des directions opposées ; au point de ne pas laisser de doute quant au fait que les deux surfaces de la lentille sont obliques par rapport à l'axe visuel et coopèrent à déformer le point focal. Cela peut aussi être vérifié en observant l'image dessinée par une lentille commune en verre, lorsqu'elle est inclinée par rapport aux rayons incidents (voir Planche IV. Fig. 28-40).

---

[91] Correction demandée dans [Young, 1801, 84].

[92] Dans son premier article, dédié déjà à la vision et qui lui vaut d'accéder à la Royal Society, Young examine déjà le fonctionnement du cristallin. A partir de la mesure des dimensions et de la distance focale d'un cristallin de bœuf, il estime son indice de réfraction (ou pouvoir réfringent) à 1,521 [Young, 1793, 174].

[93] A cet endroit qui conclue la partie de l'article consacrée par Young aux propriétés de son œil, Tscherning insère une analyse détaillée de ces mesures, comparées à celles produites tout au long du XVIIIème siècle – notamment par Hermann von Helmholtz et Marius Tscherning lui-même [Tscherning, 1894, 134-140]. Ce passage conclue — à quelques exceptions près — à la très grande cohérence des valeurs avancées par Young avec celles validées un siècle plus tard à l'aide de techniques souvent plus élaborées et reposant sur une étude plus large et une connaissance anatomique et physiologique plus avancées de l'œil.

[94] Les Figures 29 à 41 de la planche VI donnent une idée des formes décrites ici par Young.

[95] Tscherning dit pour sa part n'avoir jamais observé cette dépression ou cet aplatissement du sommet de la face antérieure du cristallin [Tscherning, 1894, 143].

   L'axe visuel étant fixé dans une direction quelconque, je peux en même temps voir un objet lumineux placé latéralement à une distance considérable de celui-ci ; mais selon les directions, l'angle est très différent. Il s'étend à 50° vers le haut, à 60° vers l'intérieur, à 70° vers le bas et à 90° vers l'extérieur. Ces limites internes du champ visuel correspondent de près {P.45} aux limites externes formées par les différentes parties du visage, quand l'œil est dirigé vers l'avant et un peu vers le bas, ce qui est sa position la plus naturelle ; bien que les limites internes soient un peu plus larges que les externes ; et toutes deux sont bien calculées pour nous permettre de percevoir le plus facilement les objets les plus susceptibles de nous concerner. L'œil du Dr WOLLASTON a un champ visuel plus large, à la fois verticalement et horizontalement mais à peu près dans les mêmes proportions, si ce n'est qu'il s'étend plus loin vers le haut. Il est bien connu que la rétine s'avance plus vers l'angle interne de l'œil que vers l'angle externe ; mais vers le haut et vers le bas son extension est presque égale, et elle est en effet plus large que les limites du champ visuel dans toutes les directions, même si l'on ne tient compte que de la réfraction par la cornée. La partie sensible semble coïncider plus exactement avec la partie colorée de la choroïde des quadrupèdes[96] : mais l'extension totale de la vision parfaite est d'un peu plus de 10 degrés ; ou, pour parler plus strictement, l'imperfection démarre à partir d'un à deux degrés de l'axe visuel, et devient presque stationnaire à la distance de 5 ou 6 degrés, jusqu'à ce que la vision soit complètement éteinte à une distance encore plus grande. L'imperfection est partiellement due aux inévitables aberrations des rayons obliques, mais principalement à l'insensibilité de la rétine : car si l'image du Soleil lui-même est reçue sur une partie de la rétine éloignée de l'axe, l'impression ne sera pas suffisamment forte pour former un spectre permanent, bien qu'un objet de luminosité très modérée produise cet effet lorsqu'il est vu directement. [97]<Il a été dit qu'une lumière faible, telle que la queue d'une comète, est plus facilement observable en vision latérale qu'en vision directe. En supposant ce fait comme certain, la raison en est probablement que les grandes masses de lumière et d'ombre sont plus distinguables quand les parties en sont un peu confuses que quand le tout est rendu parfaitement distinct ; ainsi ai-je souvent observé que le motif d'une tapisserie ou d'un tapis présentait certaines lignes quand je l'observais sans lunette ; mais que ces lignes s'évanouissaient aussitôt que la focalisation était rendue parfaite.> Il aurait probablement été incohérent avec l'économie de la nature de conférer une plus grande part de sensibilité à la rétine. Le nerf optique est déjà très gros ; et la délicatesse de l'organe fait qu'il est très susceptible d'être blessé même par une légère irritation, {P.46} et aisément sujet aux affections inflammatoires ; et afin de rendre la vue aussi parfaite qu'elle l'est, il fut nécessaire de confiner cette perfection dans d'étroites limites. Le mouvement de l'œil a une étendue de 55 degrés environ dans chaque direction ; de sorte que le champ de vision parfaite s'étend successivement par ce mouvement à 110 degrés.[98]

---

[96] Young évoque certainement ici le *tapetum lucidum*, ou « tapis clair » en français : soit un tissu réfléchissant et irisé pouvant se trouver sur la choroïde ou à l'intérieur de la rétine de certains vertébrés (chat, chien, cerf, bovins, primates, grands dauphins…), mais dont les humains ne disposent pas. Elle permet notamment de réfléchir la lumière atteignant le fond de leur œil, et d'améliorer ainsi la sensibilité de leur vision dans des conditions de faible luminosité.

[97] L'extrait suivant est ajouté dans [Young, 1807, 582].

Mais la forme globale de la rétine est d'une forme propre à recevoir sur chaque partie de sa surface l'image la plus parfaite que l'état de chaque pinceau de lumière réfracté le permettra ; et la densité variable du cristallin rend cet état mieux capable encore de dessiner une telle peinture que tout autre artifice imaginable aurait pu le faire. Pour illustrer ceci, j'ai construit un diagramme représentant les images successives d'un objet distant couvrant toute l'étendue de la vision telles qu'elles seraient formées par les réfractions successives par les différentes surfaces. En prenant l'échelle de mon propre œil, je suis obligé de substituer la série d'objets à des distances indéfiniment grandes par un cercle de 10 pouces (*25,4 cm*) de rayon ; et il est plus commode de ne considérer que ces rayons qui passent par le sommet antérieur de la lentille de l'œil ; puisque le centre véritable de chaque pinceau doit être sur le rayon qui passe par le centre de la pupille et que la courte distance de ce point au sommet de la lentille de l'œil tendra toujours à corriger la réfraction inégale des rayons obliques. La première courbe (Planche IV. Fig. 16) est l'image formée par l'intersection la plus éloignée des rayons réfractés par la cornée ; la seconde, l'image formée par l'intersection la plus proche ; la distance entre celles-ci illustre le degré de confusion de l'image ; et la troisième courbe, sa partie la plus lumineuse. Telle doit être la forme de l'image que la cornée tend à dessiner dans un œil dépourvu de lentille cristalline ; aucun remède extérieur ne peut alors non plus corriger proprement l'imperfection de la {P.47} vision latérale. Les trois courbes suivantes illustrent les images formées après réfraction par la face antérieure de la lentille de l'œil, distinguées de la même manière ; et les trois suivantes le résultat de toutes les réfractions successives. La dixième courbe est une reprise de la neuvième, avec une légère correction à proximité de l'axe, en F, où quelques rayons perpendiculaires doivent arriver du fait de la largeur de la pupille. En la comparant à la onzième, qui représente la forme de la rétine, il apparaitra que rien de plus qu'une diminution modérée de densité dans les parties latérales de la lentille de l'œil n'est nécessaire à leur coïncidence parfaite. Si la loi selon laquelle cette densité varie était plus précisément établie, son effet sur l'image pourrait être /estimé au moyen de la huitième proposition ; et probablement\[99] que l'image ainsi corrigée approcherait de très près la forme de la douzième courbe.

---

[98] Les lignes qui précèdent listent différentes propriétés du champ visuel que Young tente d'interpréter à sa manière. Les progrès de la microscopie et de la physiologie ont permis d'en élaborer une tout autre interprétation, mais aussi de confirmer la pertinence des observations de Young. En particulier, l'on sait aujourd'hui que la rétine est tapissée de deux types de photorécepteurs, baptisés cônes et bâtonnets – du fait de leurs formes respectives – et différemment répartis sur la rétine. L'essentiel des cônes est massé autour du centre du champ visuel et couvre une surface de 10° de rayon environ, appelée *macula* (ou tache jaune). C'est la partie du champ visuel où l'acuité est maximale et qui est utilisée notamment pour la lecture. La partie centrale de cette *macula* est appelée *fovea* ; c'est sur cette zone de 2 à 5° de rayon angulaire et composée exclusivement des cônes que la vision des détails est optimale. Au-delà de la *macula*, les cônes sont pratiquement absents et les bâtonnets prennent le relais. En partie du fait de la plus faible densité de leur répartition, et de leur connexion en réseaux, la vision des détails est bien plus mauvaise dans la partie périphérique de la rétine qu'en son centre. Cependant cette même interconnexion des bâtonnets, ainsi que leur plus grande concentration en pigment absorbant la lumière visible, confère une plus grande sensibilité à la lumière à cette zone périphérique de la rétine. C'est ainsi que traditionnellement les astronomes rappellent qu'il est plus facile d'observer un objet céleste ténu en ne le regardant pas directement, mais légèrement de côté. Il est intriguant que Young ne souligne pas ici la possibilité de distinguer les couleurs uniquement dans la partie centrale du champ visuel qu'il a délimitée à 10°, et leur disparition au-delà. Cette aptitude à produire une sensation visuelle colorée est le fait, on le sait aujourd'hui, de la multiplicité des types de cônes (il en existe trois types en général dans l'œil humain) dont les différentes propriétés d'absorption de la lumière visible mènent à la production par un triplet de cônes impacté par un stimulus lumineux de trois signaux nerveux d'amplitudes différentes, susceptibles de faire émerger la triplicité (teinte, saturation, luminosité) de la sensation colorée. L'unicité des bâtonnets ne permet pas cela et l'excitation de zone périphérique de la rétine ne pourra être interprétée que par une sensation de luminosité variable, corrélée à la quantité de lumière absorbée par chaque bâtonnet, dépendant indistinctement de l'intensité et du spectre du stimulus lumineux. Et si la chose est particulièrement intéressante à souligner ici, c'est que bien que cette structure cellulaire de la rétine soit parfaitement insoupçonnée à l'époque où Young écrit ces lignes, c'est lui – comme nous le verrons – qui formulera l'hypothèse de la triplicité des récepteurs de la rétine pour justifier de la sensation des couleurs dès l'année suivante [Young, 1802a, 18-21] ouvrant une voie décisive pour la résolution de ce problème.

    Afin de trouver le lieu d'entrée du nerf optique, je fixe deux chandelles à dix pouces (*25,4 cm*) de distance, me recule de seize pieds (*4,88 m*), et dirige mon œil vers un point à quatre pieds (*1,22 m*) à droite ou à gauche du milieu de l'espace qui est entre elles : elles se perdent alors en une tache lumineuse confuse ; mais toute inclinaison de l'œil ramène l'une ou l'autre d'entre elle dans le champ de vision. Dans l'œil de Bernoulli, une plus grande déviation de la direction de l'axe était nécessaire (Comm. Petrop. I. p. 314)[100] ; et la partie obscurcie paraissait plus étendue. De l'expérience relatée ici, on trouve (par la Prop. V) que la distance de l'axe visuel au centre du nerf optique est de 16 centièmes de pouce (*4,07 mm*) ; et que le diamètre de la partie la plus insensible de la rétine est d'un trentième de pouce (*850 μm*)[101]. Afin de déterminer la distance au nerf optique du point situé en face de la pupille, j'ai pris la sclérotique de l'œil humain, l'ai découpée en segments depuis le centre de la cornée vers le nerf optique, et l'ai étendue sur un plan. J'ai alors mesuré la plus longue et la plus courte {P.48} distance de la cornée à la perforation que fait le nerf, et leur différence était d'un cinquième de pouce (*5,08 mm*) exactement. A cela je dois ajouter un cinquantième (*0,51 mm*), du fait de l'excentricité de la pupille dans l'uvée, qui dans l'œil que j'ai mesuré n'était pas grande, et la distance du point faisant face à la pupille au centre du nerf sera de 11 centièmes (*2,79 mm*)[102]. D'où il apparait que l'axe visuel est cinq centièmes, ou un vingtième, de pouce (*1,27 mm*) plus loin du nerf optique que le point faisant face à la pupille. Il est possible que cette distance puisse être différente pour des yeux différents : dans les miens, l'obliquité de la lentille de l'œil et l'excentricité de la pupille par rapport à celle-ci tendra à envoyer vers lui[103] un rayon direct, sans grande inclinaison de l'œil entier ; et il n'est pas improbable que l'œil soit également légèrement tourné vers l'extérieur s'il regarde un objet situé devant lui, bien que l'inclinaison soit trop faible pour être sujette à mesure.

---

[99] Correction demandée dans [Young, 1801, 84].
[100] [Bernouilli D., 1726].
[101] Young établit donc ici le diamètre et la position de la tache aveugle de la rétine, découverte et associée à la zone d'ancrage du nerf optique par Mariotte [Mariotte, 1688, 3-5], et dont on sait aujourd'hui qu'elle est dénuée de cellules visuelles. Aujourd'hui, la position de la tache aveugle est effectivement estimée en moyenne à 4 mm du pôle postérieur du globe oculaire, mais son diamètre moyen est donné à 1,5 mm depuis le milieu du xix[ème] siècle [Helmholtz, 1867, II : 287-288]. Helmholtz justifiant la discordance des résultats de Young avec ceux obtenus par la suite par le choix malheureux de ces deux chandelles comme objets à observer, quand lui se contentait de tracer une petite croix ou une petite tache noire sur un papier blanc.
[102] Car $\frac{5,07\ mm + 0,507\ mm}{2} = 2,79$ mm.
[103] C'est-à-dire « le point faisant face à la pupille ».

Il faut également observer qu'il est très difficile d'établir les proportions de l'œil assez exactement pour déterminer avec certitude la taille d'une image sur la rétine ; la position, la courbure et la constitution de la lentille de l'œil entrainent une différence si importante dans le résultat qu'il peut y avoir une erreur de presque un dixième de l'ensemble. Ainsi, afin d'obtenir quelque confirmation par l'expérience, j'ai placé deux chandelles à petite distance l'une de l'autre, ai tourné l'œil vers l'intérieur, puis ai appliqué l'anneau d'une clé de sorte à produire un spectre dont le bord coïncidait avec la chandelle intérieure ; puis en fixant mon œil sur l'extérieure, j'ai trouvé que le spectre avançait de plus de deux septièmes de la distance entre elles. D'où la même partie de la rétine qui sous-tendait un angle de sept parties au centre du mouvement de l'œil, sous-tendait un angle de cinq à l'intersection supposée des rayons principaux (Planche III. Fig. 11) ; et la {P.49} distance de la rétine à cette intersection était de 637 millièmes de pouce (*16,18 mm*). Ceci correspond de près au calcul précédent ; la distance du point de vision la plus parfaite au centre du nerf optique ne peut alors en aucun cas être bien inférieure à celle qui lui est attribuée ici. Et dans les yeux des quadrupèdes, la partie la plus colorée de la choroïde est plus éloignée du nerf que ne l'est l'axe réel de l'œil[104].

Je me suis efforcé de représenter en quatre figures la forme de chaque partie de mon œil aussi exactement que j'aie été capable de la déterminer ; la première (Pl. IV. Fig. 17) est une section verticale ; la seconde (Fig. 18) une section horizontale ; les troisième et quatrième sont des vues de face pour différents états de la pupille (Fig. 19 et 20).

Considérant le peu d'inconfort ressenti d'une inégalité de réfraction aussi importante que celle que j'ai décrite pour la lentille de l'œil, nous n'avons pas de raison de nous à attendre une propension très exacte à corriger l'aberration des rayons latéraux[105]. Mais autant qu'on puisse l'établir par l'optomètre, l'aberration émergeant de la forme est complètement corrigée ; car quatre images ou plus de la même ligne semblent se rencontrer exactement au même point, ce qu'elles ne feraient pas si les rayons latéraux étaient plus considérablement réfractés que les rayons proches de l'axe. La forme des surfaces est parfois, et peut-être toujours, plus ou moins hyperbolique (PETIT. Mém. De l'Acad. 1725, p. 20)[106] ou elliptique : au niveau des couches internes en effet, l'angle solide de la marge est quelque peu arrondi ; mais le plus faible pouvoir réfringent des parties externes doit avoir fortement tendance à corriger l'aberration émergeant de la courbure trop importante à la périphérie du disque. Si le pouvoir réfringent avait été uniforme, peut-être aurait-elle presque aussi bien réuni les rayons latéraux d'un pinceau direct ; mais elle eût été moins bien adaptée aux pinceaux de {P.50} rayons obliques ; et l'œil aurait aussi dû être encombré d'une masse de densité plus importante que celle présentement requise, même pour les parties centrales : et si l'intégralité de la lentille de l'œil avait été plus petite, elle aurait aussi laissé entrer trop peu de lumière. Il est également possible que la remarque de M. RAMSDEN (Phil. Trans. for 1795, p.2) sur l'avantage de ne pas avoir de surface réfléchissante puisse être bien fondée[107] : mais il n'a pas été démontré qu'en traversant un milieu de densité variable on perde moins de lumière que lors d'une transition soudaine d'une partie de ce milieu à une autre ; <bien qu'une telle conclusion puisse certainement être déduite de la seule hypothèse qui fournisse dans quelque cas que ce soit une explication de la cause d'une réflexion partielle>[108]. Mais ni cette gradation, ni aucune autre précaution, n'a pour effet de rendre l'œil parfaitement achromatique. Le Dr JURIN l'avait remarqué il y a longtemps (SMITH, e. 96)[109] en observant la couleur bordant l'image d'un objet vu indistinctement. Et sur l'optomètre, le Dr WOLLASTON m'a fait remarquer l'aspect rouge et bleu des angles internes opposés des lignes qui se croisent ; et mentionna par la même occasion une expérience très élégante pour prouver le pouvoir dispersif de l'œil. A travers un prisme, il observe un petit objet lumineux qui devient bien sûr un spectre linéaire. Mais l'œil ne peut s'adapter pour faire apparaître tout le spectre comme une ligne ; car si le foyer est adapté pour collecter les rayons rouges en un point, les bleus seront trop réfractés et se répandront en une surface ; et l'inverse arrivera si l'œil est adapté aux rayons bleus ; de façon que dans l'un et l'autre cas, la ligne sera vue comme un intervalle triangulaire. L'observation est confirmée en plaçant un petit miroir concave en différentes parties du spectre prismatique et en établissant les distances extrêmes auxquelles l'œil peut collecter les rayons de différentes couleurs en un foyer. Je trouve par ce moyen que les rayons rouges provenant d'un point à {P.51} 12 pouces (*30,48 cm*) de distance sont autant réfractés que la lumière jaune ou blanche à 11 (*27,94 cm*). La différence est égale à la réfraction d'une lentille de 132 pouces (*3,35 m*) de focale[110]. Mais l'aberration des rayons rouges dans une lentille en verre crown de pouvoir réfringent moyen égal à celui de l'œil serait équivalente à l'effet d'une lentille de 44 pouces (*1,17 m*) de focale. Par conséquent, si l'on peut s'appuyer sur ce calcul, le pouvoir dispersif de l'œil considéré globalement vaut un tiers du pouvoir dispersif du verre crown, pour un angle de déviation égal. Je ne peux pas observer beaucoup d'aberration des rayons violets. Cela peut être dû en partie à leur faiblesse[111]. Je crois que c'était l'opinion de M. RAMSDEN, que puisque la séparation des rayons colorés est seulement observée là où il y a un changement soudain de densité, un corps de densité variant graduellement, tel que la lentille de l'œil, n'aurait aucun effet que ce soit sur la séparation des rayons de différentes couleurs. Si cette hypothèse se révélait être bien fondée, nous devrions attribuer la totalité de la dispersion à l'humeur aqueuse ; et son pouvoir dispersif sera de la moitié de celui du verre crown pour la même déviation. Mais nous avons un exemple avec l'atmosphère d'un changement de densité très graduel ; et pourtant M. GILPIN m'informe que les étoiles, quand elles sont près de l'horizon, paraissent très manifestement colorées[112] < ; et le Dr HERSCHEL nous a même donné les dimensions d'un spectre ainsi formé>[113]. En une saison plus favorable de l'année, il ne serait pas difficile d'établir au moyen de l'optomètre le pouvoir dispersif de l'œil et de ses différentes parties avec une précision plus grande que par l'expérience relatée ici. Le pouvoir dispersif de l'œil dans sa globalité eût-il été égal à celui du verre flint, que les distances de vision parfaite auraient varié de 12 à 7 pouces (de *30,48 cm* à *17,78 cm*) pour les différents rayons, pour une même valeur des pouvoirs réfringents moyens[114].

---

[104] Tscherning s'efforce de reproduire le calcul qui a mené Young à toutes ces valeurs [Tscherning, 1894, 149-152]. Il en déduit que Young considérait comme véritable axe de l'œil une droite passant par le centre de la pupille et divisant une section horizontale du globe oculaire en deux parties égales, et que cet axe ne passait donc pas exactement par le sommet de la cornée mais correspondrait à peu près à l'axe du cristallin.

[105] La question traitée dans le passage qui suit est celle de l'aberration sphérique de l'œil, liée au fait qu'une surface réfringente sphérique n'est jamais véritablement stigmatique et fait converger les rayons incidents vers un point de l'axe dépendant de la distance à l'axe optique du point d'incidence de ces rayons de ces rayons. Cette aberration est étonnamment faible dans l'œil humain, étant données la forte courbure de la cornée et de la face avant du cristallin, ainsi que l'ouverture importante de la pupille relativement à ces courbures. C'est en bonne partie dû au fait que l'aplatissement et la diminution de l'indice de réfraction de la périphérie du cristallin mentionnés par Young tendent à diminuer son effet [Tscherning, 1896, 153].

[106] Dans un texte dédié à la technique et à l'histoire de l'opération de la cataracte, François Pourfour du Petit mentionne brièvement la forme des faces du cristallin qu'il qualifie d'abord brièvement de sphériques, pour ensuite ajouter : « La section de cette convexité postérieure m'a paru dans plusieurs yeux plutôt parabolique que sphérique, comme on le voit dans la seconde Figure KIL. On lit dans une Thèse soutenue à Altdorff en 1678 […] que le Cristallin est plutôt hyperbolique que sphérique » [Pourfour du Petit, 1725, 14].

---

[107] La conférence Croonienne de 1795 étant dédiée à la démonstration du rôle prépondérant de la cornée dans l'accommodation, son introduction prend le temps d'écarter le rôle du cristallin dans ce mécanisme en lui accordant d'autres fonctions. En l'occurrence, Home attribue à son collègue Ramsden – artisan lunetier – l'hypothèse que la densité non uniforme du cristallin a pour fonction de compenser l'aberration sphérique nécessairement importante de la cornée ; qui plus est, par analogie avec les objectifs achromatiques à plusieurs lentillescol, dans lesquels il a pu observer que chaque lentille supplémentaire entrainait une légère perte de lumière par réflexion à ses surfaces, Ramsden infère que l'adaptation d'indice entre la périphérie du cristallin et les humeurs qui l'entourent a aussi le rôle de diminuer ces réflexions parasites et donc de permettre une correction parfaite sans perte de lumière [Home, 1795, 2-4]. Tscherning fait remarquer qu'il existe dans l'œil des images de deuxième ordre produites par des rayons qui, après avoir subi une première réflexion sur le cristallin, en subissent une seconde sur la cornée et sont ainsi renvoyés vers la rétine. Et que si l'indice du cristallin avait été uniforme, il aurait aussi dû être plus fort, ce qui aurait mené à une réflexion partielle plus importante. Les images de réflexion auraient donc été plus brillantes, et donc certainement incommodantes [Tscherning, 1894, 154].

[108] Rajouté dans [Young, 1807, II : 584] à la place de : « non plus que nous soyons suffisamment renseignés sur la cause de cette réflexion pour être en capacité de raisonner sur le sujet de manière satisfaisante ». Noter que la question de la réflexion partielle de la lumière à la surface séparant deux milieux est cruciale pour Young, puisqu'il l'a présentée dans son article précédent comme l'une des difficultés majeures s'opposant à l'acceptation d'une théorie des projectiles [Young, 1800, 125]. L'explication de la réflexion partielle qu'il évoque ici superficiellement dans sa révision de 1807 est donc certainement celle qu'il aura développée l'année suivant la publication de cette conférence, dans le cadre de sa théorie vibratoire de la lumière [Young, 1802, 30-31].

[109] James Jurin (1684-1750) était un médecin anglais, célèbre pour ses travaux sur la capillarité et la variolisation. Fervent partisan de Newton, et notamment de son optique, Jurin a rédigé *An essay upon distinct and indistinct vision* qu'il a adressé à Robert Smith et qui par conséquent se voit publié à la fin du *Compleat System of Opticks* de ce dernier, auquel renvoie la référence mentionnée par Young. Dans l'article cité [Smith R., 1738, II : 132], Jurin évoque l'apparition d'une irisation violette à la périphérie d'un cercle noir observé en vision floue (par exemple en le maintenant très près de l'œil) sur un fond blanc. Il observe que l'image de la périphérie du cercle peut être suffisamment floutée pour que son épaisseur soit telle que l'intérieur du cercle paraisse entièrement sombre. Et il ajoute que dans ce cas, le bord de la zone sombre pourra paraître teintée de violet du fait de la dispersion par l'œil de la lumière provenant de la zone blanche du papier extérieure au cercle : les rayons donnant lieu à la sensation violette étant plus déviés que ceux causant la rouge, ils déborderont sur la zone perçue comme obscure.

VII. La faculté d'accommoder l'œil à différentes {P.52} distances semble exister en des degrés très différents chez différentes personnes. La distance de vision parfaite la plus courte pour mon œil est de 26 dixièmes de pouce pour les rayons horizontaux et 29 pour les verticaux. Cette puissance est équivalente à l'addition d'une lentille de 4 pouces de focale. Le Dr WOLLASTON peut voir à 7 pouces et avec des rayons convergents ; la différence correspondant à 6 pouces de distance focale. M. ABERNETHY[115] a une vision parfaite de 3 à 30 pouces, ou une puissance égale à celle d'une lentille de 3 pouces $\frac{1}{3}$ de focale. Une jeune dame de ma connaissance peut voir à 2 pouces et à 4 : la différence étant équivalente à 4 pouces de focale. Une dame d'âge moyen à 3 et à 4 ; la puissance d'accommodation étant égale à l'effet d'une lentille de 12 pouces de focale seulement[116]. En général, j'ai des raisons de penser que la faculté diminue à certain degré au fur et à mesure que les personnes avancent dans la vie ; mais certains d'âge moyen paraissent aussi n'en disposer qu'à un très faible degré. Je considérerai à présent la plage[117] de mon propre œil comme étant probablement autour de la moyenne, et rechercherai quels changements seront nécessaires afin de la produire ; que l'on suppose que le rayon de courbure de la cornée soit diminué, ou que la distance de la rétine à la lentille de l'œil soit augmentée, ou que ces deux causes agissent conjointement, ou que la forme de la lentille de l'œil elle-même subisse une modification[118].

---

[110] La puissance de la lentille permettant de réaliser une telle correction de mise au point en lumière jaune à la distance $d_j$ et en lumière rouge à la distance $d_r$ vaut donc $\frac{1}{d_j} - \frac{1}{d_r} = \frac{1}{f}$, ou 3,6 - 3,3 = 0,3 dioptrie.

[111] Il en effet est probable que l'aberration des rayons violets n'apparaisse pas du fait du manque de luminosité du spectre observé. Young corrigera d'ailleurs ces mesures dès l'année suivante [Young, 1802b, 396-397], sans rien véritablement changer à son expérience, mais en parvenant à y inclure des mesures sur les rayons causant la sensation de violet, probablement en travaillant avec une source plus lumineuse.

[112] Il s'agit probablement de George Gilpin, qui fut assistant astronome sur le second voyage du Capitaine Cook et qui était devenu à l'époque greffier de la Royal Society et secrétaire du Bureau des Longitudes.

[113] Remarque ajoutée dans [Young, 1807, II : 585]. Young fait probablement référence ici à un catalogue d'étoiles doubles publié par William Herschel dans lequel il fait état d'un facteur d'incertitude de ses mesures dû à l'« allongement » des étoiles produit par un « pouvoir prismatique de l'atmosphère, auquel les astronomes ont peu fait attention », et qu'il estime à 7''34''' pour une étoile du Sagittaire en particulier, observée à une date particulière [Herschel, 1785, 88-89].

[114] Tscherning propose à cet endroit un tableau récapitulatif complet des mesures par Young des propriétés de son propre œil, qu'il présente comme celui ayant été le mieux examiné encore cent ans plus tard [Tscherning 1894, 156-158].

 1. Nous avons calculé que lorsque l'œil est dans un état de relaxation, la réfraction de la cornée est telle qu'elle réunit les rayons divergeant depuis un point distant de 10 pouces (*25,4 cm*) en un foyer à la distance de $13\frac{2}{3}$ dixièmes (*3,47 cm*). Afin qu'elle puisse amener au même foyer des rayons divergeant depuis un point distant de 29 dixièmes (*7,37 cm*), on trouve (par le Cor. 5, Prop. IV) que son rayon de courbure[119] doit être diminué de 31 à 25 centièmes (de *7,87 mm* à *6,35 mm*), soit très proche du rapport de 5 à 4.

 2. En supposant que le changement de vision parfaite de 10 pouces à 29 dixièmes est effectué par un recul de la rétine à une plus grande {P.53} distance de la lentille de l'œil, ceci nécessitera (par le même Corollaire) une élongation de 135 millièmes (*3,4 mm*), soit plus d'un septième du diamètre de l'œil. Pour l'œil de M. ABERNETHY, une élongation de 17 centièmes (*4,32 mm*), soit plus d'un sixième, est requise.

 3. Si le rayon de courbure de la cornée est diminué d'un seizième, soit à 29 centièmes (*7,37 mm*), l'œil doit en même temps être allongé de 97 millièmes (*2,46mm*), soit environ un neuvième de son diamètre.

 4. En supposant que la lentille cristalline change sa forme ; si elle devenait une sphère, son diamètre serait de 28 centièmes (*7,11 mm*) et, sa face antérieure conservant sa position, l'œil aurait une vision parfaite à une distance d'un pouce et demi (*3,8 cm*) (Cor. 5 et 8, Prop. IV). C'est plus du double du changement véritable. Mais il est impossible de déterminer précisément l'ampleur de la modification nécessaire sans établir la nature des courbes en lesquelles ses surfaces peuvent être changées. S'il s'agissait toujours d'un sphéroïde plus ou moins aplati, la distance focale de chaque surface varierait inversement comme le carré de l'axe[120] : mais si les surfaces, de sphériques, devenaient des portions de conoïdes hyperboliques ou d'ellipsoïdes oblongs, ou passaient de formes plus obtuses à des formes plus aiguës de cette sorte, la distance focale varierait plus rapidement. En ne tenant pas compte de l'élongation de l'axe et en supposant que la courbure de chaque surface soit changée proportionnellement, le rayon de courbure de l'antérieure doit devenir égal à /21\ environ et celui de la postérieure à /15\[121] centièmes (*5,33 mm* et *3,81 mm*).

---

[115] John Abernethy (1764-1831) est un chirurgien et physiologiste britannique réputé sur lequel Young a certainement pratiqué lui-même ces mesures.

[116] C'est ici probablement la série la plus ancienne de personnes dont on aura mesuré précisément la plage d'accommodation, quantifiée aujourd'hui par l'amplitude dioptrique d'accommodation $a = \frac{1}{d_{pp}} - \frac{1}{d_{pr}}$ exprimée donc en dioptries, où $d_{pp}$ est la distance en mètres de l'œil au point le plus proche visible nettement (ou *punctum proximum*) et $d_{pr}$ la distance en mètres au point le plus éloigné visible nettement (ou *punctum remotum*) ; la distance de focale donnée dans le texte étant dès lors égale à $\frac{1}{a}$. En résumé, pour Young $d_{pp}$ vaut 6,60 et 7,37 cm pour les rayons horizontaux et verticaux, et $a$ = 9,5 et 9,6 dioptries dans les plans horizontal et vertical ; $d_{pr}$ ayant été estimé à 17,8 cm pour les rayons horizontaux et 25,4 cm pour les verticaux [Young, 1801, 39]. Pour Wollaston, $d_{pp} = 17,78$ cm et $a = 6,6$ dioptries. Pour Abernethy, $d_{pp} = 7,62$ cm, $d_{pr} = 76,2$ cm et $a = 11,8$ dioptries. Pour la jeune dame, $d_{pp} = 5,08$ cm, $d_{pr} = 10,16$ cm et $a = 9,8$ dioptries. Enfin pour la dame d'âge moyen, $d_{pp} = 7,62$ cm, $d_{pr} = 10,16$ cm et $a = 3,3$ dioptries.

[117] Sous-entendu : « d'accomodation ».

[118] Young entre maintenant dans le vif du sujet relatif à la cause de l'accommodation qui a fait tant débat, et qu'il s'engage dans la réfutation des thèses qui ont été opposées à son premier article [Young, 1793] et dans une confirmation de son hypothèse de la déformation du cristallin par de nouvelles expériences.

[119] Sous-entendu : « de la cornée », conformément aux valeurs déjà calculées [Young, 1801, 38].

VIII. Je vais maintenant procéder à la détermination de celui de ces changements qui a lieu dans la nature ; et je commencerai par le récit d'expériences réalisées dans le but de déterminer le rayon de courbure de la cornée dans toutes les circonstances.

La méthode décrite dans la Conférence Croonienne de M. HOME de {P.54} 1795 (Phil. Trans. for 1796, p. 2) parait être bien préférable à l'appareillage de l'année précédente (Phil. Trans. for 1795, p. 13)[122] : car une différence de distance entre deux images vues dans la cornée serait beaucoup plus importante et plus visible qu'un changement de sa protubérance, et bien moins susceptible d'être perturbée par des causes accidentelles. Il est presque, et peut-être totalement, impossible de changer la focalisation de l'œil sans un léger mouvement de son axe. Les yeux sympathisent parfaitement l'un avec l'autre ; et le changement de foyer est presque inséparable d'un changement de direction relative des axes optiques ; si bien que si je dirige mes deux yeux vers un objet situé au-delà de leur foyer le plus éloigné, je ne peux m'empêcher d'amener ce foyer un peu plus près[123] : lorsqu'un axe se déplace, il n'est pas aisé de garder l'autre parfaitement au repos ; et il n'est pas impossible que pour certains yeux, une légère modification de la position de l'axe soit rendue absolument nécessaire par un changement de leurs proportions. Ces considérations peuvent partiellement expliquer la très légère différence de position de la cornée qui a été observée en 1794. Il apparait que les expériences de 1795 furent réalisées avec une précision considérable et sans aucun doute avec d'excellents instruments ; et leur échec à établir l'existence d'un changement quelconque a induit MM. HOME et RAMSDEN à abandonner dans une large mesure l'opinion qui les avait suggérées, et à supposer qu'un changement de la cornée produirait seulement un tiers de l'effet. Le Dr OLBERS de Brême, qui en l'année 1780 publia une dissertation des plus élaborées sur les changements internes de l'œil (De Oculi Mutationibus internis. Gotting. 1780. 4°) qu'il a récemment présentée à la Royal Society, avait également échoué dans ses tentatives à mesurer ce changement de la cornée, alors même que son opinion était en faveur de son existence[124].

---

[120] « of the axis » dans la version originale, c'est-à-dire : « de l'épaisseur ».
[121] Corrections apportées en [Young, 1801, 84].
[122] Les conférences Crooniennes – traditionnellement dédiées aux mouvements musculaires – sont généralement présentées en novembre, et publiées en janvier de l'année suivante dans les Philosophical Transactions. Celles de 1794 et 1795 sont données par Everard Home et toutes deux dédiées à la démonstration du rôle prépondérant de la déformation de la cornée dans le mécanisme d'accommodation de l'œil [Home, 1795 ; Home, 1796]. Dans la première, la preuve expérimentale est notamment apportée par l'observation directe, à l'aide d'un microscope disposé latéralement, d'un mouvement longitudinal de la cornée accompagnant les changements de mise au point de l'œil d'un observateur dont le visage est fixé à un cadre en bois, et observant successivement plusieurs objets par la fenêtre du cabinet de travail à travers un trou fixe. Home admet lui-même combien l'observation est malaisée, mais conclue tout de même à la mesure d'un déplacement de la cornée – en conséquence de son changement de courbure – de l'ordre de 30 μm [Home, 1795, 13-18]. Un an plus tard, Home a décidé d'approfondir le sujet et déduit cette fois les changements de courbure de la cornée de l'observateur (dont le visage est toujours fixé au même cadre en bois) à partir des modifications de l'image réfléchie par celle-ci lorsqu'il accommode à différentes distances, observée au microscope [Home, 1796, 2-3]. L'expérience ne permet cette fois pas de mesurer de déplacement franc de la cornée, mais conclue toutefois que la sensibilité de la mesure, estimée à 200 μm, est compatible avec les mesures effectuées l'année précédente. Malgré l'insuccès apparent de cette nouvelle expérience, Home et Ramsden en concluent encore que le changement de courbure de la cornée a bien lieu et qu'il est bien la cause de l'accommodation des yeux dont on a retiré le cristallin [Home, 1796, 5]. Ils ajoutent toutefois que dans un œil sain, la cornée n'est responsable que d'un tiers de la capacité d'accommodation de l'œil ; les deux autres tiers du mécanisme étant assumés conjointement par un allongement de l'œil et par un déplacement du cristallin [Home, 1796, 8-9]. Cette expérience n'ayant pas été poussée assez loin pour être conclusive, Young se propose de la réaliser à son tour dans le prochain paragraphe ; à l'identique d'abord, puis en en imaginant autant de variantes que nécessaire pour convaincre que si l'on n'observe pas de déplacement de la cornée, c'est tout simplement qu'il n'y en a pas.
[123] Soit en fixant mes yeux au-delà du *punctum remotum*, je ne peux m'empêcher d'accommoder légèrement malgré tout.

{P.55} Il y avait encore lieu cependant de répéter les expériences ; et je commençai avec un appareillage ressemblant de près à celui que M. HOME a décrit. Je disposais d'un excellent microscope achromatique, réalisé par M. RAMSDEN pour mon ami JOHN ELLIS, de cinq pouces (*12,7 cm*) de distance focale, grossissant 20 fois environ. Je lui adaptai un micromètre gradué, placé au foyer de l'œil qui n'était pas employé à regarder à travers le microscope : c'était un grand carton, divisé en quarantièmes de pouce (*635 µm*) par des lignes horizontales et verticales. Lorsque l'on comparait l'image dans le microscope à cette échelle, on prenait soin de placer la tête de façon à ce que le mouvement relatif des images sur le micromètre, causé par l'instabilité de l'axe optique, soit toujours dans la direction des lignes horizontales, et qu'il ne puisse y avoir d'erreur due à ce mouvement dans les dimensions de l'image mesurées verticalement. Je plaçai deux chandelles de façon à faire paraître leurs images en position verticale dans l'œil de M. KÖNIG, qui avait la bonté de m'assister ; puis les ayant amenées dans le champ du microscope, où elles occupaient 35 des petites divisions, je désirai qu'il fixe son œil sur des objets situés à différentes distances dans la même direction : mais je ne pus percevoir la moindre variation de la distance des images.[125]

Rencontrant une difficulté considérable à ajuster proprement le microscope, et étant capable de me fier à mon œil nu pour mesurer des distances sans erreur supérieure au 500ème de pouce (*50 µm*), je décidai de réaliser une expérience similaire sans aucun pouvoir grossissant. Je construisis un oculaire divisé à partir de deux parties de lentilles, si petites qu'elles passaient entre deux images réfléchies par mon propre œil ; puis, en regardant dans une glace, j'amenai les lieux apparents des images à coïncider et réalisai ensuite le changement[126] nécessaire pour voir des objets plus proches : mais les images {P.56} coïncidaient toujours. Je ne pouvais pas non plus observer quelque changement que ce soit des images réfléchies par l'autre œil, où elles pouvaient être vues avec plus de commodité puisqu'elles n'interféraient pas avec l'oculaire. Mais n'étant pas alors conscient de la parfaite sympathie des yeux, je pensai plus certain de limiter mon observation à celui avec lequel je voyais. Je dois remarquer qu'avec un peu d'habitude, j'ai acquis une très vive maîtrise de l'accommodation de mon œil, au point d'être capable de considérer un objet avec attention sans ajuster mon œil à sa distance.[127]

---

[124] Heinrich Olbers (1758-1840) est un médecin allemand, diplômé de la même université que Thomas Young. Plus connu pour ses observations astronomiques – et notamment pour le paradoxe qui porte son nom, qui demande comment le ciel nocturne peut paraître essentiellement noir si l'univers infini est supposé contenir un nombre infini d'étoiles – Olbers a dédié sa thèse de médecine aux mouvements de l'œil, et évoque notamment son incapacité de mesurer un déplacement de la cornée [Olbers, 1780, 27-28 ; 39]. Son opinion du mécanisme de l'accommodation privilégie toutefois l'hypothèse d'un changement de forme de l'œil [Olbers, 1780, 43].

[125] Young explore ici expérimentalement l'hypothèse d'une modification du rayon de la courbure comme étant le mécanisme permettant l'accommodation de la vision à différentes distances en observant au microscope l'image de deux flammes de bougies disposées l'une au-dessus de l'autre réfléchies sur la cornée de son patient M. König – probablement Charles Konig (1774-1851), naturaliste d'origine allemande, ayant lui aussi étudié à Göttingen, et arrivé à Londres en 1800. La distance entre les images des bougies est mesurée à l'aide d'une échelle graduée placée devant l'œil de l'expérimentateur ne regardant pas dans le microscope, à une distance telle que si ses deux yeux sont ouverts, l'image de l'échelle graduée et la scène réfléchie par l'œil du patient observée au microscope fusionnent. Si la cornée changeait de courbure lorsque le patient faisait successivement la mise au point à différentes distances, l'intervalle entre les images des deux flammes réfléchies par la cornée devrait être modifié. Mais Young n'observe rien de la sorte.

J'ai également tendu deux fils un peu inclinés l'un par rapport à l'autre en travers d'un anneau, et les ai divisés en intervalles égaux par de petites taches d'encre. J'ai ensuite fixé l'anneau, appliqué mon œil juste derrière lui et placé deux chandelles devant moi, en des positions convenables, et une troisième sur un côté afin d'illuminer les fils. Puis, en disposant un petit miroir d'abord à quatre pouces (*10,16 cm*) de distance, et ensuite à deux (*5,08 cm*), je regardais les images réfléchies dans celui-ci et observais dans chaque cas quelle partie des fils elles croisaient exactement ; avec le même résultat que précédemment.[128]

Je fixai ensuite un micromètre gradué à une distance convenable, l'illuminai fortement et l'observai à travers un trou d'épingle au moyen duquel il devenait distinct pour tout état de l'œil ; et en regardant avec l'autre œil dans une petite glace, je comparais l'image avec le micromètre de la manière déjà décrite. Je changeai ensuite la distance focale de l'œil, de sorte que les points lumineux, étant trop éloignés pour être vus parfaitement, paraissaient s'étaler en surfaces ; et je notai sur l'échelle la distance entre leurs centres ; mais cette distance était invariable.[129]

---

[126] Sous-entendu « d'accommodation ».

[127] Pour cette mesure, Young officie sur son propre œil devant lequel il place un minuscule oculaire constitué de deux morceaux de lentilles accolés. Les deux morceaux de lentille sont suffisamment petits pour laisser la lumière des bougies parvenir à la cornée et être réfléchie sur celle-ci, pour passer ensuite l'une et l'autre dans un morceau de lentille différent, permettant aux deux images finales – observées à l'aide d'un miroir – de se superposer. La superposition des deux images permettant de ne plus avoir à mesurer leur éloignement. La réussite de l'expérience repose sur la capacité de Young à porter son attention sur les deux images superposées tout en accommodant sur des objets situés à différentes distances. Young évoque au passage la « parfaite sympathie des yeux » c'est-à-dire leur propension à accommoder systématiquement tous les deux à la même distance, même quand ils ne sont pas amenés à observer la même scène simultanément.

[128] Pour cette nouvelle expérience, la finesse de la mesure de la distance entre les images des deux bougies par réflexion sur face avant de la cornée est assurée par la faible inclinaison entre les deux fils quasi-parallèles fixés tout juste devant l'œil de Young : les images réfléchies sont alignées sur une perpendiculaire à l'un des deux fils et coïncident initialement chacune avec un fil, idéalement même à une graduation marquée par Young sur le fil ; dès lors un très faible décalage de l'une par rapport à l'autre peut être aisément détecté et mesuré en évaluant à quelle autre graduation du fil (donc pour quel autre écartement) les deux images se superposent à nouveau chacune à un fil. Mais à nouveau, aucun changement de leur position relative n'est détecté.

[129] Comme auront pu le constater certaines personnes ayant des difficultés à accommoder, un tout petit trou placé devant l'œil aura la propriété de remplacer l'image stigmatique – formée sur la rétine par la cornée et le cristallin – par une image de type de celles que l'on trouve dans la chambre noire – dont le trou d'épingle joue ici le rôle de sténopé. Dès lors, la profondeur de champ de l'œil devient potentiellement infinie : d'où la possibilité ici d'observer ainsi nettement le micromètre gradué quelle que soit sa distance, quel que soit l'état d'accommodation de l'œil. Avec l'autre œil, Young observe grâce à un miroir les images des deux chandelles réfléchies par la cornée du même œil qui observe le micromètre à travers le trou d'épingle. Et comme on l'a déjà évoqué, il est possible pour un observateur de faire fusionner les deux images très différentes observées par chacun des deux yeux ; et donc de superposer l'image de l'échelle graduée avec la réflexion des deux chandelles sur la cornée. L'œil devant lequel est placé le micromètre – et dans lequel se réfléchissent les chandelles – est enfin amené à accommoder successivement à différentes distances : pour les raisons déjà évoquée, le micromètre reste nettement observé, mais du fait de la sympathie entre les deux yeux déjà évoquée plus haut,

Finalement, je traçai une échelle diagonale sur un miroir à l'aide d'un diamant (Planche III. Fig. 12) et j'amenai les images en contact avec les lignes de l'échelle. Alors, puisque l'image de {P.57} l'œil à la surface d'un miroir occupe la moitié de ses dimensions réelles, à quelque distance qu'il soit observé, sa véritable taille est toujours le double de la mesure ainsi obtenue. J'illuminai la glace fortement et pratiquai un trou dans une bande étroite de carton noir que je maintenais entre les images ; et j'étais ainsi capable de les comparer avec l'échelle, quoique leur distance apparente fût le double de celle de l'échelle. Je les voyais dans tous les états de l'œil ; mais je ne pouvais percevoir aucune variation de l'intervalle entre elles.[130]

L'adéquation de ces méthodes peut être ainsi démontrée. Appliquez une pression le long du bord de la paupière supérieure avec un petit cylindre quelconque, un crayon par exemple, et l'optomètre montrera que le foyer des rayons horizontaux est un peu allongé, alors que celui des rayons verticaux est raccourci[131] ; un effet qui ne peut qu'être dû à un changement de courbure de la cornée. Non seulement l'appareillage décrit ici, mais l'œil même sans aucune assistance, sera capable de détecter un changement notable des images réfléchies par la cornée, bien que le changement soit beaucoup plus petit que celui nécessaire pour l'accommodation de l'œil à différentes distances.

Dans l'ensemble, je ne peux pas hésiter à conclure que si le rayon de courbure de la cornée était diminué ne serait-ce que d'un vingtième, le changement serait très aisément perceptible par certaines des expériences rapportées ; et la modification globale de l'œil nécessite un cinquième.

Mais il reste une expérience bien plus précise et décisive. D'un petit microscope de botaniste, j'extrais une lentille biconvexe de huit dixièmes de pouce (*20 mm*) de rayon de courbure et de distance focale[132], fixée dans un tube d'un cinquième de pouce (*5,1 mm*) de profondeur ; tout en scellant ses bords avec de la cire, je verse dedans un peu d'eau presque

---

l'œil observant les deux images réfléchies est forcé d'raccommode lui aussi à différentes distances et ne peut donc plus observer ces images nettement ; elles lui apparaissent alors comme des surfaces lumineuses dont il mesurera l'éloignement des centres.

[130] Pour cette expérience, Young ne semble plus utiliser qu'un seul œil. La mesure repose sur le fait que la taille de l'image d'un œil qui se regarde dans un miroir, évaluée à l'aide d'une échelle située précisément dans le plan du miroir, vaut la moitié de la taille réelle de l'œil en question. En effet, l'image d'un objet par un miroir plan étant symétrique de l'objet par rapport à la surface du miroir, d'une part la distance d'un objet à son image est le double de la distance de l'objet au miroir ; d'autre part la taille de l'image est égale à celle de l'objet. Par ailleurs, la taille apparente d'un objet vu à l'œil nu étant en première approximation inversement proportionnelle à la distance de l'œil à l'objet observé, un objet situé à une distance donnée sera perçu comme étant de même taille apparente qu'un autre objet de taille double et situé à une distance double. Ainsi, dans ce cas amusant où l'œil joue à la fois le rôle du point d'observation et de l'objet dont on observe l'image dans le miroir plan, l'image de l'œil se trouve bien à une distance double du miroir, et donc la taille de l'image de l'œil mesurée grâce à l'échelle gravée à la surface du miroir est deux fois moindre que la taille de cette image. Et comme celle-ci est égale à la taille de l'œil lui-même, la mesure de l'œil donnée par l'échelle gravée sur le miroir est égale à la moitié la taille réelle de l'œil. Ceci étant posé, Young utilise à nouveau le principe du sténopé décrite précédemment, afin cette fois d'observer nettement dans le miroir gravé l'image des deux chandelles réfléchies par la cornée de l'œil, quel que soit son état d'accommodation, et de pouvoir y évaluer leur écartement à l'aide de l'échelle. Mais il ne perçoit toujours pas de variation.

[131] « elongated » puis « shortened » en anglais ; c'est-à-dire en fait que le foyer des rayons horizontaux est « éloigné » « éloignant » le *punctum remotum* (ou *proximum*) des rayons se et que celui des rayons verticaux s'est « rapproché ».

[132] La distance focale d'une lentille mince sphérique biconvexe de rayon de courbure $R$ et d'indice $n_v$ étant égal à $f = \frac{R}{2.(n_v-1)}$, on peut effectivement — dans l'hypothèse où l'indice du verre est environ égal à 3/2 — affirmer que $f \approx R$ [Morizot, 2016, 106]. Ce qui revient aussi à appliquer le résultat $e = \frac{ngh}{g+h}$ du Corollaire 7, Proposition IV, en remplaçant $e$ par $f$ ; $g$ et $h$ par $R$ ; et $m/n$ (avec toujours $n = m - 1$) par $n_v$.

froide jusqu'à ce qu'il soit rempli aux trois quarts, et l'applique ensuite à mon œil, de manière à ce que la cornée entre jusqu'à mi-chemin dans le tube et soit partout en contact avec l'eau (Planche III. Fig. 13). Mon œil devient immédiatement presbyte, et le {P.58} pouvoir réfringent de la lentille, qui est réduit par l'eau à une distance focale d'environ 16 dixièmes de pouce (*40,6 mm*) (Cor. 5. Prop. IV), n'est pas suffisant pour remplir le rôle de la cornée, rendue inefficace par l'intervention de l'eau ; mais l'ajout d'une autre lentille de cinq pouces et demi (*14 cm*) de focale restaure mon œil dans son état initial, et un peu plus[133]. J'applique ensuite l'optomètre et je trouve la même inégalité entre les réfractions horizontale et verticale que sans l'eau ; et dans les deux directions, j'ai comme avant une puissance d'accommodation équivalente à une distance focale de quatre pouces (*10,16 cm*). A première vue en effet, l'accommodation semble être un peu moindre, et capable seulement d'amener l'œil depuis l'état adapté aux rayons parallèles jusqu'à un foyer à cinq pouces de distance (*12,7 cm*)[134] ; et cela me fit un moment imaginer que la cornée put avoir quelque effet très léger à l'état naturel ; mais en considérant que la cornée artificielle était un dixième de pouce (*2,54 mm*) environ en avant du lieu de la cornée naturelle, je calculai l'effet de cette différence et le trouvai exactement suffisant pour rendre compte de la diminution de la plage de vision. Je ne peux pas déterminer la distance de la cornée à la lentille de verre au centième de pouce (*250 μm*) ; mais l'erreur ne peut pas être bien plus grande, et elle pourrait être d'un côté comme de l'autre.

    Après cela, il m'est presque nécessaire de présenter des excuses pour avoir mentionné les précédentes expériences ; mais sur un sujet si délicat, nous ne pouvons avoir une variété trop grande de preuves concordantes.

    IX. M'étant ainsi convaincu que la cornée n'est pas impliquée dans l'accommodation de l'œil, mon objectif suivant fut de rechercher si l'on pouvait détecter la moindre altération de la longueur de son axe ; car cela paraissait être la seule alternative possible : et considérant qu'un tel changement doit s'élever à un septième du diamètre de l'œil, je me flattais de l'espoir de le soumettre à une mesure. Maintenant, si l'axe de l'œil {P. 59} était allongé d'un septième, son diamètre transverse devrait être diminué d'un quatorzième, et le demi-diamètre serait raccourci d'un trentième de pouce (*0,8 mm*)[135].

---

[133] Du fait de la proximité de leurs indices, la réfraction du verre à l'eau en sortie de la lentille est moins forte que lorsqu'elle se faisait du verre à l'air. Précisément, comme démontré en Proposition II, pour le passage du verre à l'eau le rapport des réfractions est $m'/n' = 8/9$. Par conséquent, en appliquant deux fois successivement le résultat du Corollaire 5, Proposition IV, pour les surfaces sphériques de la lentille supposées de même rayon de courbure $a$, mais l'une en contact avec l'air ($m/n = 4/3$), l'autre avec l'eau, la distance focale $e$ de la lentille devient $e = \frac{m'.m.a}{m+n'} = 2.a$. Par ailleurs, la cornée étant d'un côté en contact avec l'eau, de l'autre avec l'humeur aqueuse (composée à plus de 99% d'eau) n'est plus en mesure de produire de réfraction notable — la réfraction n'ayant lieu, par définition, qu'entre deux milieux d'indices différents — et ne joue donc plus aucun rôle dans la formation de l'image rétinienne : l'œil devient donc presbyte, puisqu'il n'est plus assez convergent. Alors, l'ajout d'une lentille de 140 mm de focale, placée au contact on l'imagine avec la première, produit un doublet de distance focale équivalente $f_{eq} = \frac{f_1 f_2}{f_1 + f_2}$ = 31,5 mm environ placé à l'entrée du tube (voir Corollaire 7, Proposition IV), et permettant ici de compenser la perte de convergence due à l'inactivation forcée de la cornée. En effet, la distance focale $f_c$ d'une cornée sphérique de 7,87 mm de rayon de courbure, séparant l'air de l'humeur aqueuse d'indice $n = 4/3$, vaut $f_c = \frac{n.R}{n-1}$ = 31,7 mm environ.

[134] Au lieu de 6,6 à 7,37 cm pour Young, selon le plan de propagation des rayons.

[135] Young modélise donc ici l'allongement de l'œil par un passage d'une forme sphérique à celle d'un ellipsoïde de rotation allongé selon l'axe visuel, tout en conservant un volume constant. Cette condition implicite implique qu'un allongement d'un septième du grand axe mène à une réduction d'un quatorzième du rayon de la section verticale centrale de l'œil.

Je plaçai donc deux chandelles de façon telle que lorsque l'œil était tourné vers l'intérieur et dirigé vers sa propre image dans une glace, la lumière réfléchie depuis l'une des chandelles par la sclérotique apparaissait sur sa marge externe, de façon à la définir distinctement par une ligne lumineuse ; et l'image de l'autre chandelle était vue au centre de la cornée. J'appliquai ensuite l'oculaire double et l'échelle du miroir de la manière déjà décrite[136] ; mais aucun d'entre eux n'indiquait la moindre diminution de la distance quand la distance focale de l'œil était modifiée.

Un autre test, bien plus délicat encore, consista à appliquer l'anneau d'une clé à l'angle externe quand l'œil était autant que possible tourné vers l'intérieur, en même temps qu'il était contraint par un solide anneau de fer ovale, pressé contre lui à l'angle interne. Je forçai la clé à l'intérieur aussi loin que la sensibilité des téguments le permettait, et la maintenais coincée par une pression modérée entre l'œil et l'os. Dans cette situation, le fantôme[137] causé par la pression qui s'étendait dans le champ de vision parfaite était très précisément défini ; il n'empêchait cependant en aucune façon la perception distincte des objets effectivement vus dans cette direction, contrairement à ce que j'imaginais préalablement ; et une ligne droite entrant dans le champ de ce fantôme ovale paraissait quelque peu infléchie vers son centre (Planche III. Fig. 14) ; distorsion aisément comprise en considérant l'effet de la pression sur la forme de la rétine. En supposant maintenant que la distance entre la clé et l'anneau de fer ait été, comme elle l'était réellement, invariable, l'élongation de l'œil devait être totalement empêchée ou presque ; et au lieu d'une {P.60} augmentation de la longueur de l'axe de l'œil, la tache ovale causée par la pression aurait dû s'étendre sur un espace au moins dix fois aussi grand que la partie la plus sensible de la rétine.[138] Mais aucune circonstance de la sorte n'eut lieu : la puissance d'accommodation était aussi étendue que jamais ; et il n'y eut pas de changement perceptible, ni dans la taille ni dans la forme de la tache ovale.

A nouveau, comme on l'a déjà observé, puisque les rayons qui passent par le centre de la pupille, ou plutôt par le sommet antérieur de la lentille de l'œil, peuvent être considérés comme délimitant l'image ; et puisque la divergence de ces rayons les uns par rapport aux autres n'est que très peu affectée par la réfraction de la lentille de l'œil, on peut toujours dire qu'ils divergent depuis le centre de la pupille ; et l'image sur la rétine d'un objet donné doit être considérablement agrandie par le recul de la rétine à une plus grande distance de la pupille et de la lentille de l'œil (Cor. Prop. V). Déterminer avec exactitude la grandeur réelle de l'image n'est pas si aisé qu'il parait à première vue ; mais outre la dernière expérience relatée, qui pourrait être employée comme argument dans ce but, il y a deux autres méthodes pour l'estimer. La première est trop dangereuse pour être d'un grand usage ; mais avec les précautions convenables elle peut être tentée. Je fixe mon œil sur un cercle de cuivre placé dans les rayons du Soleil et, après quelque temps, le déplace vers le micromètre gradué ; puis, changeant la focalisation de mon œil alors que le micromètre reste à distance donnée, je m'efforce de découvrir s'il y a une différence quelconque dans la grandeur apparente du spectre sur l'échelle ; mais je ne peux en discerner aucune[139]. Je n'ai pas insisté sur cette tentative ; tout spécialement parce que je n'ai pas été capable de rendre le {P.61} spectre suffisamment distinct sans désagrément ; et aucune lumière n'est suffisamment forte pour causer une impression permanente sur quelque partie de la rétine distante de l'axe visuel. J'ai donc eu recours à une autre expérience. Je plaçai deux chandelles de manière à ce qu'elles répondent exactement à l'étendue de la terminaison du nerf optique et, en marquant précisément le point vers lequel mon œil était dirigé, je réalisai le plus grand changement possible de sa distance focale ; m'attendant à ce que, s'il y avait quelque allongement de l'axe que ce soit, la chandelle externe paraisse s'éloigner vers l'extérieur sur l'espace visible (Planche III. Fig. 15). Mais ceci ne se produisit pas ; le lieu apparent de la partie obscure était précisément le même qu'auparavant. Je ne m'engagerai pas à dire que j'aurais pu observer une très infime différence dans un sens ou dans l'autre : mais je suis persuadé que j'aurais dû découvrir une modification de moins de la dixième partie de la totalité.

---

[136] C'est-à-dire, que Young applique à la détection d'un changement de forme de l'œil, deux expériences qu'il a déjà appliquées à la détection d'un changement de courbure de la cornée : l'une utilisant deux morceaux de lentille, l'autre utilisant une échelle gravée à la surface d'un miroir.

[137] « phantom » en anglais ; pour désigner à nouveau le « phosphène », plus souvent qualifié de « spectre » dans le reste ce texte.

[138] En somme, la longueur de l'œil étant contrainte, une tentative d'élongation de celui-ci augmenterait seulement la pression de l'œil sur l'anneau le maintenant en place ; provoquant une déformation plus grande encore de la rétine et donc une augmentation de la taille du phosphène.

On peut s'enquérir si aucun changement de la grandeur de l'image ne doit être attendu de quelque autre supposition ; et il paraitra qu'il est possible que les changements de courbure soient ainsi adaptés que la grandeur de l'image floue reste parfaitement constante. En effet, en calculant à partir des dimensions que nous avons utilisées jusque-là, on s'attendrait à ce que l'image soit diminuée d'environ un /quarantième\[140] par l'augmentation maximale de la convexité de la lentille. Mais le tout dépend de la position des surfaces réfringentes et de l'augmentation respective de leur courbure, qui du fait de la densité variable de la lentille de l'œil peut difficilement être estimée avec une exactitude suffisante[141]. La pupille eût-elle été placée avant la cornée que, quelle que soit la supposition retenue, la grandeur de l'image aurait été très variable : présentement, cet inconvénient est évité par la position de la pupille ; si bien que nous avons là un exemple supplémentaire de la perfection de cet admirable organe.

{P.62} D'après les expériences décrites, il parait hautement improbable qu'un quelconque changement matériel de la longueur de l'axe ait véritablement lieu ; et il est presque impossible de concevoir par quel pouvoir un tel changement pourrait être effectué. En agissant indépendamment de l'orbite, les muscles droits, avec la substance adipeuse située au-dessous d'eux, auraient certainement tendance à aplatir l'œil : car puisque leur contraction réduirait nécessairement la circonférence ou la superficie de la masse qu'ils contiennent et arrondirait toutes ses proéminences, leurs points d'attache au nerf et à la partie antérieure de l'œil doivent alors se rapprocher (Planche V. Fig. 21, 22). Le Dr. OLBERS compare les muscles et l'œil à un cône dont les côtés sont bombés, et qui par contraction seraient ramenés à une ligne droite[142]. Mais ceci nécessiterait une force afin de maintenir la cornée comme un point fixe à une distance donnée de l'origine des muscles ; une force qui certainement n'existe pas. Dans la position naturelle de l'axe visuel, l'orbite étant conique, l'œil pourrait être quelque peu allongé, bien qu'irrégulièrement, en étant forcé vers l'intérieur de celle-ci ; mais quand il est tourné d'un côté ou de l'autre, la même action raccourcirait plutôt son axe ; et il n'y a rien non plus autour de l'œil humain qui pourrait prendre sa place. Chez les quadrupèdes, les muscles obliques sont plus larges que chez les humains ; et ils pourraient contribuer à cet effet dans bien des situations. En effet, une partie du muscle orbiculaire du globe est attachée si près du nerf qu'il pourrait également coopérer à l'action : et je n'ai aucune raison de douter de l'exactitude du Dr. OLBERS qui affirme avoir effectué une élongation considérable en attachant des fils aux muscles d'yeux de cochons et de veaux[143] ; cependant il ne dit pas dans quelle position l'axe était fixé ; et la flaccidité de l'œil après la mort pourrait avoir rendu très aisé un changement qui {P.63} serait impossible dans un œil vivant. Le Dr. OLBERS mentionne aussi une observation du Professeur WRISBERG au sujet de l'œil d'un homme qu'il croyait dépourvu de pouvoir accommodatif de son vivant et chez qui il découvrit, après sa mort, qu'un ou plusieurs de ces muscles manquait[144] : mais ce défaut d'accommodation ne fut pas du tout établi avec exactitude. J'ai mesuré la distance de l'insertion du nerf au point d'attache du muscle oblique inférieur dans l'œil humain : il était d'un cinquième de pouce (*5 mm*) ; et de moins d'un dixième de pouce (*2,5 mm*) depuis le centre de la vision ; si bien que, quoique les muscles obliques forment à peu près l'arc d'un grand cercle autour de l'œil dans certaines positions, leur action serait plus apte à l'aplatir qu'à l'allonger. Nous avons donc des raisons d'être d'accord avec WINSLOW, quand il leur attribue l'office d'aider à soutenir l'œil sur ce côté où les os sont le plus déficients[145] : ils semblent aussi correctement calibrés pour empêcher qu'il soit trop tiré vers l'arrière par l'action des muscles droits. Et même s'il n'y avait pas de difficulté à supposer que les muscles allongent les yeux en toute position, on s'attendrait cependant au moins à une petite différence dans l'étendue du changement quand l'œil est dans des positions différentes, à un intervalle de plus d'un angle droit l'une de l'autre ; mais l'optomètre montre qu'il n'y en a pas.[146]

---

[139] Karl Bahr, qui a reproduit cette expérience de Young, a pour sa part mesuré une augmentation de la taille de l'image d'un quarantième environ ; comparable à ce que Young estime quelques lignes plus bas être le résultat attendu dans le cas où l'accommodation serait due à un changement de courbure du cristallin [Bahr, 1857].

[140] Correction demandée dans [Young, 1801, 84].

[141] Les travaux de Helmholtz ont démontré dans le courant du siècle suivant que l'accommodation ne se fait pas exactement comme Young le supposait, mais que la surface antérieure du cristallin se bombe relativement beaucoup plus que sa surface postérieure. La démonstration repose notamment sur une expérience ressemblant à l'un de celles que Young réalisait pour démontrer que la courbure de la cornée restait inchangée au cours de l'accommodation : éclairant l'œil latéralement par de sources lumineuses proches, Helmholtz parvient non seulement à observer l'image en réflexion de ces deux sources sur la cornée (dont l'écartement reste inchangé), mais aussi l'image de celles-ci en réflexion sur les faces antérieure et postérieure du cristallin. Et il observe un rapprochement très significatif des images en réflexion sur la face antérieure, et beaucoup plus léger pour l'autre, lorsque la mise au point de l'œil observé passe d'une distance éloignée à une distance plus proche [Helmholtz, 1867, I : 143-145]. Et ces changements, une fois quantifiés, révèlent accessoirement avoir pour effet une légère augmentation de la taille de l'image rétinienne, dans une proportion équivalente à celle mesurée par Bahr [Helmholtz, 1867, I : 162-163].

Le Dr. HOSACK allègue qu'il était capable d'accommoder son œil à un objet plus proche en réalisant une pression sur celui-ci (Phil. Trans. for 1794. p. 212)[147] : il n'apparait pas qu'il fit

---

[142] [Olbers, 1780, 32-33]. Les Fig. 21 et 22 de la Planche V illustrent bien en fait la proposition de Olbers.
[143] [Olbers, 1780, 35].
[144] [Olbers, 1780, 37]. Heinrich August Wrisberg (1739-1808) était professeur d'anatomie à l'université de Göttingen où Young et Olbers ont étudié la médecine.
[145] Jacob Benignus Winsløw (1669-1760) est un anatomiste franco-danois à qui Young renvoie dans son précédent article sur la vision pour sa description « en générale très exacte » des principales parties de l'œil et de ses muscles [Young, 1793, 169]. Sa remarque sur les muscles obliques ici évoquée se trouve notamment dans [Winsløw, 1721, 316].
[146] Même si les idées de Young sur l'action des muscles sont incorrectes, les raisons qu'il donne contre l'hypothèse attribuant l'accommodation aux muscles externes de l'œil restent valables [Tscherning, 1894, 175].
[147] David Hosack (1769-1835) était un médecin et botaniste américain. Son article sur la vision auquel Young fait référence [Hosack, 1794] n'est pas tant intéressant pour les thèses qu'il soutient que par ce qu'il nous dit de l'état de la recherche sur la question de l'accommodation à l'époque. La vivacité et l'importance des débats sur ce sujet sont déjà bien illustrées par le fait que quatre conférences Crooniennes et une conférence Bakerienne sont dédiées à ce sujet entre 1794 et 1802. Toutefois, publié dans le même volume que la conférence Croonienne de 1794 défendant la priorité de Hunter pour la découverte du rôle du cristallin dans le mécanisme d'accommodation [Hunter, 1794], l'article de Hosack défend l'idée que celle-ci se ferait par modification de la longueur de l'œil. Mais en introduction à son article, c'est à la théorie selon laquelle l'accommodation serait due à une dilatation plus ou moins importante de la pupille qu'il décide de se confronter extensivement et en priorité,

usage de moyens très exacts pour établir ce fait ; mais si un tel effet eut lieu, une inflexion de la cornée doit en avoir été la cause.

Il n'est pas nécessaire de s'attarder sur l'opinion qui suppose une opération conjointe de changements de courbure de la cornée et {P.64} de longueur de l'axe. Cette opinion avait tiré une très grande respectabilité de la manière des plus ingénieuse et élégante dont le Dr. OLBERS l'avait traitée, et d'avoir été le dernier résultat de l'investigation de M. HOME et M. RAMSDEN. Mais chacune des séries d'expériences qui ont relatées parait suffisante pour la réfuter.

X. Reste maintenant à enquêter sur les prétentions de la lentille cristalline au pouvoir d'altérer la distance focale de l'œil. La grande objection à l'efficacité d'un changement de forme de la lentille a été tirée des expériences dans lesquelles ce qui en ont été privés paraissent posséder la faculté d'accommodation.

Mon ami M. WARE, aussi convaincu qu'il était de la méticulosité et de l'exactitude des expériences relatées dans la Conférence Croonienne de 1795, ne put cependant s'empêcher d'imaginer qu'il doit y avoir dans de tels cas un défaut de cette faculté, du fait de l'avantage évident que tous ses patients trouvèrent, après extraction de la lentille de l'œil, à utiliser deux sortes de lunettes[148]. Cette circonstance, combinée à une considération des indications très judicieusement données par le Dr PORTERFIELD pour examiner le point en question[149], me firent d'abord souhaiter répéter les expériences sur divers individus et avec l'instrument que j'ai décrit plus haut comme une amélioration de l'optomètre du Dr PORTERFIELD : et je dois ici reconnaître ma dette considérable envers M. WARE pour l'empressement et la libéralité avec lesquelles il m'introduisit à ceux de ses nombreux patients qu'il pensait le plus à même de fournir une mesure satisfaisante. Il n'est pas nécessaire d'énumérer chaque expérience particulière ; mais le résultat universel est, contrairement à l'attente avec laquelle j'entrai dans cette enquête, que dans un œil dépourvu de lentille cristalline, la {P.65} distance focale réelle est absolument immuable. Ceci apparaitra d'une sélection des observations les plus décisives.

1. M. R. peut lire à quatre pouces et à six (*10,16 cm* et *15,24 cm*) seulement, avec le même verre. Il a vu la double ligne se croiser à trois pouces (*7,62 cm*) et toujours au même point ; mais la cornée était irrégulièrement proéminente et sa vision n'était pas très distincte ; je n'avais pas non plus d'appareil convenable au moment où je l'ai vu.

J'adaptai plus tard à un petit optomètre une lentille de moins de deux pouces (*5,08 cm*) de focale, y ajoutant une série de lettres dans un ordre non alphabétique et projetées sous une forme qui soit plus lisible selon une faible inclinaison. L'excès de pouvoir grossissant avait l'avantage de rendre les lignes plus divergentes et leur croisement plus visible ; et les lettres servaient à désigner plus facilement la distance de l'intersection, et à juger en même temps

---

tant cette thèse lui semble la plus communément admise. Il ne dédie que beaucoup plus loin un long passage au discrédit de la thèse de la contraction du cristallin, prenant pour cible très spécifiquement l'article de Young [Young, 1793], mais se focalisant essentiellement sur le rejet de cette partie de la théorie selon laquelle le cristallin lui-même serait être un muscle capable de se contracter [Hosack, 1794, 201-206]. Ainsi voit-on mieux combien ce sujet pouvait être disputé à l'époque où Young écrivait ces lignes, et pourquoi il ressent le besoin manifeste de se positionner si exhaustivement dans le foisonnant paysage théorique dédié à la question.

[148] Sous-entendu donc des lunettes pour voir de loin et d'autres pour voir de près. James Ware (1746-1815) est un chirurgien londonien fameux à l'époque, spécialisé en ophtalmologie. Son expertise dans l'opération de la cataracte (voir par exemple [Ware, 1801]) fait de lui une référence de premier ordre dans le débat concernant la possibilité pour un patient démuni de cristallin d'accommoder à différentes distances. Et non seulement assura-t-il son concours à Thomas Young dans la tenue de ses investigations sur le sujet, mais aussi lui donna-t-il accès à une partie de sa patientèle, afin qu'il puisse les soumettre à l'utilisation de son optomètre.

[149] [Porterfield, 1738].

de l'étendue du pouvoir de distinguer les objets trop proches ou trop éloignés pour une vision parfaite (Planche V. Fig. 23).

 2. M. J. n'avait pas un œil très convenable pour l'expérience ; mais il semblait distinguer les lettres à 2½ pouces (*6,35 cm*) et à moins d'un pouce (*2,54 cm*). Ceci me persuada d'abord qu'il devait avoir le pouvoir de changer la distance focale : mais je me souvins ensuite qu'il avait considérablement reculé son œil pour regarder les lettres les plus proches, et qu'il avait aussi partiellement fermé ses paupières, contractant sans doute en même temps l'ouverture de la pupille ; une action qui, même dans un œil parfait, accompagne toujours le changement de focalisation. La bande coulissante[150] ne fut pas utilisée.

 3. Mademoiselle H., une jeune dame de vingt ans environ, avait une pupille très étroite et je n'ai pas eu d'opportunité d'essayer le petit optomètre : mais une fois qu'elle avait vu un objet double à travers les fentes, aucun effort ne pouvait plus le lui faire apparaitre unique à la même {P.66} distance. Elle utilisait un verre de 4½ pouces (*11,43 cm*) de focale pour les objets distants ; avec celui-ci elle pouvait lire aussi loin qu'à 12 pouces (*30,48 cm*) et aussi près qu'à cinq (*12,7 cm*) : pour des objets plus proches elle en ajoutait une autre de focale égale, et pouvait alors lire à 7 pouces et à 2½ (*17,78 cm et 6,35 cm*).

 4. HANSON, un charpentier âgé de 63 ans, avait eu une extraction de la cataracte d'un œil quelques années auparavant : la pupille était limpide et large et il voyait bien pour travailler avec une lentille de $2\frac{3}{8}$ pouces (*6,03cm*) de focale ; et pouvait lire à 8 et à 15 pouces (*20,32 cm et 38,1 cm*), mais plus commodément à 11 (*27,94 cm*). Avec le même verre, les lignes de l'optomètre paraissaient toujours se rencontrer à 11 pouces (*27,94 cm*) ; mais il ne pouvait pas percevoir qu'elles se croisaient, la ligne étant trop épaisse et l'intersection trop distante. L'expérience fut répétée plus tard avec le petit optomètre : il lut les lettres de 2 à 3 pouces (*5,04 cm à 7,62 cm*) ; mais l'intersection était toujours à 2½ pouces (*6,35 cm*). Il comprenait maintenant complètement les circonstances qu'il fallait remarquer et voyait le croisement avec parfaite distinction : à un moment, il dit qu'il était un dixième de pouce (*2,54 mm*) plus près ; mais j'observai qu'il avait reculé son œil de deux ou trois dixièmes (*5,08 mm ou 7,62 mm*) par rapport au verre, une circonstance qui rendait compte de cette petite différence.

 5. Nonobstant l'âge de Hanson, je le considère comme un sujet très correct pour l'expérience. Mais plus irréprochable encore était l'œil de Mme MABERLY. Ella a environ 30 ans et le cristallin de ses deux yeux avait été extrait quelques années auparavant, mais elle voit mieux avec le droit. Elle marche sans verres ; et avec l'assistance d'une lentille d'environ quatre pouces (*10,16 cm*) de focale elle peut lire et travailler aisément. Elle pouvait distinguer les lettres du petit optomètre d'un pouce à 2½ pouces (*2,54 cm à 6,08 cm*) ; mais l'intersection était invariablement au même point, distant d'environ 19 dixièmes de pouce (*4,83 cm*). Une partie de la capsule[151] est étirée en travers de la pupille et lui fait voir en double les objets éloignés quand elle est sans ses {P.67} verres ; elle ne peut pas non plus ramener les deux images plus proches l'une de l'autre par quelque effort que ce soit, bien que l'effort les rende plus distinctes, sans doute par contraction de la pupille. L'expérience avec l'optomètre fut conduite en présence de M. WARE, avec patience et persévérance ; aucune opinion ne fut exprimée qui aurait pu rendre son retour partial.

 Considérant la difficulté à trouver un œil parfaitement approprié aux expériences, ces preuves peuvent être considérées comme tolérablement satisfaisantes. Mais puisqu'un argument positif contrebalancera plusieurs négatifs, à condition qu'il soit équitablement fondé sur des faits, il devient nécessaire d'enquêter sur la compétence de la preuve employée pour établir le pouvoir accommodatif attribuée à l'œil de Benjamin Clerk dans la Conférence Croonienne de 1794. Et il semble que la distinction réalisée très proprement depuis longtemps par le Dr Jurin entre vision distincte et vision parfaite fournira sans hésitation la justification de l'intégralité de cette preuve.[152]

---

[150] Sous-entendu : « de l'optomètre ». C'est-à-dire la bande de carton dans laquelle a été découpée la série de deux, trois quatre ou cinq fentes parallèles à travers lesquelles l'œil peut regarder l'échelle graduée de l'optomètre (voir Planche III, Fig. 7).

[151] Sous-entendu : « du cristallin ».

Il est évident que, les pouvoirs réfringents de l'œil restant inchangés, la vision peut être rendue distincte à quelque distance donnée que ce soit au moyen d'une ouverture suffisamment petite, pourvu qu'en même temps subsiste une quantité suffisante de lumière[153]. Et il est remarquable que dans ces expériences, lorsque la comparaison avec l'œil parfait était réalisée, l'ouverture de l'œil imparfait était considérablement réduite. Benjamin Clerk, avec une ouverture de $\frac{3}{40}$ de pouce (*1,9 mm*), pouvait lire avec le même verre à $1\frac{7}{8}$ pouce (*4,76 cm*) et à 7 pouces (*17,78 cm*) (Phil. Trans. for 1795. p. 9)[154]. Avec une ouverture égale, je peux lire à 1½ pouce (*3,81 cm*) et à 30 pouces (*76,2 cm*) : et je peux maintenir l'état de repos parfait et lire avec la même ouverture à $2\frac{1}{4}$ pouces (*5,7 cm*) ; et c'est une différence aussi grande que celle qui fut observée pour {P.68} l'œil de Benjamin Clerk. Le fait que Sir Henry Englefield, aussi bien que les autres observateurs, fut très étonné de l'exactitude avec laquelle l'œil de l'homme était ajusté à la même distance au cours des essais répétés qui furent réalisés avec lui est également d'une importance non négligeable (Phil. Trans. for 1795. p. 8)[155]. Cette circonstance à elle seule rend hautement probable le fait que sa vision parfaite était confinée à des limites très étroites.

Je me suis efforcé jusque-là de montrer les inconvénients accompagnant les autres suppositions et de dissiper les objections à l'opinion d'un changement interne de la forme de la lentille de l'œil. Je ferai maintenant état de deux expériences qui, en premier lieu, s'approchent de très près d'une démonstration mathématique de l'existence d'un tel changement[156] et qui, dans un second, expliquent dans une large mesure son origine et la manière dont il est effectué.

---

[152] Rappelons que dans la première conférence Croonienne de Home, l'argument de l'inutilité de cristallin pour l'accommodation reposait essentiellement sur le constat, réalisé sur un unique patient dénommé Benjamin Clerk, d'une capacité d'accommoder demeurée intacte malgré une opération de la cataracte de l'œil de droit [Home, 1795, 5-10]. Et Young avance ici l'idée que cette observation relève probablement d'une erreur de jugement du patient et des expérimentateurs qui interprétèrent à tort sa capacité à discerner des objets situés à différentes distances (« vision distincte ») comme une capacité à faire la mise au point sur ceux ceux-ci (« vision parfaite »). Deux notions pourtant nettement distinguées par James Jurin dès les premières pages de son *Essay on Distinct and Indistinct Vision* [Smith R., 1738, II : 115-117].

[153] Il s'agit du fait déjà évoqué que l'apposition d'un tout petit trou juste devant l'œil limite son entrée à un cône très fin de rayons lumineux issu de chaque point de la scène observée, venant presque sans être dévier en dessiner une image non pas parfaite mais bien distincte, directement sur la rétine : ce petit trou réduit ainsi le fonctionnement de l'œil à celui d'une chambre noire, de profondeur de champ infinie, mais aussi de luminosité très faible.

[154] [Home, 1795, 9].

[155] Dans le but de donner plus de poids encore à ses observations, Home relate en effet dans sa conférence Croonienne de 1795 la manière dont Sir Henry Englefield (1752-1822), baronet antiquaire et astronome, membre de la Royal Society, a assisté à une partie des expériences sur Benjamin Clerk. Et combien il fut aussi surpris que Home et Ramsden « de l'exactitude avec laquelle l'œil de l'homme était ajusté à la même distance au cours des essais répétés qui furent réalisés avec lui » : ici Young reprend en fait précisément les mots de Home [Home, 1795, 8] mais dans le but justement de discréditer la plausibilité d'une observation que la formule originale présentait comme extraordinaire afin de renforcer rhétoriquement sa portée.

J'ai déjà décrit les aspects de l'image imparfaite d'un minuscule point positionné à différentes distances de l'œil à l'état de repos. Dans le but présent, je me contenterai de répéter que si le point est au-delà de la distance focale la plus éloignée de l'œil[157], il prend l'aspect généralement décrit par le nom d'étoile, la partie centrale étant considérablement la plus brillante (Planche VI, Fig. 36-39). Mais quand la distance focale de l'œil est raccourcie, l'image imparfaite est bien sûr agrandie ; et en plus de cette conséquence nécessaire, la lumière est aussi très différemment distribuée ; la partie centrale devient tout juste visible, et la marge fortement illuminée, de façon à avoir presque l'aspect d'un anneau ovale (Fig. 41). Si j'utilise la bande coulissante de l'optomètre alors que l'œil est détendu, les ombres des fentes sont parfaitement droites, divisant l'ovale dans chaque direction en sections parallèles (Fig. 42, 44) : mais quand {P.69} l'accommodation a lieu, elles deviennent immédiatement courbées, et d'autant plus qu'elles sont éloignées du centre de l'image vers laquelle leur concavité est dirigée (Fig. 43, 45). Si le point est amené bien en aval de la distance focale, le changement de l'œil augmentera l'illumination du centre aux dépens de la marge. Les mêmes aspects sont également observables quand l'effet de la cornée est supprimé par immersion dans l'eau ; et la seule façon imaginable de rendre compte de cette diversité est de supposer que les parties centrales de la lentille acquièrent un plus grand degré de courbure que les parties marginales. Si la réfraction par le cristallin restait la même, il est absolument impossible qu'un quelconque changement de la distance de la rétine produise une courbure de ces ombres qui, dans l'état de repos de l'œil, se trouvent être droites en toute part ; et le fait que ni la forme, ni la position relative de la cornée ne soient concernées, apparait de l'apposition d'eau déjà mentionnée.[158]

---

[156] La prétention de Young à démontrer mathématiquement un mécanisme biologique le démarque très certainement du travail des nombreux anatomistes dont il été fait état au fil du texte. Lui seul semble concerné par la nécessité d'associer la précision des observations anatomiques, à la rigueur de l'expérimentation et des mesures physiques et à la sanction du calcul mathématique – ou peut-être était-il seul en mesure de le faire. Les mathématiques offrant manifestement le mode le plus élevé d'administration de la preuve, comme en témoigne la préface de son cours à la Royal Institution : « Profondément impressionné par l'importance des investigations mathématiques, à la fois pour l'avancement de la science et pour l'amélioration de l'esprit, je pensai qu'il était en premier lieu mon indispensable devoir de présenter à la Royal Institution, dans le programme de mon cours, un système de philosophie naturelle connecté selon un plan rarement, sinon jamais, exécuté avant cela dans la plupart des traités. Sur toute l'étendue de la philosophie naturelle, le programme contient une démonstration stricte de chaque proposition que j'ai jugée nécessaire d'employer partant des axiomes les plus simples de mathématiques abstraites » [Young, 1807, I : 6]. C'est certainement ce même programme de fondation de ses propositions sur la base de démonstrations mathématiques qui a mené Young à introduire cette conférence *sur le mécanisme de l'œil* par une si longue section de propositions d'optique géométrique. C'est indéniablement ce qui fait la force de sa démonstration. C'est probablement aussi ce qui a pu rendre sa lecture pénible à certains spécialistes du domaine. Quoi qu'il en soit, l'ambition de Young à ce moment du texte est bel et bien de fournir une preuve « aussi proche que possible d'une démonstration mathématique » de la déformation du cristallin, parce que reposant sur le calcul optico-mathématique de la modification de forme d'images rétiniennes obtenues pour différentes courbures du cristallin, et sur la comparaison de ces résultats à l'expérience.
[157] Donc au-delà du *punctum remotum*.
[158] Les images décrites ici par Young sont à mettre sur le compte de l'aberration sphérique de l'œil, et plus spécifiquement du cristallin. Celle-ci est liée au fait déjà évoqué que les rayons issus d'un même point objet réfractés par les surfaces et lentilles sphériques (comme c'est le cas des lentilles ordinaires) ne convergent pas tous exactement vers un même point. Par exemple, un faisceau de rayons parallèles incident sur une lentille mince à faces sphériques verra ses rayons périphériques converger vers un point de l'axe situé légèrement en avant du foyer image paraxial ; on pourrait dire dès lors que la périphérie de la lentille est légèrement plus réfringente, si ce n'était pas seulement dû à un effet de sa forme, qui ne permet une formation stigmatique de

La vérité de cette explication est pleinement confirmée par l'optomètre. Lorsque je regarde à travers quatre fentes étroites sans faire d'effort, les lignes paraissent toujours se rencontrer en un point : mais lorsque je fais en sorte que l'intersection s'approche de moi, les deux lignes externes se rencontrent considérablement au-delà des deux internes, et les deux lignes du même côté se croisent à une distance encore plus grande (Planche V. Fig. 24).[159]

L'expérience ne réussira pas avec tous les yeux ; non plus que l'on puisse s'attendre à ce qu'une telle imperfection soit universelle : mais un cas est suffisant pour fonder l'argument, même aucun autre n'était trouvé. Je ne doute cependant pas que chez ceux qui ont une pupille large l'aberration soit très fréquemment observable. Pour l'œil du Dr WOLLASTON la diversité d'aspect est imperceptible ; mais M. KÖNIG a décrit les intersections exactement comme {P.70} elles m'apparaissaient, bien qu'il n'ait reçu la moindre indication de ce que j'avais observé. La réfraction latérale est la plus aisément déterminée, en substituant un morceau de carton fuselé aux fentes, de manière à couvrir toutes les parties centrales de la pupille et en déterminant ainsi le croisement le plus proche des ombres transmises par les parties marginales seulement. Quand l'intersection la plus éloignée était à 38, je pouvais l'amener à 22 parties avec deux fentes étroites ; mais avec le carton fuselé à 29 seulement. De ces données, nous pouvons déterminer assez exactement la forme en laquelle la lentille de l'œil doit être changée, en supposant que les deux surfaces subissent des modifications proportionnelles de leur courbure et en tenant pour acquises les dimensions déjà présentées : car de l'aberration latérale ainsi donnée, nous pouvons trouver (par la Prop. III) les sous-

---

l'image d'un point objet qui n'est approximativement valable que pour les rayons lumineux peu inclinés et proches de l'axe. L'aberration sphérique est donc d'autant plus importante que les surfaces réfringentes traversées seront courbées. Dans sa traduction, Tscherning propose une explication très complète du problème décrit ici par Young [Tscherning, 1894, 183-191] et remarque que l'on pourrait réaliser les mêmes observations en projetant sur un écran l'image d'un point lumineux éloigné par une simple lentille sphérique. Si l'écran est positionné après le point image, on retrouve le halo diffus avec un point lumineux au centre décrit par Young (les rayons paraxiaux s'éloignent lentement de l'axe et créent le point brillant, quand le halo est dû aux rayons périphériques divergeant plus fort et depuis un point situé avant le foyer paraxial). Si l'écran est positionné avant le point image, on retrouve le halo diffus entouré d'une bordure brillante (aucun rayon n'a encore atteint l'axe, mais les rayons latéraux croisent les rayons paraxiaux, créant ainsi un excès de luminosité à la périphérie). Surtout, la même expérience dans laquelle on introduirait au-devant de la lentille un objet tel qu'une aiguille à tricoter alignée avec un diamètre de celle-ci fait également apparaître les ombres déformées signalées par Young : les bords de l'ombre de l'aiguille restent rectilignes lorsque l'écran est situé à proximité du foyer paraxial, mais deviennent courbes sinon, avec une courbure tournée vers le centre quand l'écran est situé avant le foyer et vers la périphérie quand il est situé après. L'aberration sphérique justifie aussi ce phénomène : en modélisant la lentille par une série de zones concentriques d'autant plus réfringentes que l'on s'éloigne vers le bord, on remarque que chaque zone concentrique fait converger les rayons incidents frôlant l'aiguille vers un point d'autant plus près de la lentille que le rayon incident est écarté de l'axe. Si l'écran est placé avant le foyer, les rayons périphériques seront déjà plus proches de l'axe que les paraxiaux donc l'ombre plus étroite à la périphérie qu'au centre ; si l'écran est après le foyer, les rayons périphériques ayant croisé l'axe plus tôt et sen éloignant plus vite formeront une ombre plus large qu'au centre. Et bien sûr, si l'aberration sphérique était corrigée, cet effet disparaitrait. Ainsi, par analogie entre ces effets de l'aberration sphérique d'une lentille ordinaire et les observations qu'il fait à l'œil nu, Young se sent en mesure de déduire que l'aberration sphérique du cristallin change au cours de l'accommodation ; et même d'en déduire la déformation subie par le cristallin au cours du processus, en rapportant la variation d'aberration sphérique observée à une variation de courbure locale. C'est l'une des conséquences du travail poussé d'optique géométrique qu'il a pris le temps de présenter en introduction à ce texte.

[159] L'observation décrite ici par Young est effectivement équivalente de la précédente, puisqu'utilisant son optomètre à quatre fentes, la ligne axiale de l'instrument se trouve – comme déjà justifié – multipliée par quatre ; que ces quatre lignes se croisent en un même point (le *punctum remotum*) lorsque l'œil est au repos ; mais que dès lors que l'œil accommode sur un autre point de l'axe, les images de ces lignes, produites par des rayons lumineux ayant pénétré l'œil à différentes distances de son axe, manifestent ne plus avoir de foyer commun.

tangentes à environ un dixième de pouce (*2,54 mm*) de l'axe ; et le rayon de courbure à chaque sommet est déjà déterminé comme étant d'environ 21 et 15 centièmes de pouce (*5,3 mm et 3,8 mm*). D'où la face antérieure doit être une portion d'hyperboloïde dont le grand axe est d'environ 50 (*1,27 cm*) ; et la face postérieure sera presque parabolique. De cette manière, le changement sera effectué sans aucune diminution du diamètre transverse de la lentille. L'élongation de son axe n'excédera pas le cinquantième de pouce (*0,5 mm*) ; et selon la supposition dont nous partons, la protrusion se fera surtout au sommet postérieur. La forme de la lentille de l'œil ainsi changée sera à peu près celle de la Planche V. Fig. 26 ; l'état de repos étant à peu près comme représenté en Fig. 25. Cependant, la rigidité des parties internes ou toute autre considération dussent-elles rendre plus commode de supposer que la face antérieure soit plus changée, qu'il y aurait encore la place pour ce faire sans interférer avec l'uvée ; ou elle pourrait même forcer un peu l'uvée vers l'avant, sans altération visible de l'aspect extérieur de l'œil[160].

<On comprend aisément pourquoi, et dans quels cas, une telle imperfection de la réfraction latérale doit exister, de la façon dont la périphérie de la lentille est attachée à sa capsule. Car si l'on augmente la courbure au niveau de l'axe à quelque degré considérable, elle ne peut se prolonger bien loin vers la périphérie sans amoindrir le diamètre de la lentille et sans déchirer les ramifications qui y pénètrent depuis les procès ciliaires. Il ne parait pas non plus y avoir d'autre raison à la contraction très observable de la pupille qui accompagne toujours l'effort de voir des objets proches, que le fait que par ce moyen l'on exclue les rayons latéraux et l'on empêche le flou qui aurait émergé de l'insuffisance de leur réfraction.>[161]

{P.71} Il apparaît de cette investigation sur le changement de forme de la lentille de l'œil, que l'action que j'ai autrefois attribuée aux couches externes[162] ne permet pas d'explication du phénomène[163]. L'effet nécessaire d'une telle action serait de produire une forme proche de celle d'un sphéroïde aplati aux pôles ; et, pour ne rien dire des inconvénients accompagnant une diminution du diamètre de la lentille de l'œil, la réfraction latérale

---

[160] Même en s'appuyant sur la Proposition III, n'est pas aisé de déterminer par quel calcul exactement Young est parvenu à ces conclusions précisément chiffrées [Tscherning, 1894, 191-196]. Toujours est-il que les dernières mesures révèlent une différence d'amplitude d'accommodation entre le centre et la périphérie du cristallin (des 9,8 dioptries déjà évaluées au centre, on passe à 4,2 dioptries seulement sur les bords). Young déduit de cela – et des variations de l'aberration sphérique au cours de l'accommodation déjà mentionnées – que les surfaces s'aplatissent à la périphérie pendant l'accommodation et se rapprochent de formes coniques. Quarante ans plus tard, Alexander von Hueck (1802-1842) a observé la courbure de l'iris pendant l'accommodation en demandant à son patient de diriger son regard vers une surface lumineuse (afin que la pupille soit particulièrement resserrée) et en lui faisant faire un effort d'accommodation aussi puissant que possible [Hueck, 1841, 105]. Helmholtz observe lui aussi cette déformation, juste avant de démontrer qu'en réalité c'est la face antérieure du cristallin qui est très fortement déformée au cours de l'accommodation, quand la face postérieure est à peine modifiée [Helmholtz, 1867, I : 142].
[161] Ajouté dans [Young 1807, II : 595]. Outre le fait que Young y justifie certaines propriétés de la déformation qu'il a déterminée par l'existence de ramifications ici évoquées s'avèrent ne pas exister [Tscherning, 1894, 183], cet ajout est surtout l'occasion de porter un coup aux théories suggérant que l'accommodation serait le fruit de la contraction de l'iris (dont on a vu qu'elle avait pu être influentes [Hosack,1794]). En arguant à juste titre que la contraction de l'iris accompagnant l'accommodation à courte distance permet de limiter l'entrée de l'œil aux rayons les plus proches de l'axe – et donc de réduire l'aberration sphérique dans cette situation où elle aurait été maximale du fait de la plus forte divergence des rayons perçus comme de la plus grande courbure du cristallin – Young suggère qu'il y a bien un lien de corrélation logique entre ces deux mécanismes qui semblent effectivement aller de pair, mais pas de lien de causalité.
[162] Sous-entendu : « du cristallin ».
[163] Voir la description du mécanisme de déformation du cristallin par contraction de ses couches superficielles, aussi qualifiées de « capsule », proposée dans [Young, 1793, 172-175].

augmenterait beaucoup plus que la centrale ; et le léger changement de densité à distance donnée de l'axe ne serait du tout équivalent non plus à l'augmentation de courbure : nous devons donc supposer un mode d'action différent du pouvoir produisant ce changement. Maintenant, que nous appelions la lentille de l'œil un muscle ou non, il semble démontrable qu'un changement de forme a lieu, tel qu'il ne peut être produit par aucune cause externe ; et nous pouvons au moins l'illustrer par une comparaison avec l'action usuelle des fibres musculaires. Un muscle ne se contracte jamais sans en même temps se gonfler latéralement, et lequel de ces effets nous considérons comme primaire n'est d'aucune conséquence. Une opacité accidentelle m'a incité à donner le nom de tendon membraneux aux radiations de la lentille partant depuis son centre[164] ; mais après examen plus précis, rien de réellement analogue à un tendon ne peut y être observé. Et si l'on supposait que les parties proches de l'axe étaient entièrement de nature tendineuse, et donc immuables, la contraction devrait principalement être effectuée par les parties latérales des fibres ; de façon que, de par leur contraction, les couches deviendraient plus épaisses vers la marge, alors que le changement global de forme nécessiterait qu'elles soient plus minces ; et il y aurait contradiction entre les actions des diverses parties. Mais si nous comparons les parties centrales de chaque surface au ventre du muscle, il n'y a pas de difficulté à {P.72} concevoir que leur épaisseur augmente immédiatement, et produise une élongation immédiate de l'axe et une augmentation de la courbure centrale ; pendant que les parties latérales coopéreraient plus ou moins selon leur distance au centre, et dans des proportions un peu différentes chez les différents individus[165]. Sur la base de cette supposition, nous n'avons plus aucune difficulté à attribuer le pouvoir de changer au cristallin des poissons. Dans un grand nombre d'observations, M. Petit trouva uniformément la lentille de l'œil des poissons plus ou moins aplatie[166] : mais même si elle ne l'était pas, une légère extension de la partie latérale des fibres superficielles permettrait à ces couches plus molles de devenir plus épaisses à chaque sommet et de donner à la lentille de l'œil la forme globale d'un sphéroïde un peu oblong ; et la lentille de l'œil étant ici le seul agent de la réfraction, in altération plus faible que chez les autres animaux serait suffisante. Il vaut aussi la peine d'examiner si l'état de contraction ne pourrait pas immédiatement ajouter au pouvoir réfringent. D'après l'ancienne expérience par laquelle le Dr Goddard a tenté de montrer que les muscles deviennent plus denses quand ils se contractent[167], un tel effet pourrait naturellement être attendu. Cependant cette expérience est très incertaine, et en

---

[164] [Young, 1793, 176].

[165] Ayant démontré de façon fort convaincante que l'accommodation était le fruit d'une contraction du cristallin, non sans avoir rigoureusement écarté au préalable la majorité des hypothèses concurrentes, Young se doit absolument à présent, comme on l'a évoqué en introduction, de justifier du mécanisme musculaire à l'origine de cette contraction ; et c'est précisément le but de toute la fin de l'article. A l'instant on l'a vu rejeter ses propres suggestions passées, pour proposer maintenant l'hypothèse d'un cristallin qui lui-même serait un muscle et qui, à la manière d'un biceps se courberait en son centre en se contractant. On le verra encore, dans les pages qui suivent, s'évertuer avec patience et persévérance à découvrir des muscles intérieurs de l'œil permettant la déformation du cristallin – dans une quête, qui à une époque où la médecine ne connaissait pas encore l'existence des muscles à fibres lisses, était nécessairement vouée à l'échec. C'est en effet presque cinquante ans plus tard qu'Albert von Kölliker identifiait les fibres lisses observées sous son microscope à celles de muscles d'allure jusque-là inconnue, responsable essentiellement de mouvements involontaires [Kölliker, 1847] ; c'est seulement alors que Müller, Rouget, Brücke et Bowman pouvaient reconnaître ces fibres dans la structure du corps ciliaire bordant le cristallin, pourtant déjà parfaitement documentées avant cela ; et plus tard encore que le rôle exact de ce muscle ciliaire dans le changement de forme du cristallin a pu être déterminé.

[166] [Pourfour du Petit, 1730].

[167] Il s'agit possiblement de Jonathan Goddard (1617-1675), médecin anglais dont les archives classées de la Royal Society semblent renfermer un court texte dédié à la démonstration de la perte de volume des muscles lors de leur contraction [Goddard,1669].

effet l'opinion est généralement dispersée à son sujet, mais peut-être trop hâtivement ; et quiconque établira l'existence ou la non existence d'une telle condensation rendra un service essentiel à la physiologie en général. <Quelques expériences intéressantes sur le sujet ont été promises au public par un physiologiste très ingénieux, qui dans ses recherches a probablement employé une méthode d'investigation plus concluante. SWAMMERDAM professe avoir trouvé une telle condensation lors de la contraction d'un muscle ; mais il est évident que ce qu'il a attribué au cœur appartenait en propre à l'air seul qu'il contenait, et l'une de ses expériences, qui était dépourvue de cette source d'erreur, semble ne pas avoir montré le moindre résultat satisfaisant, bien qu'elle ait été conduite avec exactitude en enfermant le muscle dans une bouteille remplie d'eau communicant avec un tube ouvert étroit (Book of Nature, II. 126, 127).>[168]

En l'an 1719, le Dr PEMBERTON fut le premier à examiner avec systématicité l'opinion de la muscularité de la lentille cristalline (De Facultate Oculi qua ad diversas Rerum distantias se accommodat. L. B. 1719. Ap. Hall. Disp. Anat. IV. p. 301)[169]. Il se référait aux observations au microscope de LEEUWENHOEK[170] ; mais il submergea son sujet de tant de calculs complexes que bien peu ont tenté de les désenvelopper : et il fondait le {P.73} tout sur une expérience empruntée à BARROW ; qui a totalement échoué avec moi ; et je ne peux qu'être d'accord avec le Dr OLBERS quand il remarque qu'il est plus aisé de le réfuter que de le comprendre. Il argumentait en faveur d'un changement partiel de la forme de la lentille de l'œil ; et peut-être que cette opinion était plus juste que les raisons utilisées pour la soutenir. LOBE, ou plutôt ALBINUS (De quibusdam Oculi Partibus, L. B. 1746. Ap. Hall. Disp. Anat. IV. p. 301)[171], est décidément en faveur d'une théorie similaire ; et suggère l'analogie de la lentille de l'œil avec les parties musculaires des animaux transparents, dans lesquels pas même les meilleurs microscopes ne peuvent découvrir de fibres. CAMPER mentionne lui aussi l'hypothèse, avec une approbation notable (De Oculo Humano, L. B. 1742. Ap. Hall. Disp. Anat. VII. 2. p. 108,109)[172]. Le Professeur REIL a publié en 1793 une Dissertation sur la Structure de la Lentille de l'œil ; et dans un article ultérieur, annexé à la traduction de mon précédent Essai dans le journal du Professeur GREN (1794. p. 352, 354), il discutait la question de sa muscularité. Je regrette de ne pas avoir l'opportunité maintenant de me référer à cette publication ; mais je ne me souviens pas que les objections du Professeur REIL soient différentes de celles que j'ai déjà notées.[173]

---

[168] Passage ajouté dans [Young, 1807, II : 596]. Jan Swammerdam (1637-1680) est un naturaliste néerlandais auteur dans les dernières années de sa vie d'un ouvrage monumental qui ne sera publié que 60 ans plus tard à l'initiative de Boerhaave, sous le titre de *Bybel der Natuure*. C'est à son édition anglaise de 1758 que Young fait référence. Et plus spécifiquement au passage, dans lequel il mesure la variation de volume des muscles lors de leur contraction en étudiant le muscle du cœur, puis de la hanche, d'une grenouille, déposé dans un siphon de verre scellé [Swammerdam, 1758, II : 126-127]. Comme le souligne Young, bien que les résultats des deux expériences semblent diverger, Swammerdam opte pour la conclusion d'un changement de volume – opposée à celle que Young semble favoriser pour sa part – que l'on sait aujourd'hui être erronée. Plusieurs traités modernes de physiologie attribuent néanmoins rétrospectivement à Swammerdam la découverte de la conservation du volume des muscles lors de leur contraction. L'enjeu ici pour Young est l'examen rapide de l'hypothèse que la densité du muscle qu'est le cristallin puisse augmenter lors de sa contraction, et donc que sa densité réfringente soit affectée en conséquence.

[169] Henry Pemberton (1694-1771) est un médecin anglais et membre de la Royal Society. En 1719 il présente sa thèse à l'université de Leyde dédiée à l'accommodation de l'œil, dans laquelle il défend l'hypothèse de la muscularité du cristallin par un raisonnement géométrique, reposant en bonne partie sur une expérience initialement proposée par Isaac Barrow [Pemberton, 1719, 14]. On remarque que la référence donnée par Young à la thèse de Pemberton, comme celles faites à la thèse de Lobé et de Camper qui suivent, est erronée. La thèse de Pemberton est correctement intitulée bien qu'il manque un mot à son titre, mais la référence à laquelle renvoie est manifestement incorrecte, puisque c'est la même que celle donnée pour Camper. Mais Young attribue surtout ensuite à Lobé le titre de la thèse qui a été défendue par Camper en 1742, et à Camper celui de celle de Lobé. Une telle imprécision est d'autant plus surprenante que la série de recueils *Disputarum Anatomicarum Selectarum* réalisés par Albrecht von Haller desquels il extrait ces textes ne recèle pas ces erreurs, que l'on a maintenues dans le corps traduit du texte, mais que l'on a corrigées dans la bibliographie.

[170] Antoni van Leeuwenhoek (1632-1723), commerçant et savant néerlandais, améliora notablement le microscope et l'appliqua à de nombreuses observations biologiques inscrites dans le prolongement direct des travaux de Swammerdam, qui lui vaudront d'être élu à la Royal Society de Londres et à l'Académie des Sciences de Paris. Ses observations au microscope du cristallin sont notamment rapportées dans [Leeuwenhoek, 1684].

[171] Joannes Petrus Lobé (1717- ?) est un élève de Bernhard Siegfried Albinus (1697-1770), lui-même titulaire de la chaire d'anatomie de l'université de Leyde. Dans sa thèse *Sur l'œil humain* il emprunte à Albinus l'idée d'un possible changement de convexité du cristallin, comparant le cristallin à certains animaux transparents dont on observe bien la mobilité quoi qu'on soit incapable de déceler chez eux de fibres musculaires [Lobé, 1742, 34].

Considérant la sympathie de la lentille cristalline avec L'uvée et la nature délicate de son changement de forme, il y a peu de raison d'espérer que l'on ait plus de succès à stimuler l'action contractile de la lentille de l'œil à l'aide d'un quelconque stimulus artificiel que l'on en a eu avec l'uvée jusqu'à présent ; et cette contraction serait beaucoup moins visible encore sans artifice. Peu après la mort de M. HUNTER, je poursuivis l'expérience qu'il avait suggérée afin de déterminer à quel point une telle contraction pourrait être observable[174]. Mon appareil (Planche V. Fig. 27) fut réalisé par M. JONES[175]. Il consistait en un récipient de bois, noirci à l'intérieur, qu'il fallait {P.74} remplir d'eau fraîche, puis plus chaude : un miroir plan était placé en-dessous ; une perforation dans le fond était comblée par une plaque de verre ; des anneaux convenables étaient fixés pour recevoir la lentille de l'œil, ou l'œil en entier, ainsi que

---

L'analogie est audacieuse ; mais tout en réduisant le risque en l'attribuant à un autre, Young se permet ainsi d'enfoncer un coin dans des des objections déjà adressées à sa thèse de la muscularité du cristallin, selon lesquelles on ne connait pas d'exemple de muscle transparent et l'on est incapable de discerner de fibres musculaires dans le cristallin [Hosack, 1794, 202-203].

[172] Peter Camper (1722-1789) est un médecin et naturaliste néerlandais, membre de la Royal Society. Dans la partie consacrée à l'accommodation de sa thèse sur l'anatomie *De certaines parties de l'œil*, Camper liste sans conclure les différentes théories concurrentes ; et accorde néanmoins au système de Pemberton qu'il est « en partie probable, mais non suffisant » [Camper, 1746, 24].

[173] De 1790 à 1794, Friedrich Gren a édité une série de numéros du *Journal der Physik*, compilant des essais et articles scientifiques extraits des principales revues européennes, souvent traduits en allemand. Le dernier et huitième volume de cette série contient une version en allemand des *Observations sur la Vision* [Young, 1793], à laquelle est associée un texte de Johann Christian Reil (1759-1813), médecin, anatomiste et psychiatre allemand. Ce texte *Sur la structure fibreuse du cristallin* est une traduction en allemand de la dissertation inaugurale qu'il avait présentée en latin sur le sujet à l'université de Halle. Cette version se conclue par quelques pages faisant rétrospectivement référence au texte de Young, dont il n'avait pas eu connaissance, et avec lequel il s'avoue essentiellement d'accord [Reil, 1794, 352-356].

[174] Le compte rendu par Home de la conférence Croonienne prévue par John Hunter relative au cristallin faisant état de l'expérience centrale sur laquelle devait reposer sa démonstration. L'idée en était que si l'humeur cristalline était analogue à un muscle, elle devrait réagir aux mêmes stimuli que ceux-ci. Mais plusieurs types de stimulus ayant déjà échoué, Hunter envisageait d'immerger dans des eaux de différentes températures des cristallins d'animaux récemment abattus ; l'objectif étant d'observer les éventuelles modifications de l'image d'un point lumineux formée par ces cristallins, du fait de sa contraction provoquée par un changement de température de l'eau [Hunter, 1794, 22-23]. Hunter décède avant de pouvoir conclure sur ces expériences, mais Young la reprend ici, se doutant qu'elle échouerait à prouver sa thèse, mais suggérant néanmoins que l'homme auquel on l'accusait en 1793 d'avoir dérobé l'idée de la muscularité du cristallin ne disposait en fait pas des éléments pour la démontrer. Il avait donc tout à y gagner, d'autant qu'il se prévaut du risque d'affaiblir son hypothèse par l'échec de cette expérience en citant d'entrée le cas de l'iris ; dont on peut naturellement observer une contraction que l'on ne peut stimuler en aucune manière. Le montage très élaboré qu'il décrit ensuite intègre non seulement la possibilité d'observer un effet de la contraction du cristallin sous l'effet d'un changement de température de l'eau dans laquelle il baigne, mais aussi sous l'effet d'une stimulation électrique.

[175] Il s'agit probablement de William Jones (1762-1831) ou de son frère Samuel Jones (1770-1859), propriétaires de l'entreprise d'instruments scientifiques – en particulier optiques – « W. & S. Jones », établie à Londres.

des fils pour transmettre l'électricité : au-dessus d'eux, une pièce de verre peinte et dépolie destinée à recevoir l'image était soutenue par une tasseau, que l'on déplaçait grâce à un pivot, en connexion avec une échelle graduée en cinquantièmes de pouce (*0,5 mm*). Avec cet appareil je réalisai quelques expériences, assisté de M. Wilkinson dont la résidence était proche d'un abattoir[176] : mais nous ne pûmes obtenir la moindre évidence satisfaisante du changement par cette méthode ; non pas que notre attente ait été déçue. Je comprends aussi qu'un autre membre de cette Société échoua tout autant dans sa tentative de produire un changement visible de la lentille grâce à l'électricité[177].

XI.[178] La structure de la lentille est à peu près similaire chez L'homme et chez les quadrupèdes les plus communs. Le nombre de radiations et de peu de conséquences ; mais j'en trouve dix de chaque côté dans le cristallin humain (Planche VI, Fig. 46) et non trois, comme je l'avais autrefois conclu d'une observation hâtive (De Corp. Hum. Vir. Cons. p. 68)[179]. Ceux qui trouvent la moindre difficulté à repérer les fibres doivent avoir une vue bien mal adaptée aux recherches au microscope. J'ai travaillé avec la persévérance la plus obstinée à localiser les nerfs à l'intérieur de la lentille et j'ai parfois imaginé que j'y avais réussi ; mais je ne peux positivement aller plus loin qu'affirmer ma conviction totale de leur existence, et de la précipitation de ceux qui l'ont absolument niée. Les nerfs longs, qui sont très visibles entre les couches de la choroïde et de la sclérotique, se divisent chacun en deux ou trois branches, ou plus, au point où débute la zone ciliaire, et semblent effectivement fournir de {P.75} fins filaments à la choroïde au même endroit. Les branches se réunissent souvent, en une légère protubérance qui mérite à peine le nom de ganglion : ils sont ici liés et mêlés à la membrane dure et brun-blanchâtre qui couvre la substance compacte et spongieuse dans laquelle les vaisseaux des procès ciliaires s'anastomosent et se subdivisent (Planche VI. Fig. 47). La quantité de nerfs qui vont jusqu'à l'iris parait considérablement inférieure à celle qui arrive à l'endroit de la division : d'où il a peu à douter que la division soit calculée pour fournir quelques branches minuscules à la lentille de l'œil ; et il n'est pas improbable, du fait de l'aspect des parties, que certaines fibres puissent passer jusqu'à la cornée ; bien que l'on puisse plus naturellement s'attendre à ce que la tunica conjonctiva soit approvisionnée depuis l'extérieur. Mais les subdivisions qui passent probablement jusqu'à la lentille de l'œil entrent immédiatement dans un mélange de substance ligamenteuse et d'une membrane dure et brunâtre ; et je n'ai pas été capable jusque-là de les désenvelopper. Peut-être peut-on trouver des animaux chez lesquels cette substance est de nature différente ; et je ne désespère pas qu'il puisse encore être possible de les repérer chez les quadrupèdes, à l'aide d'injections pour distinguer plus facilement les vaisseaux sanguins <et d'un acide afin de blanchir les nerfs>[180]. Notre incapacité à les découvrir n'est guère un argument contre leur existence : ils doivent naturellement être délicats et transparents ; et avec la cornée nous avons un exemple de sensibilité considérable pour laquelle aucun nerf n'a encore été repéré. La capsule adhère à la substance ciliaire, et la lentille de l'œil à la capsule, en deux ou trois points principalement ; mais je confesse ne pas avoir été capable d'observer que ces points soient exactement opposés au tronc des nerfs ; de sorte que probablement, l'adhérence est principalement causée par ces vaisseaux que l'on voit parfois traverser la capsule des yeux injectés. Nous pouvons cependant {P.76} détecter des ramifications depuis certains de ces points, sur et à l'intérieur de la substance de la lentille de l'œil (Planche VI, Fig. 48), suivant généralement une direction proche de celle des fibres, et venant parfois d'un point opposé à l'une des lignes radiantes de la même surface. Mais les principaux vaisseaux de la lentille de l'œil paraissent être dérivés de l'artère centrale par deux ou trois branches à petite distance du sommet postérieur ; ce que je conçois être la cause de l'adhérence fréquente d'une partie de la cataracte à la capsule autour de ce point : ils suivent le parcours des radiations et ensuite celui des fibres ; mais il y a souvent une subdivision superficielle de l'un des rayons au point où l'un d'eux entre. Les vaisseaux venant de la choroïde paraissent principalement apporter une substance jusque-là inobservée, qui remplit la partie marginale de la capsule du cristallin, de la forme d'une zone fine, et qui produit une légère élévation, visible même à travers la capsule (Fig. 49-51). Elle est constituée de fibres plus grossières que la lentille mais dans une direction à peu près similaire ; elles sont souvent entremêlées de petits globules. Chez certains animaux la marge de la zone est crénelée, spécialement à l'arrière, où elle est plus courte : c'est observable chez la perdrix ; et chez ce même oiseau, la totalité de la surface de la lentille de l'œil est vue comme étant couverte de points, ou plutôt de globules, arrangés en lignes régulières (Planche VII, Fig. 52), de sorte à avoir un peu l'aspect d'un nid d'abeilles, mais qui sont moins uniformément disposés vers le sommet. Cette régularité est une preuve suffisante que cette apparence n'est pas une illusion optique ; bien qu'un bon microscope soit requis pour la découvrir distinctement : mais la zone peut aisément être détachée sous l'eau, et durcie dans des esprits. Son usage est incertain ; mais elle pourrait possiblement sécréter le liquide du cristallin ; et elle mérite autant le {P.77} nom de glande que la plupart des substances usuellement dénommées ainsi. En la détachant, j'ai très distinctement observé des ramifications qui passaient d'elle à la lentille de l'œil (Planche IV. Fig. 50) ; et en effet il n'est pas du tout difficile de détecter les vaisseaux connectant la marge de la lentille de l'œil à sa capsule ; et il est surprenant que M. P<small>ETIT</small> ait douté de leur existence. Je n'ai pas discerné encore clairement cette glande cristalline dans l'œil humain ; mais je déduis l'existence de quelque chose de similaire aux globules de l'aspect tacheté de l'image d'un point brillant déjà mentionné. Je ne peux en rendre compte autrement qu'en l'attribuant à un dérangement de ces particules produit par la force extérieure, et à une pression non uniforme exercée par eux sur la surface de la lentille de l'œil[181].

---

[176] Le fait que Wilkinson habitait à proximité d'un abattoir a certainement permis de fournir plus facilement les deux expérimentateurs en yeux frais d'animaux récemment abattus.

[177] Il s'agit certainement de William Charles Wells (1757-1819), médecin américain installé à Londres et membre de la Royal Society, qui en 1811 publie un article sur la vision se concluant sur le récit d'expériences qu'il avait menées en 1794, consistant à stimuler le cristallin de bœufs abattus seulement quelques secondes plus tôt de toutes les manières possibles : chimique, mécanique, électrique, dans l'eau comme dans l'air. Il rapporte qu'elles ont toutes échoué et en conclue à la fragilité de l'hypothèse de Young. Avant d'ajouter que de nouvelles expériences utilisant de la belladone pourrait contribuer à éclairer le sujet [Wells, 1811, 390-391].

[178] La dernière partie de l'article est dédiée à l'anatomie comparée du cristallin humain et des yeux de différents animaux, selon une démarche classique d'histoire naturelle consistant à chercher dans le règne animal des analogies fructueuses avec le fonctionnement du corps humain.

[179] La thèse de Young se conclut par une série de « thèses » lapidaires, sur toute une série de question médicales allant de l'usage du vin comme remède à la transmission de la variole. La sixième affirme que « Les fibres de la lentille cristalline de l'homme sont disposées dans un même ordre que chez le bœuf » [Young, 1796, 68]. Il suggère ici qu'elle doit être amendée quant au nombre de radiations qu'il observe finalement chez l'un et chez l'autre, mais qui selon lui sont sans conséquence. La Figure 46 de la Planche VI montrent bien ces dix radiations dont Young fait état.

  Chez les oiseaux et les poissons, les fibres du cristallin rayonnent uniformément, devenant d'autant plus fines qu'elles s'approchent du sommet, jusqu'à disparaître en une substance uniforme du même degré de fermeté, qui paraît être perforée au centre par un vaisseau sanguin (Planche VII. Fig. 53). Chez les quadrupèdes, les fibres, à l'endroit où elles se rencontrent en faisant un angle, ne se prolongent certainement pas au-delà de la ligne de division, comme l'imaginait Leeuwenhoek[182] ; mais il ne pas il ne parait pas y avoir de substance dissemblable interposée entre elles, excepté ces minuscules troncs de vaisseaux qui marquent souvent cette ligne. Mais puisque la masse de la lentille de l'œil dans sa globalité, autant qu'elle est mobile, est probablement dotée du pouvoir de changer sa forme, il n'y a besoin d'aucune force de cohésion, ni de lieu d'attache, pour les fibres, puisque le mouvement ne rencontre que peu ou pas de résistance. Tout muscle ordinaire retourne à sa forme naturelle aussitôt que cesse sa contraction, même sans l'assistance d'un antagoniste ; et la lentille de l'œil elle-même, quand elle est retirée de l'œil en conservant sa capsule, {P.78} a suffisamment d'élasticité pour reprendre sa propre forme dès la suppression d'une force qui l'aurait comprimée. La capsule est hautement élastique ; et puisqu'elle est fixée latéralement à la zone ciliaire, elle doit coopérer à rétablir la lentille de l'œil dans sa forme la plus plate. Si l'on demande pourquoi la lentille de l'œil n'est pas capable de devenir moins convexe aussi bien qu'elle peut le devenir plus, il peut être répondu que les parties latérales ont probablement peu le pouvoir de contraction ; et que si elles en avaient plus, elles n'auraient pas la place d'augmenter la taille du disque comme elles devraient le faire afin de raccourcir l'axe ; et les parties autour de l'axe n'ont pas de fibres arrangées de manière à le raccourcir par leur propre contraction.

---

[180] Ajouté dans [Young, 1807, 598].
[181] Il a depuis été clarifié que le cristallin n'est traversé ni de vaisseaux ni de nerfs ; que l'aspect tacheté de certaines images vient d'irrégularités de la face antérieure de la cornée ; et que ce qu'il appelle glande cristallinienne n'est pas une glande mais un renflement formé par les fibres les plus périphériques du cristallin.

----Je me considère comme étant partiellement remboursé du travail perdu à la recherche du nerf de la lentille en ayant acquis une conception plus exacte de la nature et de la position de la substance ciliaire[183]. Il avait déjà été observé que chez le lièvre et chez le loup les procès ciliaires ne sont pas attachés à la capsule de la lentille ; et si par procès ciliaires on entend ces filaments que l'on voit détachés après avoir retiré la capsule, et qui consistent en des vaisseaux se ramifiant, la remarque est également vraie pour les quadrupèdes ordinaires et, m'aventurerai-je à dire, pour l'œil humain (Vid. Hall. Physiol. V. p. 432. et Duverney, ibi citat.)[184]. <LEROI, ALBINUS et>[185] d'autres ont fait cette remarque, mais la circonstance n'est pas comprise en général. Il est si difficile d'obtenir une vue distincte de ces corps, non perturbée, que je suis en partie endetté envers le hasard de n'avoir pas été trompé à leur égard : mais ayant réalisé l'observation une fois, j'ai appris à la montrer de manière incontestable. Je retire l'hémisphère postérieur de la sclérotique, ou un peu plus, ainsi qu'autant que possible de l'humeur vitrée, j'introduis la pointe d'une paire de ciseaux {P.79} dans la capsule, retire la lentille de l'œil, et enlève la plus grande partie de la portion postérieure de la capsule et du reste de l'humeur vitrée. Je découpe ensuite la choroïde et l'uvée de la sclérotique ; et en divisant la partie antérieure de la capsule en segments depuis son centre, je les retourne sur la zone ciliaire. Les procès ciliaires apparaissent ensuite recouverts de leur pigment et parfaitement distincts à la fois de la capsule et de l'uvée (Planche VII. Fig. 54) ; on voit la surface de la capsule briller, et dans son état naturel évidemment, à proximité de la base de ces substances. Je ne nie pas que la séparation entre l'uvée et les procès s'étende un peu plus en arrière que la séparation entre les procès et la capsule ; mais la différence est négligeable et ne s'élève pas chez le veau à plus de la moitié de la longueur de la partie détachée. L'aspect des procès est totalement inconciliable avec l'idée de muscularité ; et le fait de les considérer comme des muscles attachés à la capsule est donc doublement inadmissible. Leur union latérale avec la capsule commence à la base de leur surface postérieure lisse et se poursuit presque jusqu'au point où ils sont plus intimement unis avec la terminaison de l'uvée ; de sorte que même si cette partie de la base des procès était disposée à se contracter, elle serait beaucoup trop courte pour produire le moindre effet sensible. L'utilité qui pourrait être la leur ne peut être déterminée aisément : s'il était nécessaire d'avoir un organe particulier pour la sécrétion, nous pourrions les appeler glandes, pour la filtration de l'humeur aqueuse ; mais il n'y a pas de raison de les penser destinés à ce but.

---

[182] [Leeuwenhoek, 1684, 782-786].

[183] On rapprochera avec intérêt ce témoignage de celui exprimé par Kepler dans la préface de son *Mysterium Cosmographicum* (« Je perdis beaucoup de temps à ce travail comme à un jeu, puisque nulle régularité n'apparaissait ni dans les proportions des orbes, ni dans leurs différences, et je ne tirai de là nulle autre utilité que de graver très profondément dans ma mémoire les distances mêmes, telles qu'elles sont enseignées par Copernic » [Kepler, 1596, 32], pour autant qu'il témoigne de ce que la poursuite persévérante, laborieuse et minutieuse d'une hypothèse qui plus tard se révèlera fausse, n'est souvent pas dénuée d'apports bénéfiques ; au moins en ce qu'elle permet de développer une familiarité avec le sujet d'autant plus grande que l'on y aura perdu de temps, et que l'on aura d'autant plus « perdu » de temps que l'on aura erré. Mais il nous rappelle aussi combien l'idée que les théories scientifiques sont induites de la pure observation des données est naïve. Et combien – même dans une science aussi apparemment objective que l'anatomie – c'est la théorie, ou l'hypothèse organisatrice, qui précède la donnée : puisque c'est elle qui fixe la définition même de la donnée et qui décide nécessairement de ce qui doit être recherché. Ici, des nerfs dans le cristallin ; là, des rapports de nombres entiers entre les rayons des orbes sphériques ou entre leurs différences. Et c'est bien pour cela que l'on peut perdre tant de temps à chercher quelque chose qui n'est pas. En cela on ne peut s'empêcher de rejoindre Bruno Latour, lorsqu'il suggère que « la tentation de l'idéalisme vient peut-être du mot même de *données* qui décrit aussi mal que possible ce sur quoi s'appliquent les capacités cognitives ordinaires des érudits, des savants et des intellectuels. Il faudrait remplacer ce terme par celui, beaucoup plus réaliste, d'*obtenues* » [Latour, 2007, 609].

  Le marsupium nigrum des oiseaux, et la choroïde en forme de fer-à-cheval des poissons sont deux substances qui ont parfois, avec une égale injustice, été désignées comme musculaires[186]. Toutes les fibres apparentes du marsupium nigrum ne sont, comme {P.80} HALLER l'avait très correctement affirmé, que des plis d'une membrane qui, lorsque ses extrémités sont découpées, peut aisément être dépliée sous un microscope à l'aide d'un pinceau fin, de sorte à ne plus laisser le moindre soupçon de texture musculaire. L'expérience rapportée par M. HOME (Phil. Trans. for 1796. p. 18) peut difficilement être considérée comme un argument bien fort pour attribuer à cette substance une faculté que son apparence nous permet si peu d'attendre d'elle[187]. La substance rouge de la choroïde des poissons (Planche

---

[184] Albrecht von Haller (1708-1777) a laissé une œuvre monumentale en histoire naturelle. A cette page de son ouvrage majeur de physiologie, il évoque l'observation de vaisseaux sur le pourtour du cristallin [Haller, 1766, V : 432], dont il attribue une observation antérieure au médecin français Joseph-Guichard Duverney (1648-1730) dans sa description du ligament ciliaire [Du Verney, 1761, I :150].

[185] Ajouté dans [Young, 1807, 600]. Il fait probablement ici référence à un texte de Charles Le Roy (1726-1779) médecin français membre de l'Académie Royale de Montpellier [Le Roy, 1755]. Et à une *Descriptio Oculi* attribuée à Bernhard Siegfried Albinus par le physicien néerlandais Pieter van Musschenbroek (1692-1761) [Musschenbroek, 1762, II : 744-755].

[186] Le marsupium nigrum est un pli de la choroïde qui fait saillie dans le corps vitré des oiseaux et qui sert à la nutrition du corps vitré. La substance rouge des yeux de poissons est due à un renflement des vaisseaux de la choroïde.

[187] Dans la conférence Croonienne de 1795, Everard Home ouvre l'œil d'une oie fraîchement abattue pour faire apparaître le cristallin, qu'il pousse ensuite en avant, afin de mesurer à plusieurs reprises l'élongation du marsupium nigrum. Il remarque qu'en reprenant la même expérience sur le même œil après un certain temps, le marsupium nigrum ne revient plus exactement à son état de contraction initial. Il en déduit que cette perte

VII. Fig. 55) est plus susceptible de tromper l'observateur ; sa couleur lui confère une certaine prétention[188], et j'ai commencé à l'examiner avec une prédisposition en faveur de sa nature musculaire. Mais si l'on se rappelle la couleur générale des muscles des poissons, la considération de sa rougeur n'aura plus le moindre poids. Découverte de la membrane qui couvre assez librement sa surface interne (Fig. 56), elle semble présenter des divisions transverses ressemblant un peu à celle des muscles, et se terminer de manière quelque peu similaire (Fig. 57) ; mais lorsqu'elles sont vues sous un microscope, les divisions transverses apparaissent comme des fissures et la totalité de la masse est d'une texture évidemment uniforme, sans le moindre aspect fibreux ; et le contraste devient très frappant si on la compare à une partie élémentaire de n'importe quelle sorte de muscle. De plus, elle est fermement fixée sur toute son étendue à la lamelle postérieure de la choroïde et n'a aucune attache capable de diriger son effet ; pour ne rien dire de la difficulté à concevoir quel pourrait être cet effet. Son utilité doit rester entièrement dissimulée à notre curiosité, comme celle de nombreuses autres parties de l'organisme animal.

Les coquilles osseuses des yeux des poissons, qui ont été décrites il y a longtemps <dans les Mémoires de l'Académie par MERY (II. 15),> dans les Transactions Philosophiques par M. RANBY (Phil. Trans. Vol. XXXIII. p. 223. Abr. Vol. VII. p. 435) {P.81} et par M. WARREN (Phil. Trans. XXXIV, 113. Abr. VII. 437), puis dans deux excellents Mémoires de M. PETIT sur l'œil du dindon et du hibou (Mémoires de l'Acad. 1735, p. 163, 1736, p. 166. Ed. Amst.) et dernièrement par <le Professeur BLUMENBACH (Comm. Gott. VII. 62)>[189], M. PIERCE SMITH (Phil. Trans. for 1795, p. 263.) et M. HOME (Phil. Trans. for 1796. p. 14.)[190], ne peuvent selon aucune supposition être un tant soit peu concernées par l'accommodation de l'œil à différentes distances : elles semblent plutôt être nécessaires à la protection de cet organe, large et proéminent comme il l'est et qu'aucune force ne maintient dans l'orbite, contre les divers accidents auxquels le mode de vie et le mouvement rapide de ces animaux doivent l'exposer ; et elles sont beaucoup moins susceptible de fracture qu'un anneau osseux entier de la même épaisseur ne l'aurait été. Le marsupium nigrum paraît être destiné à donner en partie de la force à l'œil pour empêcher tout changement de place de la lentille de l'œil par une force externe : il est situé de sorte à n'intercepter que peu de lumière, et ce peu est principalement ce qui serait tombé sur l'insertion du nerf optique ; et il semble être trop fermement lié à la lentille de l'œil pour même permettre la moindre élongation notable de l'axe de l'œil, bien qu'il n'empêcherait certainement pas une protrusion de la cornée.

Quant aux yeux des insectes, une remarque de POUPART mérite d'être répétée ici. Il remarque que l'œil de la libellule est creux ; qu'il communique avec un vaisseau rempli d'air placé longitudinalement dans le tronc du corps ; et qu'il est capable d'être gonflé depuis cette cavité : il suppose l'insecte est pourvu de cet appareillage pour l'accommodation de son œil à

---

d'élasticité est une preuve potentielle du pouvoir musculaire du marsupium nigrum ; bien qu'il ne lui accorde pas lui-même un crédit considérable [Home, 1796, 18].

[188] Sous-entendu : « à être un muscle ».

[189] Mentions ajoutées [Young, 1807, 601].

[190] Les coquilles osseuses que Young observe à la surface des yeux des poissons et d'autres animaux ont été décrites avant lui par le chirurgien français Jean Méry (1645-1722) sur l'aigle, le casoar et le corbeau [Méry, 1687, 15]. Par le chirurgien britannique John Ranby (1703-1773) chez l'autruche [Ranby, 1724, 225]. Par George Warren, chirurgien de Cambridge, chez l'autruche également [Warren, 1727, 115]. Par François Pourfour du Petit sur la dinde [Pourfour du Petit, 1735, 142] et le hibou [Pourfour du Petit, 1736, 138]. Par le médecin allemand – et professeur d'histoire naturelle de Thomas Young quand il étudiait à Göttingen – Johan Friedrich Blumenbach (1752, 1840) chez la chouette [Blumenbach, 1784, 57 ; 62-63]. Par Pierce Smith, alors étudiant en médecine, chez les oiseaux [Smith P., 1795, 263]. Et enfin chez les oiseaux aussi par Everard Home [Home, 1796, 14].

la perception d'objets à différentes distances (Phil. Trans. Vol. XXII. P. 673. Abr. p. 762)[191]. {P.82} Il n'y a pas de difficulté à supposer que le moyen de produire le changement des pouvoirs réfringents de l'œil peut être aussi diversifié chez les différentes classes d'animaux que le sont leurs mœurs et la configuration générale de leurs organes. <Mais un examen des yeux des libellules, des guêpes et des homards m'induit non seulement à rejeter la suggestion de Poupart, mais à m'accorder aussi avec ces naturalistes qui ont appelé à questionner les prétentions de ces organes au nom qui leur a usuellement été attribué. CUVIER a donné une présentation très juste de ce cas dans son précieux travail sur l'anatomie comparée[192] ; et ses descriptions, autant que celles de SWAMMERDAM, s'accordent en général avec ce que j'ai observé. Nous sommes prédisposés à les considérer comme étant des yeux par leur position et leur aspect général. La copieuse réserve de nerfs semble au moins prouver qu'il doit s'agir d'organes sensitifs. Chez le crabe bernard-l'hermite, SWAMMERDAM dit que les nerfs se croisent, mais ce n'est pas le cas chez l'écrevisse[193]. La couche externe est toujours transparente ; ces divisions sont généralement plus ou moins lenticulaires. De nombreux insectes n'ont pas d'autres organes ressemblant du tout à des yeux ; et quand ces yeux sont recouverts, ces insectes paraissent être totalement ou partiellement aveuglés (Hooke. Microgr. 178)[194]. Mais d'un autre côté, beaucoup d'insectes sont dépourvus de ces yeux, et parmi ceux qui les ont, beaucoup en ont d'autres également, incontestablement plus adaptés à la vision. Les parties avoisinantes de la peau dure ou de la carapace, lorsque l'on retire la croûte qui les tapisse, sont souvent aussi transparentes qu'eux. Les antennes d'apis longicornis, comme M. KIRBY m'en a informé le premier, ont un peu le même aspect réticulé, mais pas assez pour fonder un quelconque argument quant à leur usage[195]. Cette couche réticulée est toujours complètement tapissée d'un mucus obscur et opaque qui paraît parfaitement inadapté à la transmission de lumière ; or il n'y a rien de ressemblant à une humeur transparente dans toute la structure non plus : et la convexité des parties lenticulaires n'est en aucune manière suffisamment importante pour amener les rayons lumineux vers un foyer très proche ; en fait, chez les homards la surface externe est parfaitement uniforme et la surface interne est seulement divisée en carrés par une texture striée adhérant à elle. Il n'y a rien d'analogue en quoi que ce soit à une rétine et il ne peut y avoir formation d'une image telle que celle qui se forme dans les yeux de tous les autres animaux, sans même excepter les vers : il ne semble pas y avoir lieu non plus d'accorder avec BIDLOO qu'il y a sous le centre de

---

[191] François Poupart (1661-1709) est un médecin français membre de l'Académie royale des sciences qui adresse en 1701 à Martin lister de la Royal Society une lettre sur les libellules qui observe que des canaux émanent de leurs yeux qui pourraient servir à les gonfler d'air dans le but d'accommoder [Poupart, 1701, 676]. Dans la version de 1801 suit la phrase « Je n'ai pas encore eu d'opportunité d'examiner {P.82} l'œil de la libellule ». Mais elle est supprimée en 1807 [Young, 1807, II : 601], comme le justifiera la suite.
[192] Young fait ici référence aux *Leçons d'anatomie comparée* de Georges Cuvier, pas encore publiées lors de la première édition de la conférence, mais disponible à l'heure où Young ajoute ces lignes. Il y dédie évidemment un article aux yeux des insectes et des crustacés [Cuvier, 1805, II : 442-445].
[193] Le *Livre de la Nature, ou l'Histoire des Insectes* de Swammerdam est aussi un trésor d'anatomie comparée, dans lequel est notamment évoqué l'œil du bernard-l'hermite [Swammerdam, 1758, I : 91-92].
[194] Hooke remarque que lorsqu'il coupe ce qu'il appelle les yeux des crabes, homards et crevettes, ceux-ci se heurtent aux rochers alentours et ne réagissent plus quand il passe la main devant eux. Il en déduit que ce qui se passe avec ces « crustacés marins » se produirait de même avec les « insectes crustacés » [Hooke, 1665, 178].
[195] William Kirby (1759-1850) membre de la Société Linnéenne et de la Royal Society est considéré comme l'un des pères de l'entomologie. Il publie en 1802 une monographie en deux volumes sur les abeilles d'Angleterre, dont il a dû parler avant cela avec Young, et dans lequel il décrit les antennes d'*apis longicornis* comme étant constituées « de minuscules lentilles, et probablement de même forme hexagonale » [Kirby, 1802, I : 48].

chaque hexagone une perforation laissant passer la lumière[196]. Si ce sont des yeux, leur manière de percevoir la lumière doit plutôt ressembler au sens de l'audition qu'à celui de la vision, et ils ne doivent convoyer qu'une idée imparfaite des formes des objets[197]. Et l'on peut remarquer que les coléoptères, qui n'ont d'autres yeux, volent beaucoup la nuit et ont une vue proverbialement médiocre. Les stemmates[198], qui sont en général au nombre de 3, 6, 8 ou 12, ont bien plus indiscutablement l'aspect d'yeux. Chez les guêpes, ils consistent en apparence en une lentille biconvexe épaisse fermement fixée à la carapace, parfaitement transparente et très dure à l'extérieur, mais plus molle à l'intérieur ; derrière semble se trouver une humeur vitreuse, et probablement derrière elle y a-t-il une rétine. Ici nous devons considérer la lentille cristalline comme unie à la cornée, sans uvée ni humeur aqueuse. Dans les yeux réticulés, il n'y a rien qui ressemble à une lentille cristalline. Les stemmates n'ont jamais le moindre mouvement mais ils sont capables, conjointement, d'envisager un champ visuel très large ; et il est possible que la partie postérieure de la lentille des stemmates puisse avoir un pouvoir de changer sa convexité pour percevoir des objets à différentes distances.>[199]

*Je demande l'autorisation de corriger ici une remarque de mon précédent article relative aux légères radiations latérales que je supposais s'avancer depuis la périphérie de l'iris (Phil. Trans. for 1793. p. 178). Après examen plus poussé, je trouve qu'elles sont occasionnées par des réflexions des cils.*[200]

XII. Je vais maintenant récapituler finalement les principaux objets et résultats des recherches que j'ai pris la liberté de détailler si complètement à la Royal Society. Premièrement, la détermination du pouvoir réfringent d'un milieu variable et son application à la constitution de la lentille cristalline. Deuxièmement, la construction d'un instrument pour déterminer après simple inspection la distance focale exacte de tout œil et le remède à ses imperfections. Troisièmement, montrer l'ajustement exact de toutes les parties de l'œil pour voir distinctement la plus grande étendue d'objets possible au même instant. Quatrièmement mesurer la dispersion collective des rayons colorés dans l'œil. Cinquièmement, en immergeant l'œil dans l'eau, démontrer que son accommodation ne dépend pas d'un quelconque changement de courbure de la cornée. Sixièmement, en confinant l'œil aux extrémités de son axe, prouver qu'aucune altération substantielle de sa longueur ne peut avoir lieu. Septièmement, examiner quelle déduction peut être tirée des expériences réalisées jusque-là sur des personnes dépourvues de la lentille ; poursuivre ces recherches selon les principes suggérés par le Dr PORTERFIELD ; et confirmer son opinion quant à {P.83} l'incapacité absolue de telles personnes de changer l'état réfringent de l'organe. Huitièmement, déduire de l'aberration des rayons latéraux un argument décisif en faveur d'un changement de forme du cristallin ; établir, par la quantité de cette aberration, la forme dans laquelle la lentille semble être projetée dans mon propre œil, et le mode selon lequel ce changement doit être produit dans celui de toute autre personne. Et je caresse l'espoir de ne pas être jugé trop prompt à qualifier cette série d'expériences de suffisamment démonstratives[201,202].

---

[196] Govert Bidloo (1649-1713) est un médecin néerlandais, professeur d'anatomie et de médecine à l'université de Leyde et membre de la Royal Society.

[197] L'analogie et ce contraste entre l'ouïe et la vue est un schème qui flotte dans la tête de Young à cette époque. Ils sont déjà présents au début de cet article [Young, 1801, 25] et nous verrons qu'ils nous semblent être à l'origine du raisonnement qui mène Young à imaginer la triplicité des récepteurs visuels de la rétine [Young, 1802a, 18-21]. La logique de cette idée transparait d'ailleurs ici, puisque l'un des éléments qui semble justifier la structure de la rétine, c'est justement la capacité de chacun de ses points de convoyer toue une gamme de sensations différentes, quand l'oreille génère elle aussi une large gamme de sensations, mais chacune produite en un point différent de sa partie réceptrice, ne permettant pas dès lors de rendre véritablement une « image sonore » de l'environnement. En somme, la cochlée – du fait de sa structure si élaborée – a nécessairement un volume important, ne permettant pas de résolution spatiale. C'est en cela que Young peut comparer la perception de la lumière par les homards à notre sens de l'ouïe plutôt qu'à celui de la vision. Et nous pensons que dans sa *Théorie de la Lumière et des* Couleurs il se contentera de franchir le pas dans l'autre sens, en déduisant que la rétine est nécessairement tapissée de récepteurs extrêmement simples (beaucoup plus simples que la cochlée) du fait de sa résolution spatiale bien supérieure à celle de l'oreille. Il ne lui reste plus alors qu'à expliquer comment des récepteurs très petits et simples produisent une gamme aussi variée de sensations.

[198] Le mot employé ici en anglais par Young est « stemmata », soit en français « stemmates : n.m.pl. Yeux simples, parfois groupés latéralement, souvent disposés circulairement chez les larves holométaoles ; de structure variable suivant les différents groupes. Un ocelle latéral ; œil simple, ocelle » [Séguy, 1967]. Mais il est probable que Young ne fasse pas la distinction entre les différentes catégories d'ocelles, ou yeux simples – par opposition aux yeux « réticulés », ou composés – que l'on trouve chez certains insectes. La description qu'en donne Young ensuite en est raisonnablement correcte, si ce n'est qu'ils peuvent être en nombre plus varié que ce qu'il affirme ; qu'ils ne sont pas fermés par une lentille biconvexe mais plutôt par une simple cornée, indéformable, rarement capable de former de véritables images sur ses quelques centaines ou milliers de photorécepteurs.

[199] Ce passage apparait dans [Young, 1807, 601-602].

[200] Ce paragraphe est absent de la réédition de 1807 [Young, 1807, 602]. Dans son article de 1793, Young entreprend de répondre à une série de questions sur la vision, à commencer par celle de la cause des radiations que l'on observe autour d'une bougie lorsqu'on l'observe avec les yeux presque fermés. Il attribue alors une partie de ces radiations à des réflexions sur le bord de la pupille [Young, 1793, 178] ; mais il se rétracte donc en 1801 pour les attribuer à des réflexions sur les cils. Cette réponse étant directement corrigée dans la réédition des *Observations sur la Vision* de 1807, il n'était plus nécessaire de faire paraître ce paragraphe à cet endroit.

{P. 85} EXPLICATION DES FIGURES.

Planche II. Fig. 1. Voir Page 28. Prop. III.
Fig. 2. Voir Page 28. Prop. IV.
Fig. 3. Voir Page 28. Prop. V.
Fig. 4-6. A propos de l'optomètre. Voir Page 34.
Planche III. Fig. 7. La forme des extrémités de l'optomètre lorsqu'il est fait de carton. Les ouvertures des épaulements sont là pour tenir la lentille : les extrémités carrées passent en-dessous et sont attachées ensemble.
Fig. 8. L'échelle de l'optomètre. La ligne médiane est divisée en pouces depuis l'extrémité basse. La colonne suivante montre le numéro de la lentille concave requise pour un œil à vue courte ; en regardant à travers la bande coulissante et en observant le nombre en face duquel apparait l'intersection quand elle est la plus éloignée. En observant le lieu d'intersection apparente le plus proche, le nombre requis se trouvera dans l'autre colonne à condition que l'œil ait un pouvoir d'accommodation moyen. À l'autre extrémité, la ligne médiane est graduée pour étendre l'échelle de pouces au moyen d'une lentille de quatre pouces (*10,16 cm*) de focale ; les nombres négatifs impliquant que les rayons tels que ceux qui proviennent d'eux convergent vers un point situé de l'autre côté de la lentille. L'autre colonne montre la distance focale des verres convexes requis par les yeux pour lesquels l'intersection la plus proche apparaît en face des lieux respectifs de ces nombres.
Fig. 9. Une vue latérale de l'optomètre, de la moitié de sa taille.

---

[201] Tscherning juge « curieux de remarquer que Young, parmi ses découvertes, ne cite pas celle qui a eu jusqu'ici la plus grande importance pour l'ophtalmologie pratique, à savoir la découverte de l'astigmatisme » [Tscherning, 1894, 240]. C'est que le texte de Young n'est précisément pas un précis d'ophtalmologie pratique. Et que l'histoire des théories scientifiques montre bien l'impossibilité de leurs créateurs à dire l'avenir et anticiper la descendance de leurs théories ; tant leur élaboration relève de la résolution de problèmes contextuels, dont le sens et la pertinence disparaissent rapidement, quand les lois et conclusions générales de la théorie continueront longtemps d'être utilisées.
[202] Young profite de l'impression de sa conférence initialement présentée oralement pour glisser à sa suite une douzaine de corrections que nous ne reprendrons pas ici, puisque nous les avons déjà intégrées au fil du texte.

Fig. 10. L'aspect des lignes à travers la bande coulissante.

Fig. 11. Méthode de mesure de la grandeur d'une image sur la rétine. Voir page 48.

{P.86} Fig. 12. Échelle diagonale dessinée sur un miroir.

Fig. 13. La méthode d'application d'une lentille avec de l'eau à la cornée.

Fig. 14. L'aspect du spectre produit par pression ; et l'inflexion de lignes droites vues dans les limites du spectre.

Fig. 15. Une illustration de l'agrandissement de l'image qui serait la conséquence d'un allongement de l'œil : les images des chandelles qui, dans un cas tombent sur l'insertion du nerf, dans l'autre cas tombent en-dehors de celle-ci.

Planche IV. Fig. 16. Les formes successives de l'image d'un grand objet distant, tel qu'elle serait délimitée par chaque surface réfringente dans l'œil ; pour montrer comment cette forme coïncide finalement avec la rétine. EG est la distance entre les foyers des rayons horizontaux et verticaux dans mon œil.

Planche V. Fig. 17. Section verticale de mon œil droit vu depuis l'extérieur ; deux fois la taille naturelle.

Fig. 18. Section horizontale, vue de dessus.

Fig. 19. Vue de face de mon œil gauche quand la pupille est contractée ; taille naturelle.[203]

Fig. 20. La même vue quand la pupille est dilatée.

Fig. 21. Contour de l'œil et de ses muscles droits à l'état de repos.

Fig. 22. Changement de forme qui serait la conséquence de l'action de ces muscles sur l'œil et sur la substance adipeuse située derrière celui-ci.

Fig. 23. Échelle du petit optomètre.

Fig. 24. Aspect de quatre images d'une ligne vue par mon œil quand son foyer est le plus court.

{P.87} Fig. 25. Contour de la lentille lorsqu'elle est détendue ; d'après une comparaison des mesures de M. Petit avec le phénomène de mon propre œil, et selon la supposition qu'après la mort on le trouve à l'état de repos.

Fig. 26. Contour de la lentille suffisamment modifiée pour produire la plus courte distance focale.

Fig. 27. Appareil pour déterminer la distance focale de la lentille dans l'eau.

Planche VI. Fig. 28. Diverses formes de l'image dessinée par un pinceau de rayons cylindriques réfracté obliquement par une surface sphérique, lorsqu'ils sont reçus par des plans situés à des distances progressivement plus grandes.

Fig. 29. L'image d'un minuscule objet lumineux maintenu très proche de mon œil.

Fig. 30. Le même aspect lorsque l'on frotte l'œil.

Fig. 31–37. Différentes formes de l'image d'un point brillant à des distances de plus en plus grandes ; le foyer le plus parfait étant comme la Fig. 33, mais beaucoup plus petit.

Fig. 38. Image d'un point très éloigné vu par mon œil droit.

Fig. 39. Image d'un point éloigné vu par mon œil gauche ; étant plus obtus à une extrémité, probablement du fait d'une moindre obliquité de la surface postérieure de la lentille cristalline.

Fig. 40. Combinaison de deux figures similaires à la cinquième variété de Fig. 28 ; Afin d'imiter la Fig. 38.

---

[203] Tscherning remarque que « les mesures que Young indique dans le texte, concordent très bien avec la fig. 19, mais pas avec la fig. 20. Il faut admettre que cette dernière représente l'œil droit, ou que la figure ait été retournée par méprise » [Tscherning, 1894, 122-124].

Fig. 41. Aspect d'un point lumineux distant lorsque l'œil est adapté à un objet très proche.

Fig. 42,44. Ombre de fils parallèles dans l'image d'un point distant, lorsque l'œil est détendu.

Fig. 43,45. Les mêmes ombres rendues courbes par un changement de forme de la lentille cristalline.

{P.88} Fig. 46. L'ordre des fibres du cristallin humain.

Fig. 47. La division des nerfs dans la zone ciliaire ; la sclérotique étant retirée. L'un des nerfs de l'uvée est vu passant par devant et se subdivisant. Du veau.

Fig. 48. Ramifications depuis la marge de la lentille cristalline.

Fig. 49. La zone du cristallin faiblement aperçue à travers la capsule.

Figure 50. La zone est soulevée depuis cet endroit avec les ramifications la traversant jusque dans la lentille de l'œil.

Fig. 51. La zone du cristallin, détachée.

Planche VII. Fig. 52. La zone crénelée et les globules régulièrement arrangés sur le cristallin de la perdrix.

Fig. 53. L'ordre des fibres dans la lentille d'oiseaux et de poissons.

Fig. 54. Les segments de la capsule du cristallin retournés, pour montrer les procès ciliaires détachés. Du veau.

Fig. 55. Partie de la choroïde de la morue, avec sa substance rouge. L'artère centrale pend librement depuis l'insertion du nerf.

Fig. 56. La membrane couvrant l'intérieur de cette substance, soulevée par le tube soufflant de l'air.

Fig. 57. L'aspect de la substance rouge après l'extraction de la membrane.

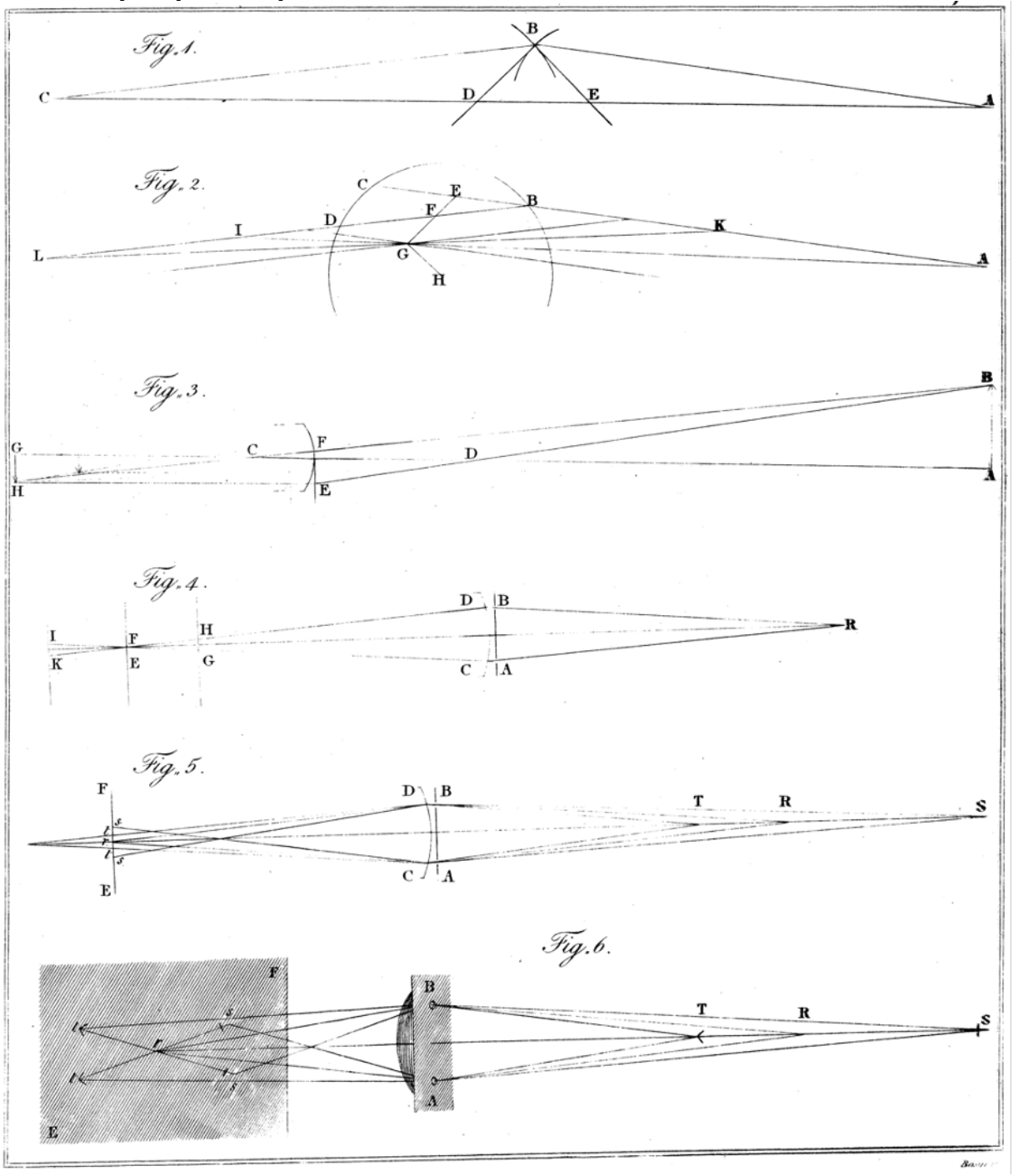

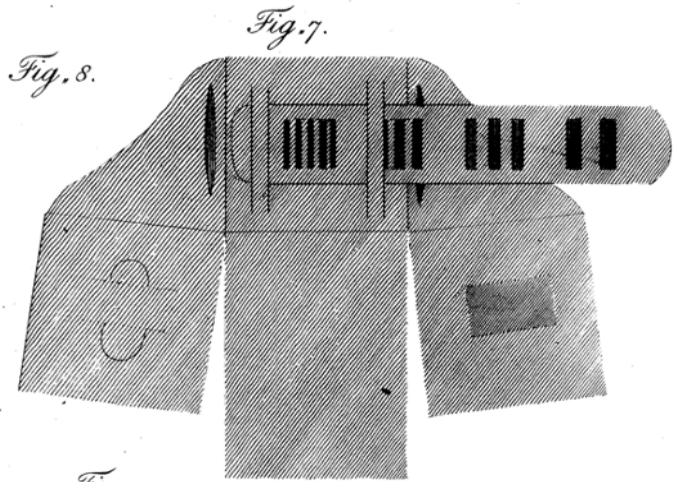
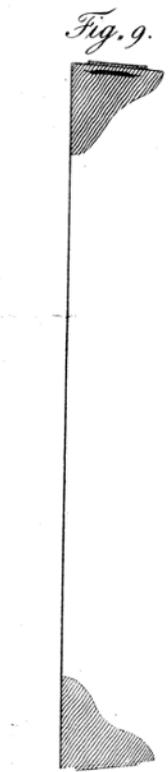
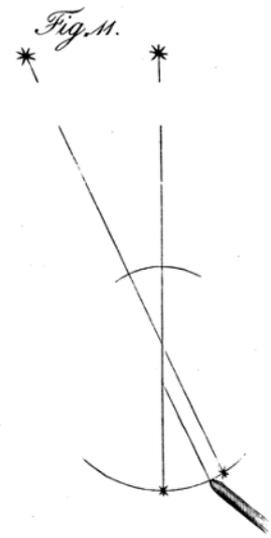
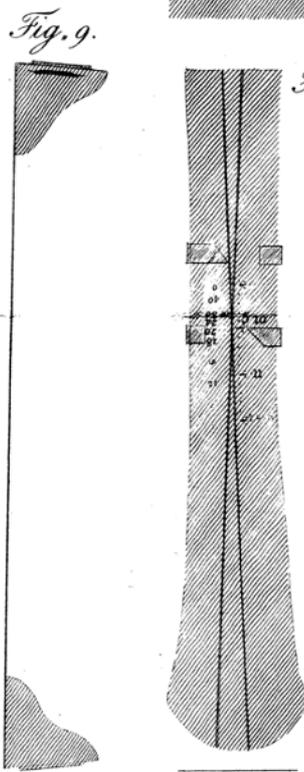
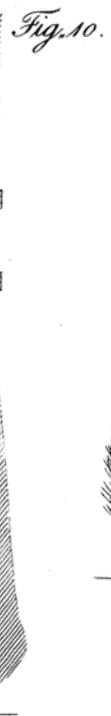
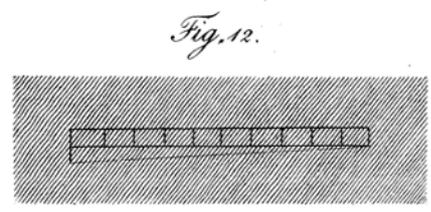
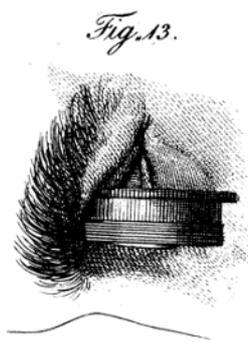
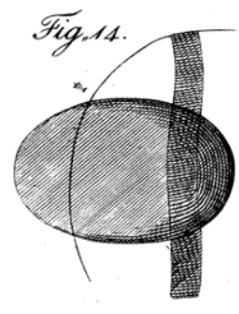
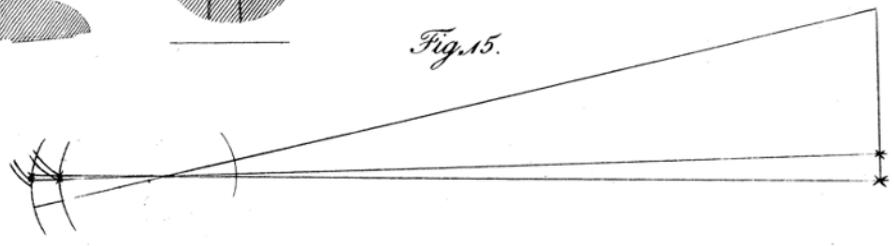

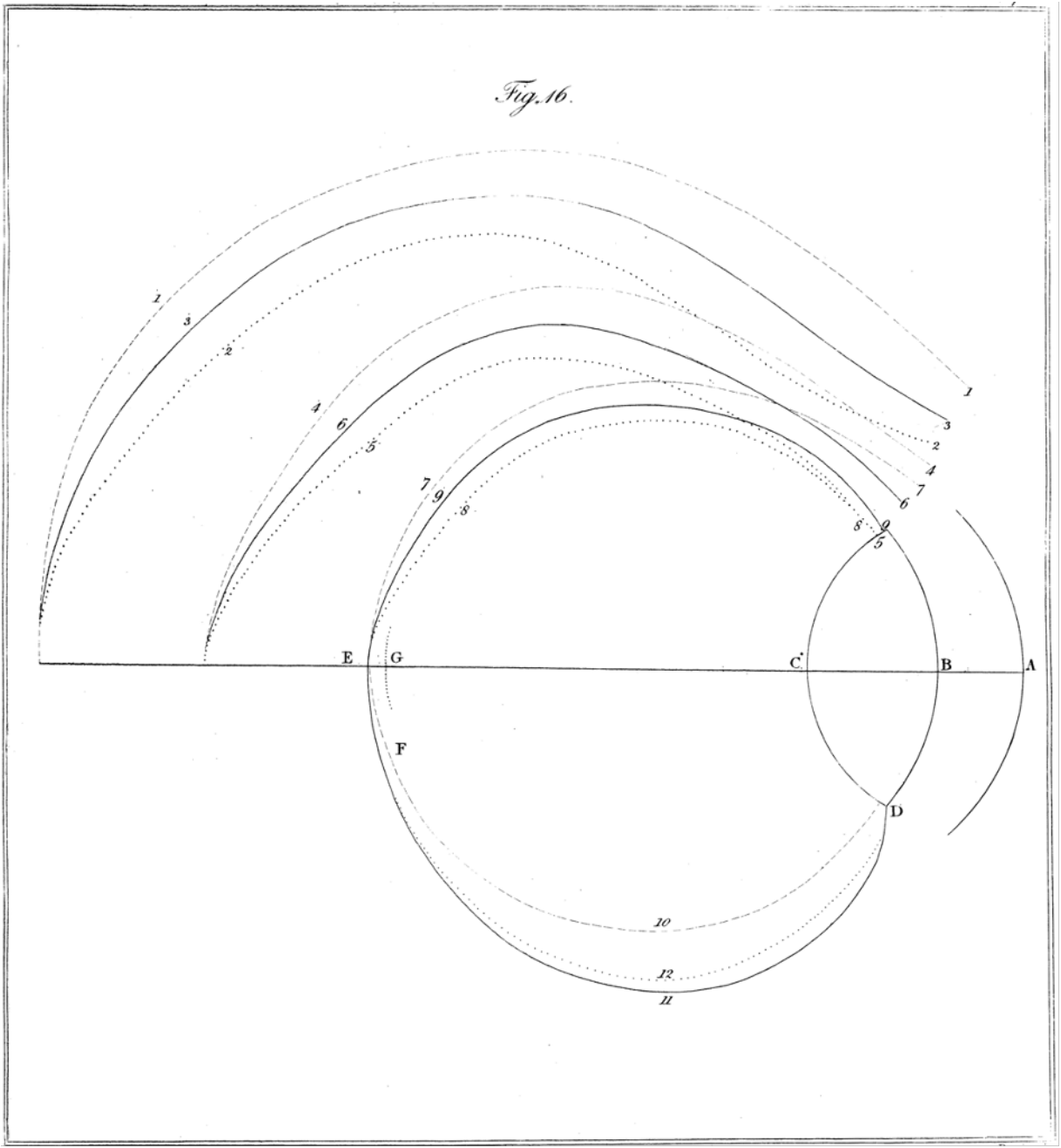

Fig. 16.

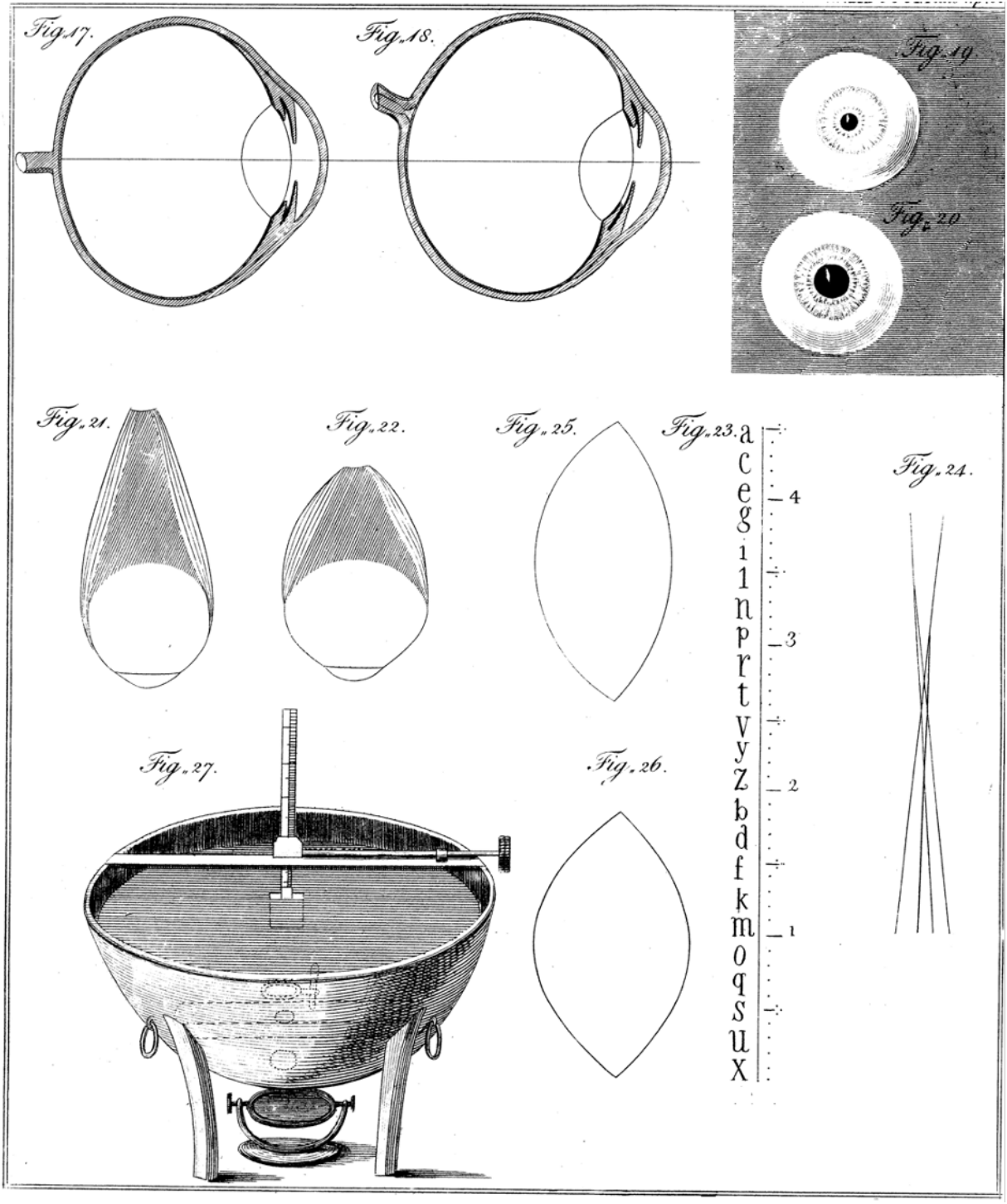